\documentclass[useAMS,usenatbib,breaklinks=true]{mnras}
\usepackage{graphicx}
\usepackage{times}

\usepackage{mn2e-breakabs}

\usepackage{amssymb}
\usepackage{amsmath}
\usepackage{hyperref}
\hypersetup{
    colorlinks=true,
    linkcolor=blue,
    citecolor=blue,
}

\title[Structure growth rate measurement from the DR14 quasar power spectrum]{The clustering of the SDSS-IV extended Baryon Oscillation Spectroscopic Survey DR14 quasar sample: structure growth rate measurement from the anisotropic quasar power spectrum in the redshift range $0.8<z<2.2$}

\author[H. Gil-Mar\'in et al.]{H\'ector Gil-Mar\'in$^{1,2}$\thanks{hector.gilmarin@lpnhe.in2p3.fr}, Julien Guy$^{2,3}$,\,Pauline Zarrouk$^{4}$,  Etienne Burtin$^{4}$,  \and Chia-Hsun Chuang$^{5,6}$, Will J. Percival$^{7}$, Ashley J. Ross$^8$, Rossana Ruggeri$^{7}$, \and Rita Tojerio$^{9}$, Gong-Bo Zhao$^{7,10}$, Yuting Wang$^{10}$, Julian Bautista$^{7,11}$,  Jiamin Hou$^{12}$, \and Ariel G. S\'anchez $^{12}$ , Isabelle P\^aris$^{13}$, Falk Baumgarten$^{14,15}$, Joel R. Brownstein$^{11}$,  \and Kyle S. Dawson$^{11}$, Sarah Eftekharzadeh$^{16}$, Violeta Gonz\'alez-P\'erez$^7$, Salman Habib$^{17}$, \and Katrin Heitmann$^{17}$,   Adam~D.~Myers$^{16}$,  Graziano Rossi$^{18}$, Donald P. Schneider$^{19}$,  \and Jeremy L. Tinker$^{20}$ \& Cheng Zhao$^{21}$\\
$^1$ Sorbonne Universit\'es, Institut Lagrange de Paris (ILP), 98 bis Boulevard Arago, 75014 Paris, France \\
$^2$ Laboratoire de Physique Nucléaire et de Hautes Energies, Universit\'e Pierre et Marie Curie, 4 Place Jussieu, 75005 Paris, France \\
$^3$ Lawrence Berkeley National Laboratory, 1 Cyclotron Road, Berkeley, CA 94720, USA \\
$^4$ IRFU,CEA, Universit\'e Paris-Saclay, F-91191 Gif-sur-Yvette, France\\
$^5$ Leibniz-Institut f\"ur Astrophysik Potsdam (AIP), An der Sternwarte 16, D-14482 Potsdam, Germany \\
$^{6}$ Kavli Institute for Particle Astrophysics and Cosmology \& Physics Department, Stanford University, Stanford, CA 94305, USA\\
$^7$ Institute of Cosmology \& Gravitation, Dennis Sciama Building, University of Portsmouth, Portsmouth, PO1 3FX, UK \\
$^8$ Center for Cosmology and Astro-Particle Physics, Ohio State University, Columbus, Ohio, USA \\
$^{9}$ School of Physics and Astronomy, University of St Andrews, St Andrews, KY16 9SS, UK\\
$^{10}$ National Astronomy Observatories, Chinese Academy of Science, Beijing, 100012, P.R. China \\
$^{11}$ Department Physics and Astronomy, University of Utah, 115 S 1400 E, Salt Lake City, UT 84112, USA \\
$^{12}$ Max-Planck-Institut f\"ur Extraterrestrische Physik, Postfach 1312, Giessenbachstr., 85748 Garching bei M\"unchen, Germany \\
$^{13}$ Aix-Marseille Universit\'e, CNRS, LAM (Laboratoire d'Astrophysique de Marseille), 38 rue F. Joliot-Curie 13388 Marseille Cedex 13, France\\
$^{14}$ Leibniz-Institut f\"ur Astrophysik Potsdam (AIP), An der Sternwarte 16, D-14482 Potsdam, Germany \\
$^{15}$ Humboldt-Universit\"at zu Berlin, Institut f\"ur Physik, Newtonstrasse 15, D-12589 Berlin, Germany \\
$^{16}$ Department of Physics and Astronomy, University of Wyoming, Laramie, WY 82071, USA.\\
$^{17}$ HEP and MCS Divisions, Argonne National Laboratory, Lemont, IL 60439, USA \\
$^{18}$ Department of Astronomy and Space Science, Sejong University, Seoul 143-747, Korea \\
$^{19}$ Department of Astronomy and Astrophysics \&  Institute for Gravitation and the Cosmos, The Pennsylvania State University, University Park, PA 16802, USA\\
$^{20}$ Center for Cosmology and Particle Physics, Department of Physics, New York University, New York, NY 10003, USA \\
$^{21}$ Tsinghua Center for Astrophysics and Department of Physics, Tsinghua University, Beijing 100084, China \\
 }

\date{Accepted XXX. Received YYY; in original form \today}

\pubyear{2016}

\begin{document}
\label{firstpage}
\pagerange{\pageref{firstpage}--\pageref{lastpage}}
\maketitle
\begin{abstract}
We analyse the clustering of the Sloan Digital Sky Survey IV extended Baryon Oscillation Spectroscopic Survey Data Release 14 quasar sample (DR14Q). We measure the redshift space distortions using the power spectrum monopole, quadrupole and hexadecapole inferred from 148,659 quasars between redshifts 0.8 and 2.2 covering a total sky footprint of 2112.9 deg$^2$. We constrain the logarithmic growth of structure times the amplitude of dark matter density fluctuations, $f\sigma_8$, and the Alcock-Paczynski dilation scales which allow constraints to be placed on the angular diameter distance $D_A(z)$ and the Hubble $H(z)$ parameter. 
At the effective redshift of $z_{\rm eff}=1.52$, $f\sigma_8(z_{\rm eff})=0.420\pm0.076$,  $H(z_{\rm eff})=[162\pm 12]\, (r_s^{\rm fid}/r_s)\,{\rm km\, s}^{-1}{\rm Mpc}^{-1}$, and  $D_A(z_{\rm eff})=[1.85\pm 0.11]\times10^3\,(r_s/r_s^{\rm fid})\,{\rm Mpc}$, where $r_s$ is the comoving sound horizon at the baryon drag epoch and the superscript `fid' stands for its fiducial value. The errors take into account the full error budget, including systematics and statistical contributions.  These results are in full agreement with the current $\Lambda$-Cold Dark Matter ($\Lambda$CDM)  cosmological model inferred from Planck measurements. 

Finally,  we compare our measurements with other eBOSS companion papers and find excellent agreement, demonstrating  the consistency and complementarity of the different methods used for analysing the data.

\end{abstract}

\quad

\section{Introduction}

The large-scale structure of the Universe encodes a significant amount of information on how the  late-time Universe has evolved since the accelerated expansion became the dominant component of the cosmos at $z\lesssim2$. One way to access this information is through spectroscopic observations of dark matter tracers, such as galaxies, quasars or inter-galactic gas. Measuring the correlation function of these tracers allows to infer the distribution of dark matter on the Universe and to constrain cosmological parameters such as the matter density of the Universe, namely $\Omega_m$, how gravity behaves at large scales, or to put constraints in the total neutrino masses and its effective number of species.

Two complementary approaches to extract such information are the Baryon Acoustic Oscillations (BAOs) and Redshift Space Distortions (RSDs). The BAO technique  measures the BAO peak position of the observed tracer to infer the evolution of the Universe since the epoch of recombination, when the BAO peak was imprinted in the matter distribution. The BAO signal was detected on the galaxy distribution for the first time in the Sloan Digital Sky Survey (SDSS) \citep{Eis05} and in the 2-degree Field Galaxy Redshift Survey (2dFGRS) \citep{Cole05}. The RSD technique \citep{Kaiser:1987} examines the information of the radial component of the peculiar velocity field and the corresponding distortion in the position of tracers in redshift-space. Such distortions contain information about how  gravity  behaves at inter-cluster scales ($\gtrsim10~\rm{Mpc}$) as well as the  total  matter  content of the Universe. Since the distortions caused by the peculiar velocity field are coherent with the growth of structure, the RSD technique is sensitive to the matter content and to the model of gravity of the Universe. 

The extended-Baryon Oscillation Spectroscopic Survey (eBOSS) \citep{Dawsonetal:2016}, part of the SDSS-IV experiment \citep{Blantonetal:2017} has been constructed, in part, to measure redshifts for approximately $500,000$ quasars at $0.8<z<2.2$ (\citealt{Myersetal:2015}, including spectroscopically confirmed quasars previously observed in the SDSS-I/II/III). Compared to previous SDSS large-scale projects, the eBOSS quasar sample presents relatively low number density of objects, which for the current data release 14 (DR14, \citealt{Abolfathietal:2017}), oscillates typically between $1\times$ and 2 $\times 10^{-5}\,[{\rm Mpc}/h]^3$. However, eBOSS will compensate for this drawback by covering a large volume of the Universe in a redshift-range which has been barely unexplored to date by any spectroscopic survey. 

The selection of quasars in eBOSS uses two different techniques: {\it i)} a `CORE' sample uses a Bayesian technique called XDQSOz \citep{Bovyetal:2012} which selects from the SDSS optical {\it ugriz} imaging combined with mid-IR imaging from the WISE satellite; {\it ii)} a selection based on variability in the multi-epoch imaging from the Palomar Transient Factory (e.g. \citealt{Palanque-Delabrouilleetal:2016}). A full description of these selection techniques is presented in \cite{Myersetal:2015}, alongside the characterisation of the final quasar sample, as determined by the early data. These early data were observed as a part of SEQUELS (Sloan Extended QUasars, ELG and LRG survey),  part of SDSS-III and -IV, which acted as a pilot survey for eBOSS \citep{Dawsonetal:2013,Rossetal:2011}.

Recently, \cite{Ataetal:2017} measured the isotropic BAO scale using the same DR14 quasar sample (DR14Q). In the present paper we describe a complementary analysis based on RSD which extends the anisotropic signal to the previous BAO analysis. In particular, we measure the power spectrum monopole, quadrupole, and hexadecapole from the DR14Q sample in the redshift range $0.8<z<2.2$. We perform the following complementary analyses: {\it i)} we examine the whole redshift bin and perform the measurement of parameters of cosmological interest at the effective redshift, $z_{\rm eff}=1.52$; {\it ii)} we explore the cosmological constraints by setting the ratio of parameters $\alpha_\parallel/\alpha_\perp$ to be 1 (see Eqs. \ref{eq:apara} and \ref{eq:aperp} for definitions) or to leave them as free parameters; {\it iii)} we use three different redshift estimates, based on different features of the quasar spectra; {\it iv)} we separate the full redshift range in three overlapping redshifts bins, {\it lowz} between $0.8\leq z\leq1.5$; {\it mid-z} between $1.2\leq z\leq1.8$ and  {\it high-z}  between $1.5\leq z\leq2.2$, and measure cosmological parameters in each of these three redshift bins, where the correlation among the parameters at different redshift bins is also computed. In all cases, we focus on measuring the logarithmic growth of structure times the amplitude of dark matter density fluctuations, $f\sigma_8(z)$. For those analyses where $\alpha_\parallel$ and $\alpha_\perp$ are treated as free independent parameters, we also measure the angular diameter distance, $D_A(z)$ and Hubble parameter, $H(z)$.

This paper is structured as follows. In \S~\ref{sec:data} we describe the dataset used in the paper, including how the actual quasars have been targeted, their redshifts estimated, and also the techniques to produce the quasar mocks used in this paper. In \S~\ref{sec:methodology} we present the methodology of our analysis, how the power spectrum multipoles have been measured, and the theoretical model used for measuring the cosmological parameters. In \S~\ref{sec:meas} we present the power spectrum multipoles measurements and how they compare to the mocks and to the best-fitting models. In \S~\ref{sec:systematictests} we perform systematic and robustness tests, using mocks and N-body simulations, in order to evaluate the systematic error budget. \S~\ref{sec:results} displays the final results in terms of cosmological parameters measured from the quasar sample using the analyses described above, and  \S~\ref{sec:cosmo} displays the cosmological implications of our findings.  This paper is presented alongside several companion papers which perform complementary and supporting analyses on the same DR14Q sample. \cite{Houetal:2017} and \cite{Zarrouketal:2017} perform a reciprocal RSD analysis to the one presented in this paper, but in configuration space instead of Fourier space.  \cite{Zhaoetal:2017} and \cite{Ruggerietal:2018b} perform  RSD analyses using a redshift weighting technique, which accounts for a redshift evolution of the cosmological parameters across the considered redshift bin. A more detailed description of these works, along with a comparison on the predicted cosmological parameters, is presented in \S~\ref{sec:consensus}.  Finally in \S~\ref{sec:conclusions} we present the conclusions of this paper.

\section{dataset}\label{sec:data}

We start by describing the DR14Q dataset features in detail, along with the mock catalogues used in this work.

\subsection{SDSS IV DR14 quasar sample}

We review the imaging data that have been used to define the observed quasar sample, which is later selected for spectroscopic observation, how the spectroscopy for each quasar target is obtained, and how the quasars redshifts are measured.  

All the eBOSS quasar targets selected for the DR14Q catalogue (\citealt{Parisetal:2017}) are based on the imaging from SDSS-I/II/III and theWide Field Infrared Survey Explorer (WISE, \citealt{WISE}). We briefly describe these datasets below. SDSS-I/II catalogues \citep{Yorketal:2000} imaged a  7606 deg$^2$ northern and 600 deg$^2$ southern parts of the sky in the {\it ugriz} photometric pass bands \citep{Fukugitaetal:1996,Smithetal:2002,Doietal:2010}, and were released as part of the SDSS DR7 \citep{Abazajianetal:2009}. The SDSS-III catalogues \citep{Eisensteinetal:2011} observed additional photometry in the SGC area, increasing the contiguous footprint up to 3172 deg$^2$, and were released as part of DR8 \citep{Aiharaetal:2011}. Further astrometry improvement of these data was presented in DR9 \citep{Ahnetal:2012}. All the photometric data were collected on the 2.5-meter Sloan Telescope \citep{Gunnetal:2006}, located at the Apache Point Observatory in New Mexico in the USA,  using a drift-scanning mosaic CCD camera \citep{Gunnetal:1998}. The eBOSS project does not add any extra imaging area to that released in DR8, although it takes advantages of upgraded photometric calibrations of these data, so-called ``uber-calibration''   \citep{Padmanabhanetal:2008,Schlaflyetal:2012}, released under the name of SDSS DR13 \citep{Albaretietal:2016}. In addition, the WISE satellite \citep{WISE} observed the full sky using four infrared channels centred at $3.4\mu m$ (W1), $4.6\mu m$ (W2), $12 \mu m$ (W3) and $22 \mu m$ (W4), and the eBOSS quasar sample makes use of W1 and W2 band for its targeting.

The quasar target selection criteria for eBOSS is presented in \cite{Myersetal:2015}. Objects that fulfill  this criteria and without any previously known and secure redshift measurements are flagged as ``QSO EBOSS CORE'', selected for spectroscopic observation and assigned an optical fibre. The spectroscopic observation are performed using the BOSS double-armed spectrographs \citep{Smeeetal:2013}, which cover the wavelength range $3,600\leq \lambda [{\rm \AA}] \leq 10,000$, with $R=1500$ up to 2600. The description on how the pipelines process the data from a CCD-level to a 1D spectrum level, and eventually to the measurement of the redshift are described in \cite{ Albaretietal:2016} and \cite{Boltonetal:2012}. The sources of redshifts are divided into three classes: {\it i)} Legacy, where the quasar redshifts are obtained by SDSS I/II/III via non-eBOSS related programs, {\it ii)} SEQUELS, where the redshifts are obtained from the Sloan Extended QUasar, ELG and LRG program (SEQUELS, \citealt{sequels}), {\it iii)} eBOSS, for those previously unknown quasar redshifts obtained by the eBOSS project. The eBOSS quasar redshifts, represent more than 75\% of the redshifts in the current DR14Q catalogue. For further details on the imaging data, target selection criteria and the final construction of the DR14Q catalogues we refer the reader to \citealt{Parisetal:2017}.

\subsubsection{Redshift measurements}\label{sec:redshifts}

One of the main challenges of using quasars as dark matter tracers is the reliability of their spectral classification and consequently their redshift estimation. Although the typical quasar spectrum has wide and prominent emission lines, the existence of quasars outflows may produce systematic shifts in the location of the broad emission lines, which may lead to uncorrected errors in the measurements of their redshifts \citep{Shenetal:2016}. Therefore, having an accurate measurement of quasar redshift is key for achieving the scientific goals of SDSS-IV/eBOSS. For the present DR14Q catalogue we use a number of different redshift estimates to test the impact of these  potential systematics in the final scientific outcome. 

The large number of quasar targets in the current DR14Q catalogue makes the systematic visually inspection procedure (used in the previous SDSSIII/BOSS Ly$\alpha$ analyses) unfeasible. However, the observations taken on the sub-program SEQUELS were all visual inspected, which tested the performance of the automated classification used in the whole DR14Q. The automated pipeline was able to securely classify 91\% of the quasar spectra targeted for clustering studies; less than 0.5\% of these classifications were found to be false when visually examined \citep{Dawsonetal:2016}. Among the remaining 9\% of objects, which the automated pipeline failed to report a secure classification, approximately half were identified as quasars when they were visually inspected. As described in \citealt{Parisetal:2017}, the DR14Q combines automated pipeline together with visual inspections results, providing  a variety of value-added information, containing three automated redshift estimates that we consider in this paper, $z_{\rm PL}$, $z_{\rm PCA}$ and $z_{\rm MgII}$. 
\begin{itemize}
\item The $z_{\rm PL}$ automated classification uses a Principle Component Analysis (PCA) decomposition of galaxy and quasar templates \citep{Boltonetal:2012}, alongside a library of stellar templates, to fit a linear combination of four eigenspectra to each observed spectrum. The reference sample for these redshift estimates are visually inspected quasars from SEQUELS. 
\item The $z_{\rm PCA}$  automated classification uses a PCA decomposition of a sample of quasars with redshifts measurements at the location of the maximum of the MgII emission line, fitting a linear combination of four eigenvectors to each spectrum. In addition, this classification accounts for the potential presence of absorption lines, including broad ones, and it is trained to ignore them.
\item The $z_{\rm MgII}$ automated classification uses the maximum of the MgII emission line at $2799{\rm \AA}$. This broad emission line is in principle less susceptible to the systematic shifts produced by astrophysical phenomena; when a robust measurement of this line is present, it offers a minimally-biased estimate of the systemic redshift of the quasar. Consequently, this method  produces an extremely low number of redshift failures (less than 0.5\%). On the other hand, this method is more susceptible to variations in the signal-to-noise ratio.  When this emission line is not detected in the spectrum of the quasar, the $z_{\rm MgII}$ automated classification uses the $z_{\rm PL}$  prescription.  
\end{itemize}
A comparison of the performance of these redshift estimates is presented in table 4 of \citealt{Parisetal:2017} along with visually inspected redshifts. For the DR14Q we adopt as a standard redshift estimate $z_{\rm fid}$, which consist of any of the three options described above depending on the particular object (see \citealt{Parisetal:2017} for further details), which provides the lowest rate of catastrophic failures. In order to test the robustness of the different redshift estimates, we run at the same time our science pipeline code on the DR14Q using $z_{\rm fid}$, $z_{\rm PCA}$ and $z_{\rm MgII}$, as we did for the BAO analysis in \cite{Ataetal:2017}. 

\subsubsection{DR14Q catalogue details}

The DR14Q catalogue used in this paper (\citealt{Parisetal:2017}) comprises 158,757 objects between $0.8\leq z \leq 2.2$ that the automatic pipeline has classified as quasars. 20,641 of these objects were also visually inspected and confirmed to be quasars and their redshifts were also determined. 148,659 of these quasars have a secure spectroscopic redshift determination and are the objects used in this paper. The remaining objects 10,098, either did not received a spectroscopic fibre or the redshift could not be determined accurately, as we describe below in more detail. 

5,188 objects were photometrically identified as potential quasars, but did not receive a spectroscopic observation. The fibre allocation is designed to that maximise the number of fibres placed on targets considering the constraints of the physical size of the fibres, which correspond to an angle in the focal plane of 62\arcsec, which at $z=1.5$ corresponds to 0.54 Mpc. The fibre-assignment algorithm is therefore sensitive to the target density of the sky, so highly populated regions tend to be covered by several tiles. This overlap of tiles locally resolves some collision (1015 quasars redshifts are identified at less then 62\arcsec  angular separation, 677 in the northern Galactic hemisphere and 338 in the southern). In section \ref{weights:sec} we describe how the unobserved quasar due to fibre collisions are treated. 

\begin{figure}
\centering
\includegraphics[trim={0 90 0 60},clip,scale=0.23]{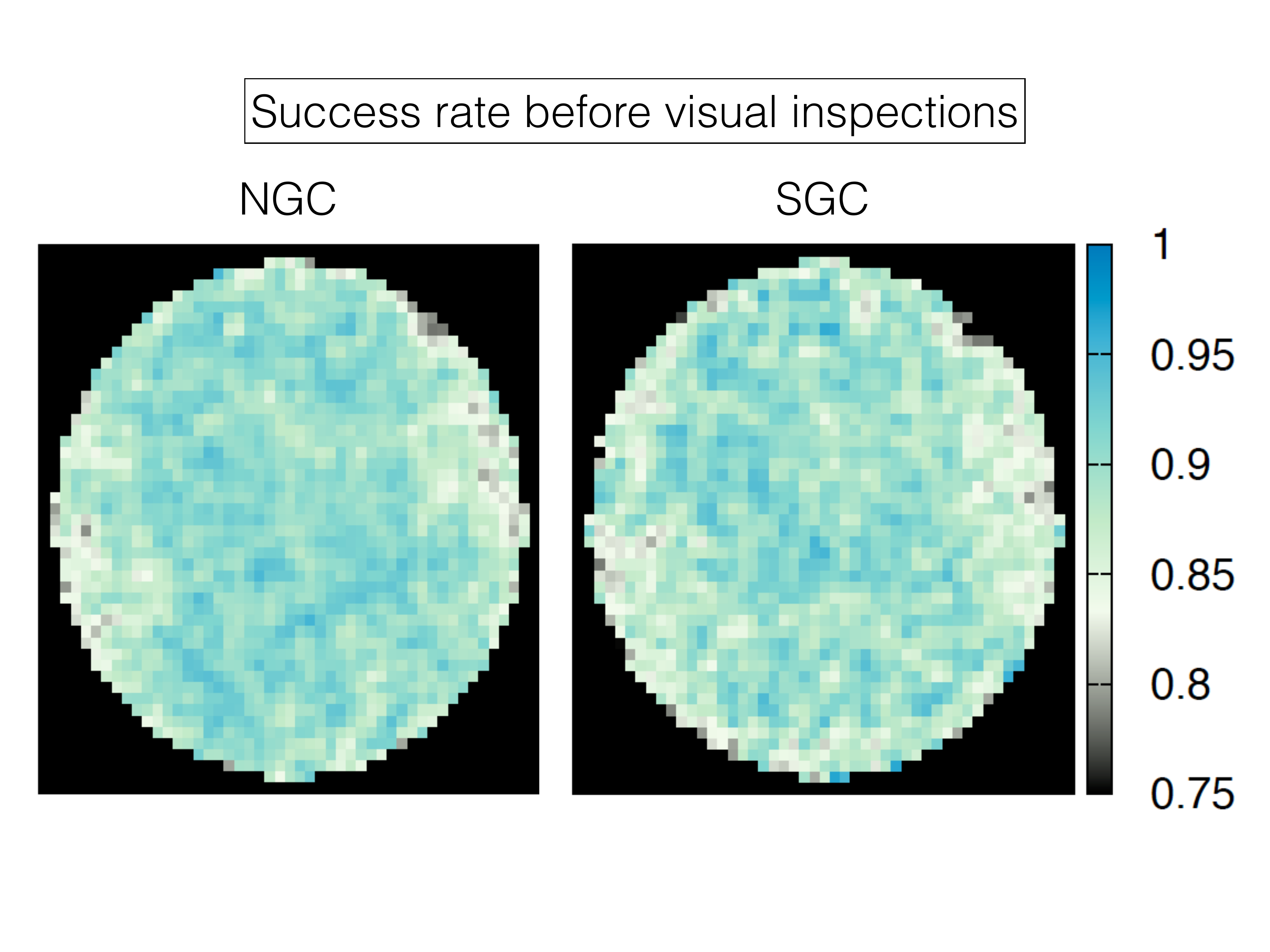}
\includegraphics[trim={0 90 0 0},clip, scale=0.23]{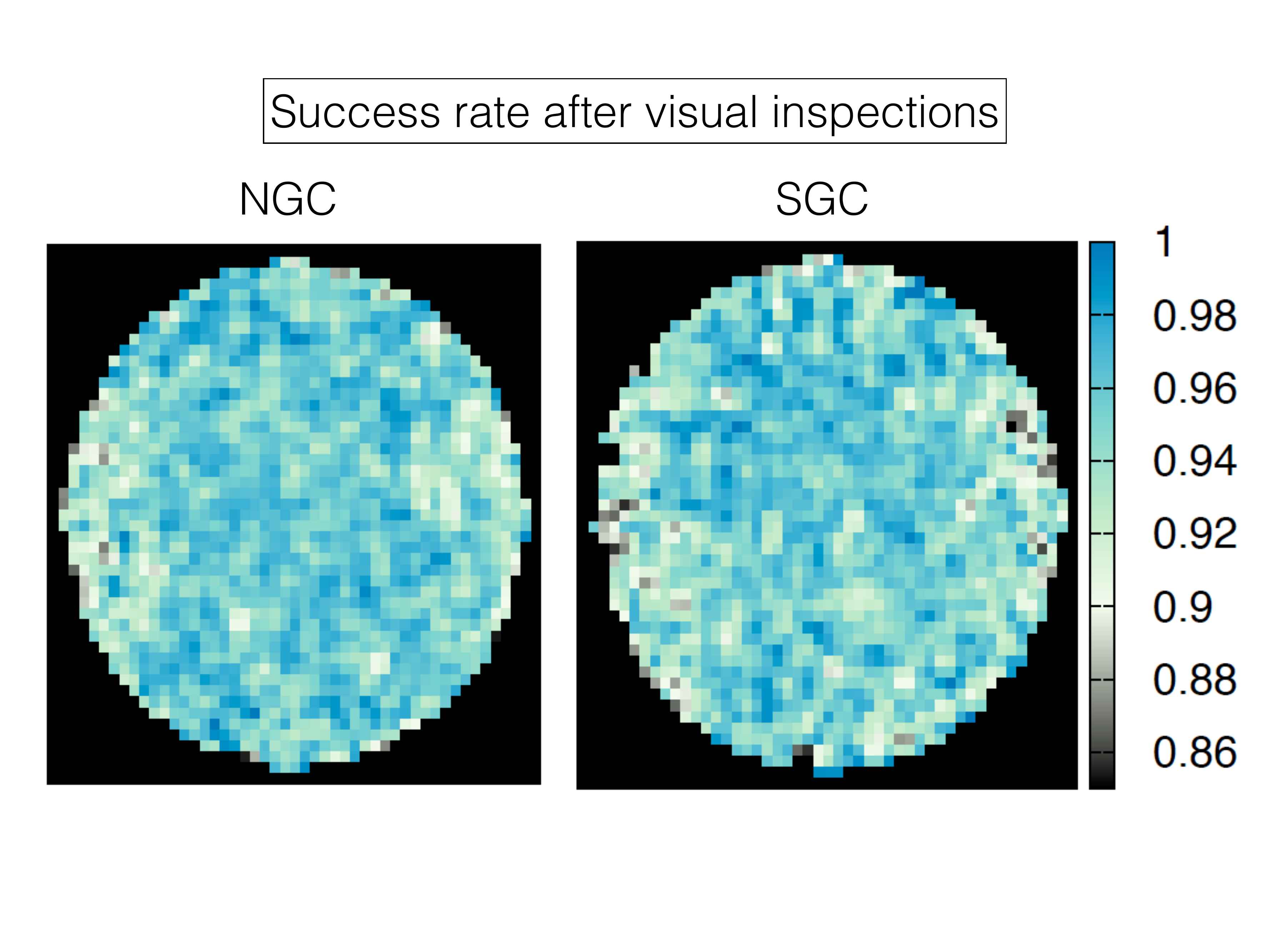}
\caption{Redshift success rate as a function of plate position for the DR14Q catalogue after and before visual inspections, top and bottom panels, respectively. The higher failure rate (lower success rate) in the edges of the plate across the $x$-axis is caused by the less sensitive areas of the detector associated to those plate regions. The higher failure rate in the SGC plates is associated to a poorer photometrie conditions in the SGC compared to those in the NGC. For each tile of the survey, the $x$-axis of the plate is aligned along the iso-declination lines, and the $y$-axis along the iso-right ascension lines, in such a way that the top areas of the plate in the figure correspond to objects with higher declination than the lower areas of the plate. }
\label{fig:failurerate}
\end{figure}

4,910 objects were securely classified by the automated pipeline as quasars, but their redshifts could not be securely determined and did not receive a visual inspection. The distribution of these objects is not uniform across the plate position. We refer to these objects as  ``redshift failure quasars''. In section \ref{weights:sec} we describe how we treat these objects in our analysis. 
Fig.~\ref{fig:failurerate} displays the success rate of securely measuring the redshift of a quasar (number of successfully identified redshifts over total number of objects) as a function of the fibre location in the plate. For each tile in the survey, the vertical axis is aligned to lines of constant declination. The top panels show the success rate produced by the automated pipeline (without any visual inspection), whereas the bottom panels display the success rate after a fraction of the objects were visually inspected. The non-uniform distribution of failure rates across the plate is produced by the non-uniform efficiency of the detectors which record the spectra. The fibres positioned on holes on the left and right edges of the plate are most frequently fibres on the edges of the fibre slit in the spectrographs, corresponding to edges of the spectrograph camera focal plane for which the optical aberrations are larger. The variation of the sensitivity of the spectrograph across its position can reach 5\% \citep{Laurentetal:2017}. 

The observed objects are distributed along an angular footprint (see Fig.~\ref{fig:footprint}) with an effective area of $2112.9\, {\rm deg}^2$, with three disconnected regions: 1 in the Northern Galactic Cap (NGC) whose effective area is $1214.6\, {\rm deg}^2$, and two in the Southern Galactic Cap (SGC), with a total area of $898.3\, {\rm deg}^2$. 
The sub-region of the SGC with declinations $<10\,{\rm deg}$ has an area of $412.2\, {\rm deg}^2$, and the other one has a area of $486.1\, {\rm deg}^2$.

In total, the DR14Q sample contains an effective volume\footnote{We follow the effective volume definition by eq. 5 of \citealt{Tegmark97}.} of $0.246\,{\rm Gpc}^3$ which corresponds to an associated comoving volume of $\sim32\,{\rm Gpc}^3$. The large difference between these two volumes is caused by the factor $\{P_0\bar{n}(r)/[1+P_0\bar{n}(r)]\}^2$ in the effective volume definition. In the case we had a high density number of objects,  $P_0\bar{n}\gg1$, both effective and comoving volume would be similar, as $\{P_0\bar{n}(r)/[1+P_0\bar{n}(r)]\}^2\rightarrow 1$. On the other hand, for the DR14Q sample we have $P_0\sim6\times10^3\,[h^{-1}{\rm Mpc}]^3$ and $\bar{n}\sim 10^{-5}\,[h{\rm Mpc}^{-1}]^3$, and therefore, $P_0\bar{n}\ll1$, indicating that we are in a shot noise dominated regime and the two definitions are substantially different.  The effective volume should be interpreted as the fraction of the associated comoving volume utilised for measuring the power at the wave number whose $P(k)$ is $P_0$. Therefore in terms of Fisher information, the covariance matrices scale according the effective volume. 

All quoted distances in this work correspond to comoving and all quoted volumes are effective volumes according to \cite{Tegmark97} unless mentioned otherwise.

\begin{figure}
\centering
\includegraphics[scale=0.3]{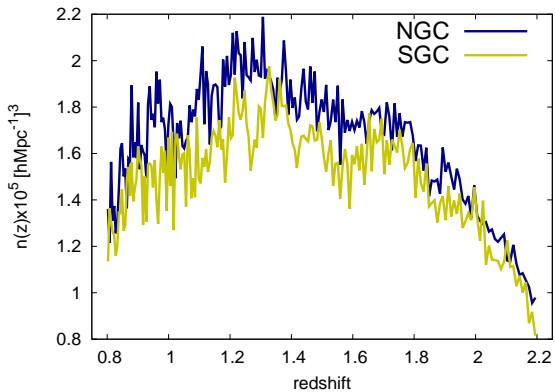}
\caption{Mean density of 148,659 quasars in the DR14Q catalogue as a function of redshift, for the NGC and SGC regions in blue and yellow lines, respectively. The slight difference between the two regions is caused by differences in the target efficiency.}
\label{fig:density}
\end{figure}

\begin{figure*}
\centering
\includegraphics[trim={450 0 -450 0},scale=0.3]{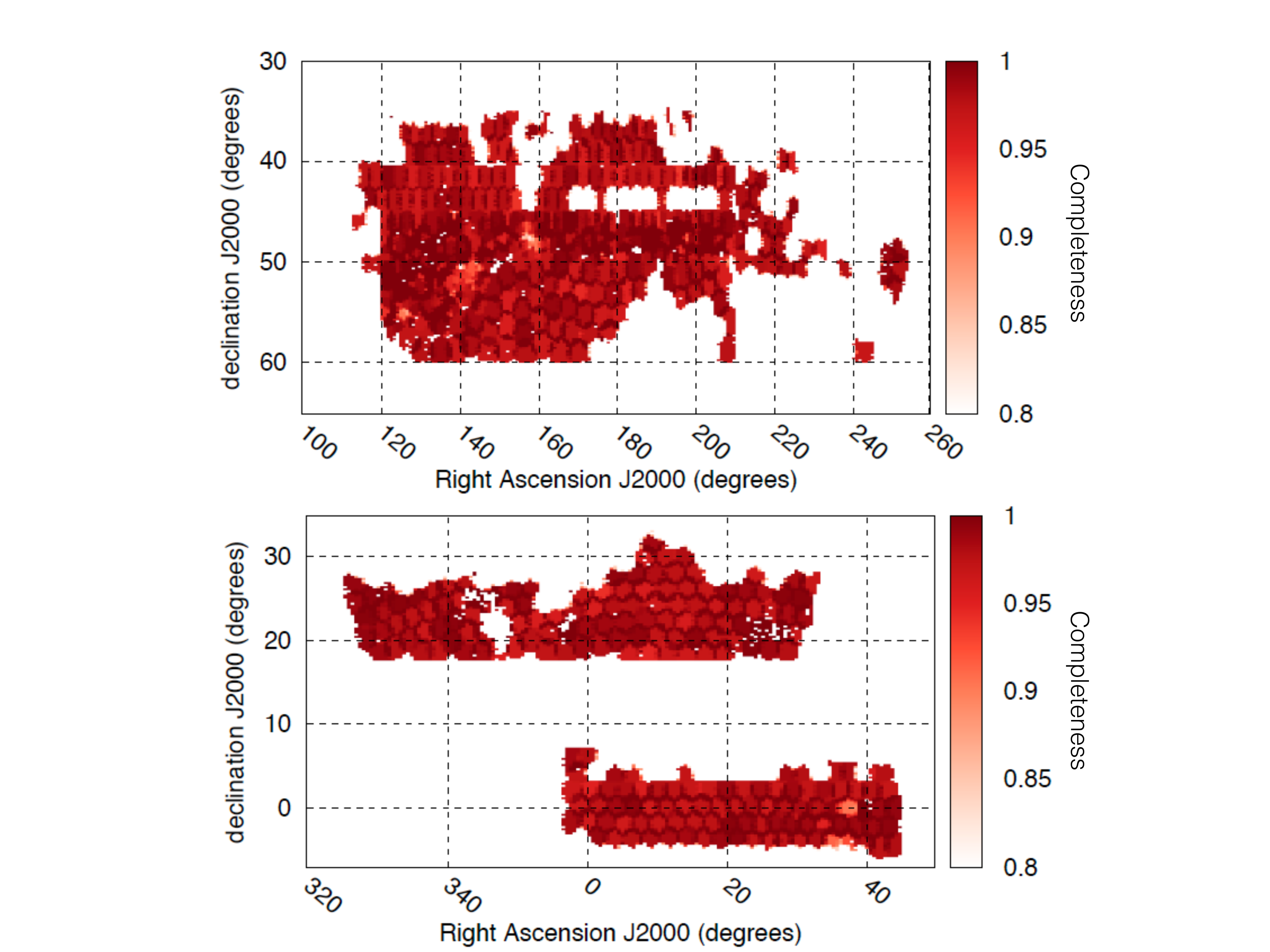}
\includegraphics[trim={-450 -17 450 800},scale=0.3]{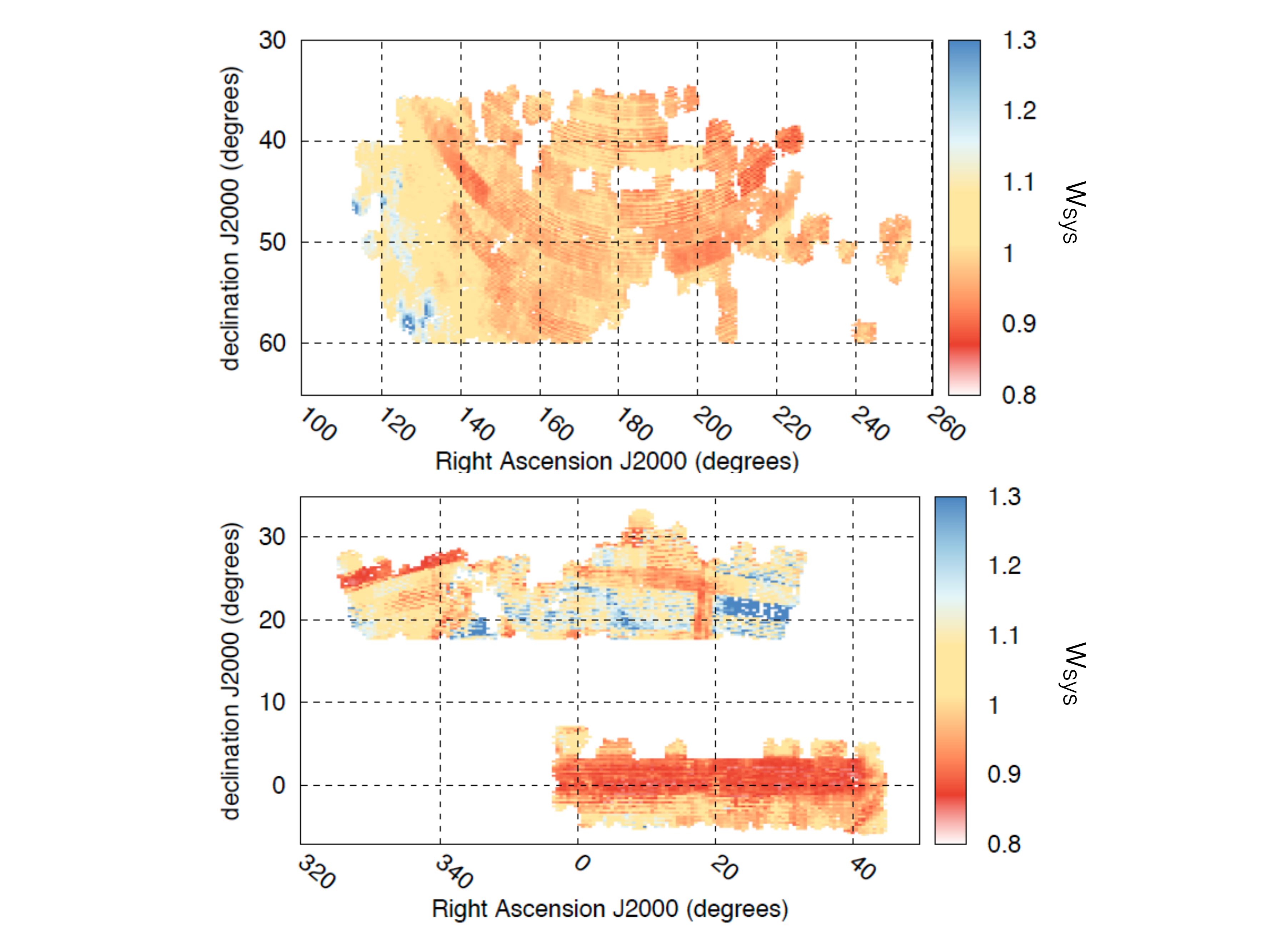}
\caption{Angular footprint of the DR14Q sample for the NGC (top panels) and SGC (bottom panels), where the colour mapping indicates the completeness, $C_{\rm eBOSS}$ (see Eq. \ref{eq:completeness}), and the imaging weight, $w_{\rm sys}$, (see Eq. \ref{eq:wsys}), in the left and right panels, respectively.}
\label{fig:footprint}
\end{figure*}

\subsection{Weights}\label{weights:sec}

The observed density of quasars varies across the analysed redshift range (see. Fig.~\ref{fig:density}). In order to compensate for the different signal-to-noise ratio produced by these variations we weight each observed quasar according to the measured mean density of quasars at that redshift, $\bar{n}(z)$. We refer to this weight as FKP-weight, $w_{\rm FKP}$, and it is defined as \citep{Feldmanetal:1994}, 
\begin{equation}
w_{\rm FKP}(z)\equiv\frac{1}{1+\bar{n}(z)P_0},
\end{equation}
where $P_0=6,000\,[h^{-1}{\rm Mpc}]^3$ is the amplitude of the power spectrum at $k=0.14\,h{\rm Mpc}^{-1}$, which is the typical scale at which the BAO signature in the DR14Q has the highest signal. Ultimately, the statistical gain brought by the FKP-weight is small thanks to the relatively small variation of the mean density across the observed redshift range.

\subsubsection{Spectroscopic weights}\label{sec:spectroweights}
The spectroscopic completeness is mainly affected by two effects, the fibre collisions and redshift failures. We have briefly described these processes above.

The physical size of the optical fibres prevents the observation of two quasars at an angular scale lower than 62\arcsec angular separation using a single tile. This effect is partially mitigated by overlapping tiles in those regions of the sky where the concentration of targets is high. However, we still miss a small fraction of quasars due to this effect.  We account for this effect by up-weighting the lost target to the nearest neighbour with a valid redshift and spectroscopic classification (always within 62\arcsec  unless it has been flagged as a redshift failure). 
This weight is denote as $w_{\rm cp}$, which is 1 by default for all those quasars that have not been up-weighted, and an integer $>1$ for the cases of fibre collisions. In total, 4\% and 3\% of the eBOSS quasar targets are flagged as fibre close pairs in the NGC  and SGC, respectively. A fraction of these up-weighted quasars are true companions of the lost target in physical distance. In these cases the up-weight is physically motivated: we displace the lost target by a small cosmological distance (few Mpc) along the line-of-sight (LOS), which barely distorts the clustering signal. However, for the cases where two targets are not true companions, and the LOS projected distance is large (hundreds of Mpc), moving targets along the LOS does produce a spurious clustering signal along the LOS with respect to the clustering across the LOS. More complex prescriptions based on the probability distribution of the close-pairs along the LOS have been recently presented in the literature \citep{ChangHoonetal:2017}. In this work we do not implement these techniques, which may have a subdominant contribution with respect to the statistical errors, and leave their implementation for future data releases. 

We have shown above that the efficiency in which the redshift of a quasar is inferred depends on its position in the plate. In previous data releases of the BOSS survey, the fraction of objects that were classified as redshift failures was less than $1\%$. For the DR14Q the percentage of failures has increased up to 3.4\% and 3.6\% in the NGC and SGC, respectively, due to the more challenging task of measuring the redshift of a quasar at $z\simeq 1.5$, compared to, e.g., a LRG at $z\simeq 0.5$. In the recent BAO analysis of the DR14Q data \citep{Ataetal:2017} we opted to correct the redshift failures with a similar procedure as the one used to correct for fibre collisions: up-weighting the lost target to the nearest neighbour with a valid redshift and spectroscopic classification, what we designate $w_{\rm noz}$. However, later we will show that this prescription produces a spurious signal in the LOS dependent quantities, such as the quadrupole and hexadecapole, which are later transmitted to systematic shifts on the $f\sigma_8$ value. 

In order to avoid this kind of signal contamination  we opt for a more complex prescription to deal with the redshift failures. We measure the probability of obtaining a redshift failure classification as a function of the plate position by stacking all the measured quasars with good redshift classification divided by the total number of observed quasars. The resulting pattern of success rate is shown in the bottom panels of Fig.~\ref{fig:failurerate} for the NGC and SGC plates, as indicated. We then assign the following weight to all the targets with a valid redshift and spectroscopic classification based on their position on the focal plate, $(x_{\rm foc},\,y_{\rm foc})$, 
\begin{equation}
w_{\rm foc}\equiv\frac{1}{P_{\rm success}(x_{\rm foc},y_{\rm foc})}.
\label{eq:wfoc}
\end{equation}
Those quasars that are placed in regions where $P_{\rm success}<1$ are up-weighted to take into account that, {\it on average}, there are quasars at those  specific regions of the plate that are classified as redshift failures. Later in \S~\ref{sec:systematictests} we will compare how these two prescriptions for correcting the redshift weights perform on a controlled sample, using mock quasars. 
 
The total spectroscopic weight that we apply to the DR14Q catalogues is, 
\begin{equation}
w_{\rm spec}=\frac{w_{\rm cp}}{P_{\rm success}(x_{\rm foc},y_{\rm foc})}.
\end{equation}

\subsubsection{Imaging weights}

We make use of the imaging weights defined in \cite{Laurentetal:2017} and applied to the DR14Q catalogues in \cite{Ataetal:2017}. These weights are required in order to remove the spurious dependency on the $5\sigma$ depth magnitude, known as `depth', and Galactic extinction. \cite{Laurentetal:2017} found that quasars are more securely identified where the value of the depth larger, and Galactic extinction is the variable that most affects differences in depth among the SDSS imaging bands, as they were almost simultaneously observed. The most important observational systematics identified in \cite{Laurentetal:2017} were those related to the depth in the $g$-band magnitude and Galactic extinction, which used the map determined by \cite{Schlegel_Finkbeiner_Davis:1998}. 

The weights used in this paper are the same as those described in section 3.4 of \cite{Ataetal:2017}. Unlike the weights presented in \cite{Laurentetal:2017}, the weights used here are derived from the full DR14 set and the weights are separately defined for the NGC and SGC. As in previous works \citep{Rossetal:2012,Rossetal:2017,Laurentetal:2017}, these weights are derived based on linear fits: first the dependency with the depth and then with Galactic extinction. The total imaging weight is the product of the depth and extinction weights,
\begin{equation}
\label{eq:wsys}w_{\rm sys}=\frac{1}{(A_d+dB_d)(A_e+eB_e)},
\end{equation}
where $d$ is the $g$-band depth and $e$ the Galactic extinction. The best-fitting coefficients, $A_i$ and $B_i$, are the same as those quoted in section 3.4 of \cite{Ataetal:2017}, and are different for NGC and SGC. The right panels of Fig.~\ref{fig:footprint} represent the value of $w_{\rm sys}$ associated to each quasar for the NGC and SGC patches. The $w_{\rm sys}$ is related to the observational quality of each imaging observation, and therefore varies during the observation season. Those areas of the sky observed along the same nights may have similar observational conditions, and the strips in the $w_{\rm sys}$ map are related to the sky scanning followed by the imaging telescope (see \citealt{Gunnetal:1998}). 

Along with the $\rm FKP$ and spectroscopic weights, we weight each object in the DR14Q catalogue with,
\begin{equation}
\label{eq:wtot} w_{\rm tot}=w_{\rm FKP}w_{\rm sys}w_{\rm spec}.
\end{equation}

\subsubsection{Targeting completeness and veto mask}

We define the target completeness of the eBOSS quasar survey by computing the ratio among the objects that have passed the target selection algorithm, $N_{\rm obs}$, over the total number of targets per sector\footnote{Sector is defined as the union of spherical polygons defined by a unique intersection of spectroscopic tiles. See table 1 of \citealt{Reidetal:2016} for further details and definitions.}, $N_{\rm tot}$. The difference among these two quantities is therefore the number of unobserved targets,  $N_{\rm mis}$, which  accounts both for those quasars that have not yet been observed and those that will remain unobserved by SDSS-IV because of a fibre collision with {\it another} target class (the fibre collision among quasar targets are already accounted by the $w_{\rm spec}$ weight). A summary of the different types of targeted objects contained in $N_{\rm obs}$ is described in table 1 of \cite{Ataetal:2017}. Thus,  we define a quasar targeting
completeness per sector as,
\begin{equation}
\label{eq:completeness}C_{\rm eBOSS}=\frac{N_{\rm obs}}{N_{\rm obs}+N_{\rm mis}}.
\end{equation}
The quantity $C_{\rm eBOSS}$ {\rm does not} take into account the targets missed by either fibre collisions or redshift failures, as they are already corrected by up-weighting prescriptions, as described above in \S~\ref{sec:spectroweights}. The $C_{\rm eBOSS}$  quantity is colour mapped along with the survey angular footprint in the left panels of Fig.~\ref{fig:footprint}. The edges of the survey generally contain low values of $C_{\rm eBOSS}$, as those objects are assigned to
tiles-to-be observed by eBOSS in the forthcoming data releases.  
The target completeness of Legacy targets is always 1, as this sample is 100\% complete and has already been observed. We sub-sample the Legacy targets in order to match the $C_{\rm eBOSS}$ value in each sector, following the same procedure used in BOSS \citep{Reidetal:2016}, where 861 and 348 Legacy targets are removed in sectors $C_{\rm eBOSS}>0.5$, in the NGC and SGC, respectively. On the other hand, SEQUELS observations are similar to eBOSS ones, and therefore we treat them in the same way, without any distinction in the DR14Q catalogue. Only sectors with $C_{\rm eBOSS}>0.5$ are included in the final DR14Q catalogue, which discards $<300$ and $<100$ objects in the NGC and SGC footprint, respectively. We also exclude sectors for which the fraction of quasars with secure redshift (redshift completeness sector) is below 0.5, which only represent 20 objects over the two Galactic hemispheres.

We apply a veto mask to the DR14Q catalogue in order to exclude sectors in potentially problematic regions. For the DR14Q catalogues we veto areas under the same conditions than those in BOSS DR12 \citep{Reidetal:2016}. These veto conditions include bad photometric fields, cuts on seeing and on Galactic extinction. Further details on the veto mask areas are described in section of 3.2 of \cite{Ataetal:2017}, we do not repeat them here.

\subsection{DR14Q synthetic catalogues}\label{mock:sec}
 
In this paper we employ three types of synthetic catalogues, constructed to reproduce the observed DR14 quasar sample. We generically refer to them as `mocks',
although they are generated with different techniques and are thus characterised by distinct properties. The first two types of mock catalogues are indicated as the
`Extended Zel'dovich mocks' (or '\textsc{ez} mocks`;  \citealt{Chuangetal:2015}) and the `Quick Particle Mesh' mocks (or `\textsc{qpm} mocks'; \citealt{Whiteetal:2014}). They both consist of hundreds of realisations and are constructed with approximate methods to avoid performing computationally expensive N-body simulations. We use these mocks to estimate the covariance matrix of measured quantities from actual data catalogues, to test our pipeline codes that extract cosmological parameters from the data, and to compute the correlation among parameters inferred at different redshift bins. Tests on our pipeline codes are further refined by a third set of high-fidelity mocks, constructed instead from a high-resolution N-body simulation (the \textsc{OuterRim} simulation, \citealt{Heitmannetal:2014}) . In what follows, we provide a brief description of the main features of all of these mock catalogues.

\begin{table*}
\caption{Expected values of cosmological parameters for the \textsc{qpm} and \textsc{ez}-mocks at different redshift ranges, when analysed using the fiducial cosmology model. }
\begin{center}
\begin{tabular}{ccccccc}
\hline
\hline
type & $z$-range & $z_{\rm eff}$ & $\alpha_{\rm iso}$  & $\alpha_\parallel$ & $\alpha_\perp$ & $f(z)\sigma_8(z)$  \\
\hline
\textsc{ez} & $0.8 -1.5$ & 1.19 & 1.00072   & $1.00179$ & $1.00018$ & $0.41582$ \\
\textsc{ez} &$1.2 - 1.8$ & 1.50 & 1.00100   & $1.00213$ & $1.00043$ & $0.38050$ \\
\textsc{ez} &$1.5 - 2.2$ & 1.83 & 1.00122  & $1.00237$ & $1.00064$ & $0.34642$ \\
\textsc{ez} &$0.8 - 2.2$ & 1.52 & 1.00101  & $1.00215$ & $1.00045$ & $0.37836$ \\
\hline
\textsc{qpm} &$0.8 - 2.2$ & 1.52 & 1.00108  & $1.00108$ & $1.00108$ & $0.36432$ \\
\hline\hline
\end{tabular}
\end{center}
\label{table:mocksalphas}
\end{table*}%

\subsubsection{\textsc{qpm} mocks}\label{sec:qpmmocks}

The \textsc{qpm} mocks follow the procedure described in \cite{Whiteetal:2014}. Briefly, a low-resolution particle mesh gravity solver is used to evolve a density field in time, partially capturing the non-linear evolution of the field, but with insufficient spatial resolution to resolve virialised dark matter haloes. Particles are sampled from the field to approximate the distribution of the small scale densities of haloes, mimicking the one-point and two-point distribution of haloes and their mass and bias functions. We have adjusted the parameters of \cite{Whiteetal:2014} that map the local density into the halo mass in order to account for the actual redshift range of the catalogue, also extending this mapping to lower mass haloes, required by the halo occupation distribution (HOD) of quasars. 

We parametrise the HOD of quasars through the 5-parameter HOD presented in \cite{Tinkeretal:2012}, which divides objects into central and satellite quasars. The HOD parameters are determined by matching  {\it i)} the peak of the $n(z)$ curve observed (see Fig.~\ref{fig:density}) and {\it ii)} the measured large scale quasar bias, $b_Q=2.45$ in \cite{Laurentetal:2017}. This approach also allows the estimation of the fraction of haloes with a quasar object in their centres, usually named the duty cycle. The best-fitting parameters suggest that the satellite fraction is around 0.15 (see Fig. 9 of \citealt{Ataetal:2017}), although there is some expected degeneracy between the satellite fraction and the duty cycle, which remains unknown. 

We simulated 100 cubic boxes of side $L_b = 5120\,h^{-1}{\rm Mpc}$, which we remapped to fit the volume of the full-planned survey using the code \textsc{make survey} (\citealt{Carlson_White:2010} and \citealt{Whiteetal:2014}). Since the DR14Q catalogues correspond to a smaller volume than the mocks, we can use different parts of the \textsc{qpm} cubic box to produce different realisations. We identify four configurations with less than 1.5\% overlap, which allow us to generate 400 \textsc{qpm} realisations per Galactic cap. Since the same 100 cubic boxes are used for the NGC and SGC we need to combine them by shifting the indices of the four realisations produced out of each cubic box. After this action, the overlap among NGC and SGC could be as high as 10\%, although we identified pairs of configurations where the overlap is less than $2\%$. The veto mask and the survey geometry of both Galactic caps are applied also using the code \textsc{make survey}, which down-samples the redshift distributions to match the observed one (Fig.~\ref{fig:density}). Finally, we apply a Gaussian smearing which accounts for the spectroscopic redshift errors \citep{Dawsonetal:2016}, whose Gaussian width is, $\sigma_z=300\,{\rm km}\,s^{-1}$ for $z<1.5$ and $\sigma_z=[400\times(z-1.5)+300]\,{\rm km}\,s^{-1}$ for $z\geq1.5$. Comparisons among \textsc{qpm} mocks and DR14Q measurements are displayed later in the bottom panel of  Fig.~\ref{plot:measurements2}.

The underlying cosmological model in which the density field has been generated and evolved follows a flat $\Lambda {\rm CDM}$ with the following parameters, $\mathbf{\Delta}^{\rm \textsc{qpm}}=\{\Omega_m, \Omega_bh^2,h,\sum m_\nu,\, \sigma_8,n_s\}=\{0.31,0.022,0.676,0,0.8,0.97\}$, where the subscripts $m$, $b$ and $\nu$ stand for the matter, baryon and neutrino, respectively, $h$ is the
standard dimensionless Hubble parameter, $\sigma_8$ is the amplitude of dark matter perturbations, and $n_s$ is the spectral index.  Additionally, other derived parameters, such as the Hubble parameter, the angular and isotropic-BAO diameter distances, and the sound horizon at drag redshift,  are displayed in Table~\ref{table:dv}.

\subsubsection{\textsc{ez} mocks}

Following the methodology described by \cite{Chuangetal:2015}, we generated 1000 \textsc{ez} mock realisations for each Galactic cap, matching the DR14Q footprint and redshift evolution. These mocks are produced via the Zel'dovich approximation of the density field, which is able to account for non-linear effects and also halo bias. In particular, non-linearities and halo bias are modelled through effective free parameters directly calibrated from DR14Q measurements, independently treating the NGC and SGC regions. 
Using this technique we are able to rapidly  generate catalogues which reproduce the 2- and 3-point correlation functions of the desired sample. Each light-cone mock is constructed from seven redshift shells generated from \textsc{ez} mock cubic volumes of $L_b=5000h^{-1}{\rm Mpc}$ at different epochs using \textsc{make survey}. Each of these cubic boxes is computed using different internal parameters, but they share the same initial Gaussian density field, making the background density field continuous. More details on the generation of the \textsc{ez} mocks can be found in section 5.1 of \cite{Ataetal:2017}. Comparisons among \textsc{ez}-mocks and DR14Q measurements are displayed later in the top panel of  Fig.~\ref{plot:measurements2}.

The underlying cosmological model of the \textsc{ez} mocks follows a flat $\Lambda {\rm CDM}$ with the following parameters, $\mathbf{\Delta}^{\rm \textsc{ez}}=\{\Omega_m, \Omega_bh^2,h,\sum m_\nu,\, \sigma_8,n_s\}=\{0.307115,0.02214,0.6777,0,0.8288,0.96\}$.  Other derived parameters, such as the Hubble parameter, the angular and isotropic-BAO diameter distances, and the sound horizon at drag redshift,  are displayed in Table~\ref{table:dv}.

\subsubsection{\textsc{OuterRim} N-body mock}\label{sec:OR}

We perform an accurate systematic test of our pipeline code using a small set of high-fidelity mocks, constructed from a high-resolution N-body simulation.
Unlike \textsc{ez} and \textsc{qpm} mocks, synthetic catalogues directly constructed from N-body simulations fully capture the non-linear signal of the clustering at all scales of interest, and are thus more reliable to assess the validity of our pipeline. Clearly, N-body simulations are expensive to run, but they do contain the correct non-linear dark matter evolution field, and they may be able to resolve dark matter haloes with sufficiently small mass to host quasars,  depending on their actual resolution power. In this work, we use the \textsc{OuterRim} N-body simulation (\textsc{or}, \citealt{Heitmannetal:2014}), a cubic box of size $L_b=3000\,h^{-1}{\rm Mpc}$ with $10240^3$ dark matter particles with a force resolution of $6\,h^{-1}{\rm kp}$, implying a mass resolution per particle $m_{\rm part} = 1.82\times10^9\,h^{-1}M_\odot$; hence, dark matter haloes with sufficient mass to host quasars (i.e., $M = 10^{12.5}M_\odot$) are well-resolved.

\begin{figure}

\includegraphics[scale=0.3]{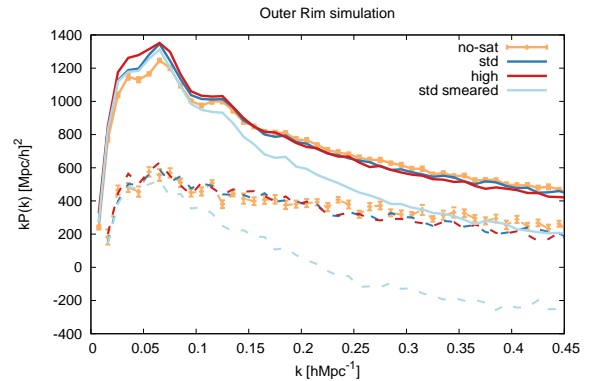}
\caption{\textsc{OuterRim} N-body simulation power spectrum monopole (solid lines) and quadrupole (dashed) lines, computed as the mean of 20 realisations. The colours represent different satellite fractions used, {\it no-sat} with $f=0$ (orange lines), {\it std} with $f=0.13$ (dark-blue lines), and {\it high} with $f=0.22$ (red lines), with no smearing. The light blue lines correspond to the smearing case for the $f_{\rm std}$ satellite fraction. At large scales increasing the satellite fraction increases the amplitude of the monopole, consistent with an enhancement of the linear bias parameter. At small scales the satellites induce a non-linear damping term consistent with the expected by a intra-halo velocity dispersion. This effect is saturated when the redshift smearing effect is included, making it difficult to distinguish among the cases with different fractions at small scales (not plotted for clarity).}
\label{ORplot}
\end{figure}

We construct the \textsc{or}-skycut from a single snapshot at $z=1.433$, applying the same HOD parametrisation used in the \textsc{qpm} mocks \citep{RTetal:2017}, except for the fraction of satellite quasars, which we fix at distinct values to test its effect. The concentration of each halo is determined from its mass using the \cite{Ludlowetal:2014} prescription. The positions and velocities of the satellites are drawn from a NFW profile \citep{Navarroetal:1996}. Finally, the fraction of satellites is chosen to be 0\% ($f_{\rm no-sat}$), 13\% ($f_{\rm std}$) and 22\% ($f_{\rm high}$) and  the fraction used on the \textsc{qpm} mocks HOD is 15\%. Finally, sky geometry cuts are applied so that the final \textsc{or}-skycut derived from the \textsc{or} cubic box covers an angular area of $1888\,{\rm deg}^2$, and the downsampling of objects is performed to match the redshift distribution of the data.  Taking advantage that the DR14Q measurements are shot noise dominated due to the low density of objects, and that the duty cycle for quasars is low, we draw 20 realisations out of the same single parent box, which we consider to be independent. Additionally, to each configuration we do and do not apply a Gaussian smearing in order to mimic the effect of spectroscopic redshift errors, in the same manner done for the \textsc{qpm} mocks. Using this procedure, we generate 20 independent realisations for $3\times 2$ cases, although the realisations are not independent across the different HOD or smearing parameters. Fig.~\ref{ORplot} displays the mean of the 20 measurements of the monopole and quadrupole signal of the \textsc{or}-skycut, for the different satellite fractions, and for the smeared for the $f_{\rm std}$ case.

The underlying cosmological model of the \textsc{or} simulations follows a flat $\Lambda {\rm CDM}$ with the following parameters, $\mathbf{\Delta}^{\rm \textsc{or}}=\{\Omega_m, \Omega_bh^2,h,\sum m_\nu,\, \sigma_8,n_s\}=\{0.26479,0.02258,0.71,0,0.8,0.963\}$, which is consistent with the WMAP7 cosmology \citep{Komatsuetal:2011}.	

Bear in mind that the \textsc{or}-skycuts are derived from a single snapshot at $z=1.433$, which does not match the effective redshift derived from the DR14Q range ($0.8\leq z \leq 2.2$). Since here we are interested in using the \textsc{or} just to perform systematic tests on the model, it is not really important that DR14Q and \textsc{or} match perfectly the redshift range. Because of this freedom, we reduce the redshift range of the \textsc{or}-skycut to be $0.8<z<2.0$, which has an effective redshift of $z_{\rm eff}=1.43$, matching the cubic snapshot epoch. With these parameters, the expected value for $f\sigma_8$ is 0.38216, and the expected values for the $\alpha$s are 1, as we analyse the \textsc{or}-skycut using the simulated cosmology as fiducial cosmology. 

\subsubsection{Synthetic observational features}

We include the fibre collision and redshift failures in the \textsc{ez} and \textsc{qpm} mocks in order to {\it i)} have a more realistic covariance matrices which match the actual number of observed targets, {\it ii)} quantify the systematic shifts (if any) that the weights described in the \S\ref{sec:spectroweights} produce in the cosmological parameters of interest. 

We start by imprinting the same tile distribution of the data in the mocks. In practice, the tile distribution of the data is applied in order to minimise the number of untargeted objects by overlapping the tiles in the densest regions of the survey, which makes the tiling process cluster-dependent. We do not follow the same procedure on the mocks, which would require us to run the same algorithm for every mock, producing a different tiling pattern each time. For simplicity we apply the DR14Q tiling distribution. 

We start by assigning each mock particle to a specific plate. In the case the mock particle falls in an overlap region, it is randomly assigned to an overlapping plate, but with higher probability of falling to the plates whose centre is closer. The collision pair effect is applied to those particles within 62\arcsec  and which both fall into non-overlapping regions (and to those particles that have not been already removed by the close pair selection algorithm). One particle is removed and the other is assigned a +1 $w_{\rm cp}$ weight. The redshift failure effect is applied following the pattern of bottom panels of  Fig.~\ref{fig:failurerate}. We assign the plate coordinates $(x_{\rm foc},\,y_{\rm foc})$ to each mock particles and from those a probability of failing ($1-P_{\rm success}$). The particles tagged as failure are removed from the catalogue. At the end of these two processes, the remaining particles are assigned a $w_{\rm foc}$ weight according to the same pattern, as it is done for the DR14Q catalogue. 

These two processes do not change the effective number of particles ($N_{\rm eff}=\sum_i w_{\rm cp}w_{\rm foc}$), although they remove actual particles from the mocks. Since the covariance matrix of the DR14Q sample is dominated by shot noise, by producing the mocks with the same number of particles that the DR14Q catalogue, the covariances derived from the mocks contain the same level of shot noise.  

\section{Methodology}\label{sec:methodology}

\subsection{Fiducial Cosmology}
We analyse all the \textsc{ez}, \textsc{qpm} mocks and data in a flat, $\Lambda$CDM cosmological model with $\mathbf{\Delta}^{\rm fid}=\{\Omega_m, \Omega_bh^2,h,\sum m_\nu,\, \sigma_8,n_s\}=\{0.31, 0.022, 0.676, 0.06eV, 0.8, 0.97\}$, which matches the fiducial cosmology used for the BOSS DR12 analysis \citep{Alametal2016} and for the eBOSS DR14Q BAO analysis \citep{Ataetal:2017}. The cosmology of the mocks is similar to the chosen fiducial cosmology and,  as a consequence,  the expected shift in the dilation scale factors parameter is $\leq1\%$ 
. 

Table~\ref{table:mocksalphas} displays the expected values for the dilation scale factors and $f\sigma_8$, for \textsc{qpm} mocks and \textsc{ez} mocks when analysed under the fiducial cosmology model. Since the \textsc{ez} mocks light-cone is produced using snapshots at different epochs, we display the expected parameters at the different redshift ranges that are later used in the analysis of the data.

\subsection{Power Spectrum estimator}

We start by defining the function \citep{Feldmanetal:1994}, 
\begin{equation}
F( { r})=w_{\rm tot}[n_{\rm qso}( { r}_i)-\alpha_{\rm ran} n_{\rm ran}( { r}_i)]/I_2^{1/2},
\end{equation}
where $w_{\rm tot}$ is the total weight applied to the quasar sample (see Eq. \ref{eq:wtot}), $n_{\rm qso}$ and $n_{\rm ran}$ are the number density of quasars and random objects, respectively, at position $ { r}_i$, and $\alpha_{\rm ran}$ is the ratio between the weighted number of quasars and randoms. In this work we use 40 times density catalogue for the random catalogue applied to the actual dataset ($\alpha_{\rm ran}=0.025$) and 100 times for the randoms applied to the mocks ($\alpha_{\rm ran}=0.01$). The higher number of random objects in the mock catalogues ensures that the derived-covariance matrix is not dominated by the shot noise of the random catalogue. For \textsc{ez} and \textsc{qpm} mocks, the $w_{\rm tot}$ contains no imaging weight ($w_{\rm sys}= 1$ in Eq. \ref{eq:wtot}).  The normalisation factor, $I_2^{1/2}$, normalises the amplitude of the observed power spectrum in accordance with its definition in a distribution of objects with no survey selection,
\begin{equation}
I_2\equiv \int_{\mathcal{S}}\, d\Omega \int dr\, \langle w_{\rm sys}w_{\rm spec} n_{\rm qso}\rangle^2(r) w_{\rm FKP}^2(r),
\end{equation}
where  $\int_{\mathcal{S}}\, d\Omega$ is the angular integration over all the survey surface of the sky, and results being the effective area of the survey in steradians, and $\langle w_{\rm sys}w_{\rm spec} n_{\rm qso}\rangle$ is the mean number density of quasars. This integration is performed by sampling the mean number density of quasars in radial shells, where we used redshifts bins equivalent to $6.5\,h^{-1}{\rm Mpc}$ in the numerical integration over redshift

In order to measure the power spectrum multipoles of the quasar distribution we begin by assigning the objects of the data and random catalogues to a regular Cartesian grid. This approach allows the use of Fourier Transform (FT) based algorithms. In order to avoid spurious effects of the Cartesian grid we developed a convenient interpolation scheme to convert particle position in grid over-density field.

We embed the full survey volume into a cubic box of side $L_b=7200 h^{-1}{\rm Mpc}$, and subdivide it into $N_g^3=1024^3$ cubic cells, whose resolution and Nyqvist frequency are $7\,h^{-1}{\rm Mpc}$ and $k_{\rm Ny}=0.447\,h{\rm Mpc}^{-1}$, respectively. We assign the particles to the cubic grid cells using a $5^{\rm th}$-order B-spline mass interpolation scheme, where each data/random particle is distributed among $6^3$ surrounding  grid-cells. Additionally, we interlace two identical grid-cells schemes displaced by 1/2 of the size of the grid-cell; this allow us to reduce the aliasing effect below 0.1\% at scales below the Nyqvist frequency (\citealt{HockneyEastwood81}, \citealt{Sefusattietal:2016}).

We follow the Yamamoto estimator \citep{Yamamotoetal:2006}, and in particular the implementation presented by \cite{Bianchietal:2015} and \cite{Soccimarro:2015}, to measure the power spectrum multipoles accounting for the effect of the varying LOS. We proceed by defining the following functions,
\begin{equation}
A_n( { k})=\int d { r}\,(\hat { k}\cdot\hat { r})^n F( { r})e^{i { k}\cdot  { r}}.
\end{equation}
Measuring the monopole, quadrupole, and hexadecapole requires one to consider those cases with $n=0,\,2$. The case $n=0$ can be trivially computed using FT based algorithms, such as \textsc{fftw}\footnote{Fastest Fourier Transform in the West: http://fftw.org}. The $n=2$ case can also be decomposed into 6 FTs by expanding the scalar product between $ { k}$ and $ { r}$ and extractin the $k$-components outside the integral, as it is shown in eq. 10 of \cite{Bianchietal:2015}. From the $A_n$ functions the power spectrum monopole, quadrupole, and hexadecapole read,
\begin{eqnarray}
P^{(0)}( k) &=&\frac{1}{I_2}\int\frac{d\Omega_k}{4\pi}  |A_0( { k})|^2-P_{\rm noise}, \\
P^{(2)}( k)&=& \frac{5}{2I_2}\int\frac{d\Omega_k}{4\pi}  A_0( { k})\left[ 3A^*_2( { k})-A_0^*( { k}) \right], \\
P^{(4)}( k)&=&\frac{9}{8I_2}\int\frac{d\Omega_k}{4\pi} \{ 35A_2 [ A_2^*-2A_0^*]+3|A_0|^2 \}.
\end{eqnarray}
Unless stated otherwise, we perform the measurement of the power spectrum binning $k$ linearly in bins of $\Delta k=0.01\,h {\rm Mpc}^{-1}$ up to $k_{\rm max}=0.30\,h{\rm Mpc}^{-1}$. The resulting power spectrum multipoles for the DR14Q sample are displayed in red and blue symbols in Fig.~\ref{plot:measurements}.

\subsection{Modelling}

The theoretical model used in this paper to describe the power spectrum multipoles is identical to the one used in previous analyses of the BOSS survey for the redshift range $0.15<z<0.70$ (\citealt{BispectrumDR11} and \citealt{RSDDR12}), so we briefly present the model without details to avoid repetition. We refer the reader to the references of this section for a further description. 

\subsubsection{Bias Model}

We assume the Eulerian non-linear bias model presented by \cite{McDonald_Roy:2009}. 
The model has four bias parameters: the linear bias $b_1$, the non-linear bias $b_2$, and two non-local bias parameters, $b_{s^2}$ and $b_{3{\rm nl}}$. As in previous works, we assume $b_1$ and $b_2$ to be free parameters of the model. The remaining two non-local bias parameters can be constrained by assuming that the bias model is local in Lagrangian space, which sets $b_{s^2}$ and $b_{3{\rm nl}}$ as a function of $b_1$: $b_{s^2}=-4/7\,(b_1-1)$ \citep{Baldaufetal:2012} and $b_{3{\rm nl}}=32/315\,(b_1-1)$ \citep{Saitoetal:2014}. 

\subsubsection{Redshift Space Distortions}
We model the redshift space distortions in the power spectrum multipoles following the approach presented by \cite{TNS} (TNS model). We assume that there is no velocity bias between the galaxy field and the underling dark matter field, at least on the scale of interest for this paper. The TNS model provides a prescription for the redshift space power spectrum in terms of the real space quantities: the matter-matter, velocity-velocity and the cross matter-velocity non-linear power spectra. These non-linear quantities are computed using the resumed perturbation theory at 2-loop order as described in \cite{GilMarinetal:2012}. All these non-linear power spectrum quantities are fuelled with the linear matter power spectrum computed using \textsc{camb} \citep{CAMB}. The power spectrum multipoles encode the coherent velocity field through the redshift space displacement and the logarithmic growth of structure parameter, $f\equiv \frac{d\log D(z)}{d\log a(z)}$. The effect of this parameter is to increase the clustering along the LOS with respect to the transverse direction, boosting the amplitude of the isotropic power spectrum and generating an anisotropic component.  

We include a Lorentzian damping factor term of the form, 
\begin{equation}
D(k,\mu;\sigma_P)=(1+[k\mu\sigma_P]^2/2)^{-2},
\end{equation}
which multiplies the theoretical LOS-dependent power spectrum, $P(k,\mu)$. Here, $\mu$ is the cosine of the angle between the galaxy-pair direction and the LOS, and $\sigma_P$ is a free parameter, which may depend on redshift. The physical motivation for this damping factor is to include the effect of Finger-of-God (FoG, \citealt{FoG}): the velocity dispersion of the satellite quasar inside the host dark matter haloes, which damps the power spectrum at small scales. However, other observational features are also included in this parameter, such the spectroscopic redshift errors, whose effect is to produce a broadening of the observed redshift distribution and which has been previously discussed in \S~\ref{sec:qpmmocks} in the context of the \textsc{qpm} mocks. We remind that, the approach described by the equation above is phenomenological, and that other choices for the damping term factor (such as a Gaussian term) are also possible. We choose the Lorentzian damping factor over the Gaussian, as in previous works \citep{BispectrumDR11,RSDDR12} has demonstrated to better reproduce the mock and actual data signal.  

We also consider that the shot noise contribution in the power spectrum monopole may differ from the Poisson sampling prediction. We parametrise this potential deviation through a free parameter, $A_{\rm noise}$, which modifies the amplitude of shot noise, but does not introduce any scale dependence,
\begin{equation}
P_{\rm noise}=(1-10^{-3}A_{\rm noise})P_{\rm Poisson},
\end{equation}
where $P_{\rm Poisson}$ is the Poisson prediction,
\begin{equation}
P_{\rm Poisson}=I_2^{-1} \int d { r}\, \langle n_{\rm qso} w_{\rm sys}w_{\rm spec} \rangle (w_{\rm sys}w_{\rm spec}+\alpha_{\rm ran}),
\end{equation}
and the $10^{-3}$ factor has been conveniently included to make the best-fitting value of $A_{\rm noise}$ close to unity. $A_{\rm noise}=0$ would be consistent with the pure Poisson case, $A_{\rm noise}>0$ with a sub-Poissonian case, typically attributed to halo exclusion (\citealt{MoWhite96} and \citealt{CasasMiranda2002}), and  $A_{\rm noise}<0$ to a super-Poissonian case. For the DR14Q dataset we find $P_{\rm Poisson}=70390.2\,[h^{-1}{\rm Mpc}]^3$. In this paper we consider that the non-Poissonian shot noise only affects the power spectrum monopole, having no effect on the higher order multipoles. We have checked that including the term $[A_{\rm noise}P_{\rm Poisson}]$ in the galaxy power spectrum modelling, $P_g(k,\,\mu)$,  through an additive term on $P_{\delta\delta}$\footnote{This is the approach followed in \citealt{Beutleretal:2014} (see eq. 40).}, and therefore having an impact on all the multipoles, does not change the cosmological parameters, although it produces some changes on the best-fitting bias parameters, $A_{\rm noise}$ and $\sigma_P$. 

\subsubsection{The Alcock-Paczynski effect}
The Alcock-Paczynski effect (AP effect, \citealt{AP}) is produced when converting the observed redshift of galaxies into comoving distance using a different cosmological model than the actual one. As a consequence, an anisotropic signal component in the power spectrum is induced, as the distortion is different in the radial direction with respect to the transverse direction: along the LOS the observed signal is proportional to the inverse of Hubble parameter, $H$; across the LOS the distortion is proportional to the angular diameter distance, $D_A$. When a fiducial model is assumed to convert redshifts into comoving distances, the AP effect can be described through the parallel and perpendicular dilation scales, 
\begin{eqnarray}
\label{eq:apara}\alpha_{\parallel}&\equiv&\frac{H^{\rm fid}(z)r_s^{\rm fid}(z_d)}{H(z)r_s(z_d)}, \\
\label{eq:aperp}\alpha_\perp&\equiv&\frac{D_A(z)r_s^{\rm fid}(z_d)}{D_A^{\rm fid}(z)r_s(z_d)},
\end{eqnarray}
 where, $r_s(z_d)$ is the sound horizon at the baryon-drag epoch and the `fid' index stands for the fiducial cosmology (the one assumed to convert redshifts into distances). 
 
 The dilation scale factors, $\alpha_\parallel$ and $\alpha_\perp$ describe how the true frequencies $k'$ have been distorted into the observed ones $k$, ($k_\parallel=\alpha_\parallel k'_\parallel$ and $k_\perp=\alpha_\perp k'_\perp$), by the effect of assuming an incorrect  cosmological model. 
  
 \begin{table}
\caption{Values of the BAO isotropic distance, $D_V$, the angular diameter distance $D_A$, the Hubble parameter $H$ and the sound horizon at drag redshift, $r_s(z_d)$ for the fiducial, \textsc{ez}- and \textsc{qpm}-mock true cosmology at their effective redshifts,  $z_{\rm eff}=1.52$ for fiducial and \textsc{ez} mocks, and $z_{\rm eff}=1.51$ for \textsc{qpm} mocks. Units of $D_V$, $D_A$ and $r_s$ are in Mpc, whereas $H$ is in ${\rm km}\,s^{-1}{\rm Mpc}^{-1}$}
\begin{center}
\begin{tabular}{|c|c|c|c|c}
\hline
\hline
 & $D_V(z_{\rm eff})$ & $H(z_{\rm eff})$ & $D_A(z_{\rm eff})$ & $r_s$ \\
 \hline
 fiducial & $3871.0$ & $160.70$ & $1794.7$ & $147.78$\\
  \textsc{qpm}-mocks & $3858.5$ & $159.85$ & $1794.4$ & $147.62$\\
  \textsc{ez}-mocks & $3871.8$ & $160.48$ & $1794.1$ & $147.66$ \\
 \hline
 \hline
\end{tabular}
\end{center}
\label{table:dv}
\end{table}%

 Alternatively to $\alpha_\parallel$ and $\alpha_\perp$, we can work with the following combination of variables, 
 \begin{eqnarray}
 \label{eq:aiso}\alpha&\equiv&\alpha_\parallel^{1/3}\alpha_\perp^{2/3}, \\
 \epsilon &\equiv&(\alpha_\parallel/\alpha_\perp)^{1/3}-1.
 \end{eqnarray}
 \cite{Rossetal:2015} demonstrated that $\alpha$ is the optimal variable to be constrained when the information of the monopole is considered alone in the absence of redshift space distortions. This is the situation when the BAO peak position is fitted from monopole without RSD: under these conditions we formally refer to this variable as $\alpha\equiv\alpha_{\rm iso}$. This is the reason why $\alpha_{\rm iso}$ is usually called the isotropic shift, and it is equivalent to the observed BAO shift in the isotropic power spectrum and correlation function (eq. 12 of \citealt{Ataetal:2017}). The parameter  $\epsilon$ corresponds to the anisotropic shift due to the AP dilation scales, and most of its signal arises from the power spectrum quadrupole. In the appendix \ref{sec:APtest} we show that even when the monopole and quadrupole are both taken into account, the combination $\alpha_\parallel^{1/3}\alpha_\perp^{2/3}$ is sufficiently close to the optimal direction in the $\alpha_\parallel-\alpha_\perp$ plane (given the statistical errors), and that therefore, $\alpha_{\rm iso}|_{\epsilon=0}\simeq\alpha_\parallel^{1/3}\alpha_\perp^{2/3}$ is a valid approximation.
 
The measurement on $\alpha_{\rm iso}$ sets constraints on a particular combination of the Hubble parameter and the angular diameter distance, $D_V$, which we usually refer as the spherically average BAO distance,
\begin{equation}
D_V(z)=[cz(1+z)^2H^{-1}(z)D^2_A(z)]^{1/3}.
\end{equation}
and 

\begin{equation}
\alpha_{\rm iso}=\frac{D_V(z) r_s^{\rm fid}(z_d)}{D_V^{\rm fid}(z) r_s(z_d)}.
\end{equation}
Since in this paper we always use (at least) the power spectrum monopole and quadrupole, the measurement of $\alpha_{\rm iso}$ under the prior condition $\epsilon=0$ provides constraints on $f\sigma_8$ and $D_V$ measurements which are not independent from Planck \citep{Planck15}. Indeed, the $\epsilon=0$ condition implies $H(z)D_A(z)=H(z)^{\rm fid}D_A(z)^{\rm fid}$, where the fiducial values are very close to the Planck cosmology. We will return to this point when presenting the results, making clear the prior information of each cosmological derived quantity. 

Table~\ref{table:dv} displays the fiducial values for $D_A$, $H$ and $D_V$ for the different cosmologies used in this paper at their effective redshifts.

\subsubsection{Survey Geometry}

The last step to be included in the model is the effect of the window function produced by the non-uniform distribution of quasars, both angularly and radially. We account this effect by following the procedure described by \cite{Wilsonetal:2017}. In practise the window has two main effects, {\it i)} to reduce the observed power at large scales, being more sever as one increases the value of the $\ell$-multipole is, {\it ii)} to increase the covariance among adjacent $k$-modes, specially at large scales. We follow the same formalism used by \cite{Beutleretal:2016}, which is fully described in Appendix~\ref{sec:surveygeometry}.

\subsubsection{Free parameters of the model}

In addition to the free parameters of the model described above, we also marginalise over the amplitude of the linear power spectrum through the amplitude of the dark matter fluctuations filtered with a top-hat filter of $8\,{\rm Mpc}$, $\sigma_8(z)$. This parameter is highly degenerate with other parameters such as the bias parameters and the logarithmic growth of structure. Thus, we set $\sigma_8$ to the fiducial value of our cosmology in the non-linear terms of the model, and constrain the combination of $\sigma_8$ times the bias parameters or the logarithmic growth of structure from the large scale modes. 

To summarise, the full power spectrum model described in the sections above has seven free parameters: two bias parameters, $b_1\sigma_8$ and $b_2\sigma_8$; two nuisance parameters, $A_{\rm noise}$ and $\sigma_P$; and three cosmological parameters, $f\sigma_8$, $\alpha_\parallel$ and $\alpha_\perp$. For those cases where $\epsilon$ is set to 0, $\alpha_\perp=\alpha_\parallel\equiv\alpha_{\rm iso}$, and the number of free parameters is reduced by one. 

\subsection{Parameter Estimation}\label{sec:parameterestimation}

We start by defining the likelihood distribution, $\mathcal{L}$, of the vector of parameters of interest, $p$, as a multi-variate Gaussian distribution,
\begin{equation}
\mathcal{L}\propto e^{{-\chi^2(p)/2}},
\end{equation}
where $\chi^2(p)$ is defined as,
\begin{equation}
\chi^2(p)\equiv { D_p} C^{-1}  { D_p}^T,
\end{equation}
where $D_p$ is the difference between the data and the model when the $p$-parameters are used, and $C$ represents the covariance matrix of the data vector, which we approximate to be independent of the $p$-set of parameters. 

The covariance matrix is computed using a large number of mock quasar samples described in \S\ref{mock:sec}. For our fiducial results we use the 1000 realisations of the \textsc{ez} mocks, unless otherwise noted. Due to the finite number of mock realisations when estimating the covariance, we expect a noise term to be present which requires a correction to the final $\chi^2$ values. We apply the corrections described in \cite{Hartlap07}. Such corrections represent a $\sim15\%$ factor in the $\chi^2$ values; we use 1000 mock realisations to estimate the full covariance of 84 $k$-bins, including monopole quadrupole and hexadecapole.  We do not apply any extra corrections, such the ones described in \cite{Percival14}, which have a minor contribution to the final errors. 

Using a \textsc{simplex} minimisation algorithm (\citealt{NelderMead}, \citealt{NR}), we  explore the surface of the likelihood function to find the best-fitting value for each of the $p$-parameters and its $1\sigma$ marginalised error. We ensure that the minima found are global and not local by running the algorithm multiple times with different starting points and different variation ranges.

As mentioned above, the value of $\sigma_8$ is set constant to its fiducial value in the non-linear terms of the model, so the parameter $f\sigma_8$ is effectively fitted. We have checked that due to the high degree of degeneracy between $f$ and $\sigma_8$, the impact of following this procedure does not change our results for physical values of $\sigma_8$.

In order to compute the full likelihood surface of a set of parameters, we also run Markov-chains ({\textsc{mcmc}-chains). We use a simple Metropolis-Hasting algorithm with a proposal covariance and ensure its convergence performing the Gelman-Rubin convergence test, $R-1<10^{-3}$, on each parameter. We apply the flat priors listed in Table \ref{table:generalpriors} otherwise stated.

\begin{table}
\caption{Flat priors ranges on the parameters of the model. The priors on $\alpha_\parallel$ and $\alpha_\perp$ are modified when the three redshift bins are analysed as described in Table \ref{table:priors}. Since $\sigma_8$ is fixed to its fiducial value during the fit (see text) the priors are effectively applied to the parameters $f$, $b_1$ and $b_2$. In order to obtain the priors on $f\sigma_8$, $b_1\sigma_8$ and $b_2\sigma_8$ the priors need to be re-scaled by the fiducial value of $\sigma_8$, $0.397$. }
\begin{center}
\begin{tabular}{|c|c|}
\hline
\hline
Parameter & flat-prior range \\
\hline
$\alpha_{\rm iso}$ & $[0,\,2]$ \\
$\alpha_{\rm \parallel}$ & $[0,\,2]$ \\
$\alpha_{\rm \perp}$ & $[0,\,2]$ \\
$f$ & $[0,\,5]$ \\
$b_1$ & $[0,\,5]$ \\
$b_2$ & $[-10,\,10]$ \\
$\sigma_P$ & $[0,\,30]$ \\
$10^{-3}A_{\rm noise}$ & $[-1,\,1]$ \\

\hline
\end{tabular}
\end{center}
\label{table:generalpriors}
\end{table}%

\section{Measurements}\label{sec:meas}
In this section we present the measurement of the power spectrum multipoles of the DR14 quasar sample, as well as the performance of the model and the mocks. We start by discussing the measurements in the whole redshift bin, $0.8\leq z\leq2.2$; and we later divide the full redshift range into three overlapping redshift bins: {\it lowz}, $0.8\leq z \leq 1.5$; {\it midz}, $1.2\leq z \leq 1.8$; {\it highz}, $1.5 \leq z \leq 2.2$. Following this second approach we are in principle sensitive to redshift-evolution quantities, such as $b_1(z)\sigma_8(z)$ and $f(z)\sigma_8(z)$. In \S~\ref{sec:cosmo} we will explore how the two different approaches, the single and multiple redshift bins, performed when constraining cosmological parameters. 

\subsection{Single redshift bin}\label{sec:singlez}

\begin{figure}
\centering
\includegraphics[scale=0.30]{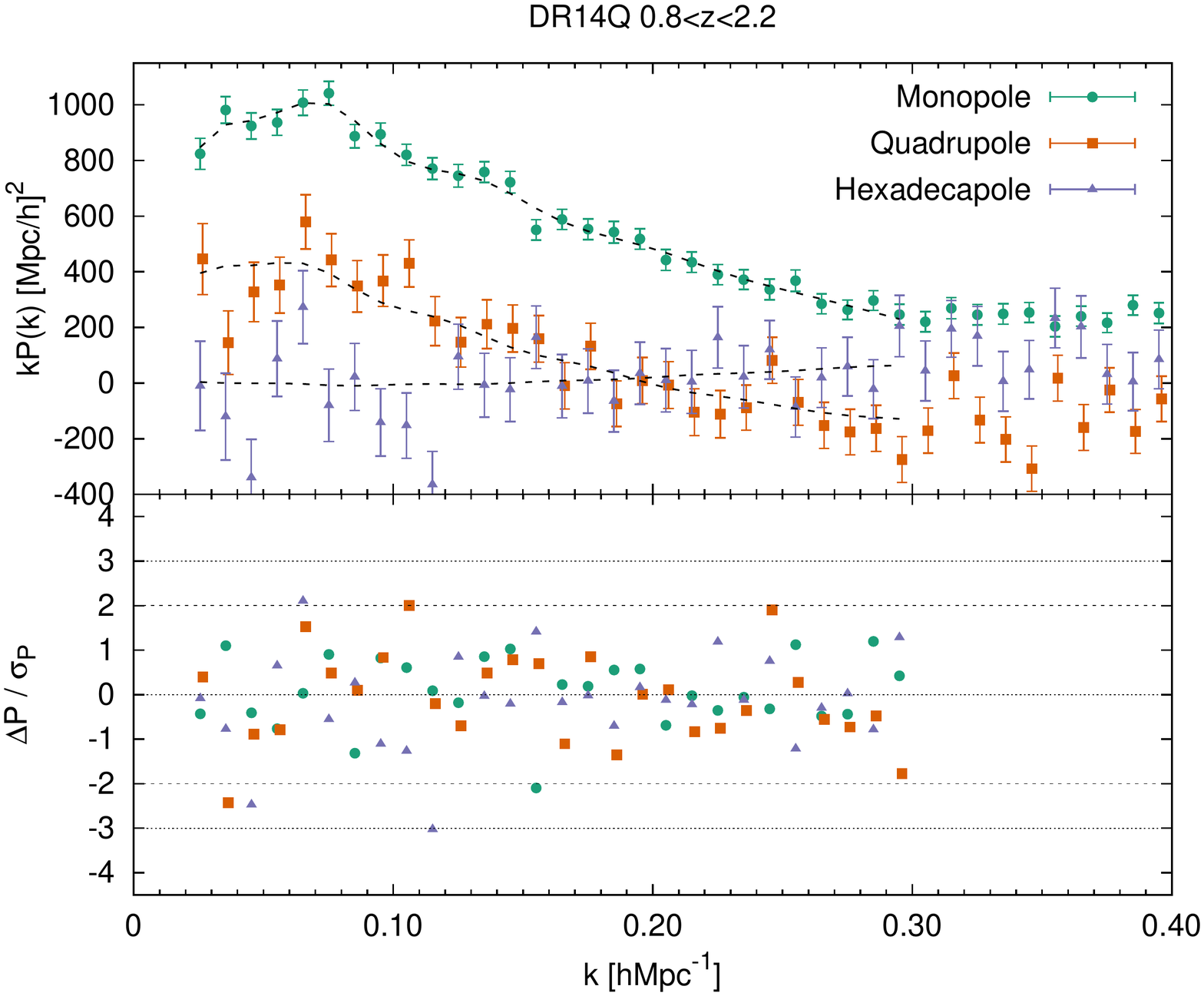}

\includegraphics[scale=0.30]{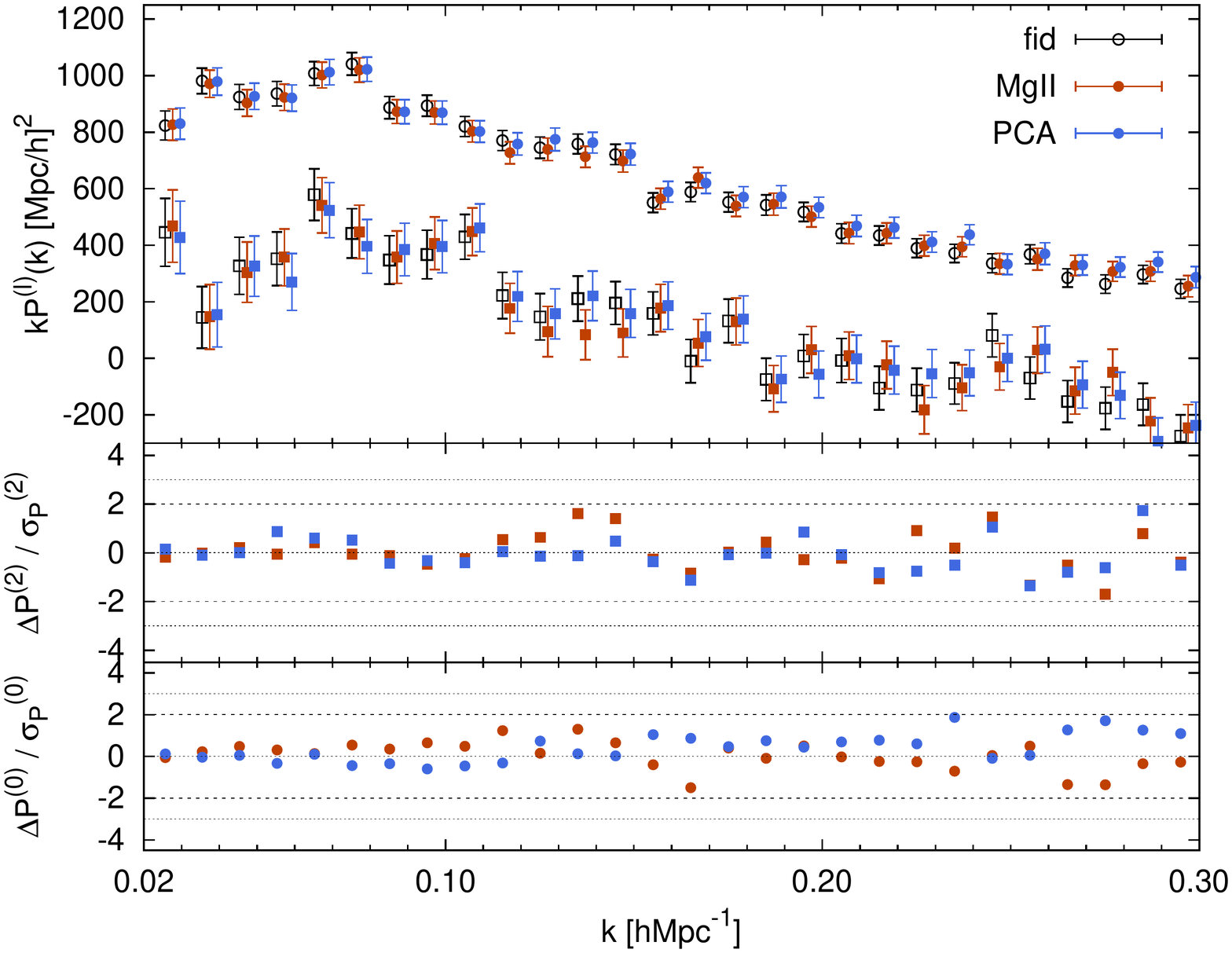}

\caption{{\it Top panel}: The DR14 quasar power spectrum monopole (green), quadrupole (orange) and hexadecapole (purple) in the redshift range $0.8\leq z\leq2.2$, including both NGC and SGC sky patches. The displayed error-bars are the {\it rms} of 1000 realisations of the \textsc{ez}-mocks. The dashed black lines represent the best-fitting model for the $k$-range $0.02\leq k\,[h{\rm Mpc}^{-1}]\leq0.30$. The bottom sub-panel show the differences between the model and data, divided by the diagonal errors, using the same colour scheme. The  2 and $3\sigma$ confidence levels are marked with dashed and dotted black lines, respectively. {\it Bottom panel}: Monopole and quadrupole measurement of the data for the different redshift estimates described in \S~\ref{sec:redshifts}: $z_{\rm fid}$ in black symbols, $z_{\rm Mg II}$ in red symbols, and $z_{\rm PCA}$ in blue symbols. The bottom sub-panels display the difference with respect to the fiducial redshift estimate relative to the statistical errors for the monopole and quadrupole. For clarity we do not display the results for the hexadecapole, where the degree of agreement is similar to the other two multipoles.  } 
\label{plot:measurements}
\end{figure}

The top panel of  Fig.~\ref{plot:measurements} displays the DR14Q measured power spectrum monopole (green circles), quadrupole (orange squares) and hexadecapole (purple triangles) in the redshift range of $0.8\leq z \leq 2.2$. The error-bars correspond to the diagonal elements of the covariance matrix estimated from the {\it rms} of the 1000 realisations of the \textsc{ez} mocks. The black dashed lines represent the best-fitting theoretical model. Although the power spectrum has been measured in the range $0<k\,[h{\rm Mpc}^{-1}]<0.40$, only those $k$-bins in the range $0.02<k\,[h{\rm Mpc}^{-1}]<0.30$ have been used to fit the theoretical model, therefore the black dashed lines only cover this specific range. The lower sub-panels display the difference between the measured power spectrum and the best-fitting theoretical model divided by the $1\sigma$ error. The associated $\chi^2$ with this fit is $84.0/(84-7)$. The contribution from the monopole-only data points is $\chi^2_{P^{(0)}}=20.1/(28-7)$, from the quadrupole $\chi^2_{P^{(2)}}=30.2/(28-6)$, and from the hexadecapole $\chi^2_{P^{(4)}}=34.6/(28-4)$\footnote{The degrees of freedom  are just 28-6 and 28-4, respectively, because $A_{\rm noise}$ does not contribute to the shape of the quadrupole or hexadecapole, nor do the bias parameters ($b_1$, $b_2$) contribute to the shape of the hexadecapole.}. Ignoring the covariance between the multipoles would reduce the $\chi^2$ by  just $0.9$, suggesting that the three power spectrum multipoles are barely correlated (see Fig.~\ref{plot:covariance1} in Appendix~\ref{ap:cov} for a further description of the correlation among $k$-bins).

The bottom panel of Fig.~\ref{plot:measurements} displays the impact of changing the redshift estimate of the DR14Q sample from its fiducial methodology, $z_{\rm fid}$ to the one based on the maximum of the MgII line, $z_{\rm MgII}$ (red symbols) and to the one based on the PCA decomposition technique which also uses the position of the MgII line, $z_{\rm PCA}$ (blue symbols). The three redshift estimates display a consistent behaviour for both the monopole and quadrupole, not revealing any specific systematic trend, and the differences on specific $k$-modes are always below $2\sigma$. We must bear in mind that these measurements must be correlated up to some extent, and therefore we cannot quantify in terms of $\chi^2$ the agreement among them, nor their correlation, as we lack different redshift estimates for the mocks. Producing mocks which capture such behaviour would require a simulation of realistic quasar spectra at a given redshift for each particle in the mocks, which is beyond the scope of this paper. For simplicity we do not show the hexadecapole measurements, as the degree of agreement is similar to the one found in the monopole and quadrupole. In \S~\ref{sec:results} we will present the cosmological derived parameters based on these three redshift estimates. 

The top panel of Fig.~\ref{plot:measurements2} shows the power spectrum monopole and quadrupole measured on the NGC and SGC separately. Since the two samples are well disconnected they can be considered fully independent. The different colours and symbols distinguish between the NGC and SGC region and the power spectrum multipole, as indicated. The best-fitting theoretical model is indicated by a solid black line for NGC, and a dashed black line for SGC. The middle and lower sub-panels display the difference between the model and the data divided by the corresponding $1\sigma$ error. For clarity we do not indluce the results on the hexadecapole. 

We observe that the data from the SGC presents a slightly higher amplitude than in the NGC, especially for $k\gtrsim0.15\,h{\rm Mpc}^{-1}$ and in the quadrupole. This effect translates into a best-fitting theoretical model with higher bias in the SGC quadrupole. In \S~\ref{sec:results} we will quantify this discrepancy and conclude that these differences are not statistically  significant. As for the combined NGC+SGC sample, there is no $k$-bin which deviates more than $3\sigma$ with respect to the prediction of the model, and just three points at more than $2\sigma$. The $\chi^2$ values for NGC and SGC are $64.5/(84-7)$ and $76.6/(84-7)$, respectively. For the NGC, the separate contribution for the monopole, quadrupole, and hexadecapole are, $\chi^2_{P^{(0)}}=19.8/(28-7)$, $\chi^2_{P^{(2)}}=22.0/(28-6)$ and $\chi^2_{P^{(4)}}=25.0/(28-4)$; and for the SGC  $\chi^2_{P^{(0)}}=25.6/(28-7)$, $\chi^2_{P^{(2)}}=24.0/(28-6)$ and $\chi^2_{P^{(4)}}=27.6/(28-6)$.

\begin{figure}
\centering
\includegraphics[scale=0.3]{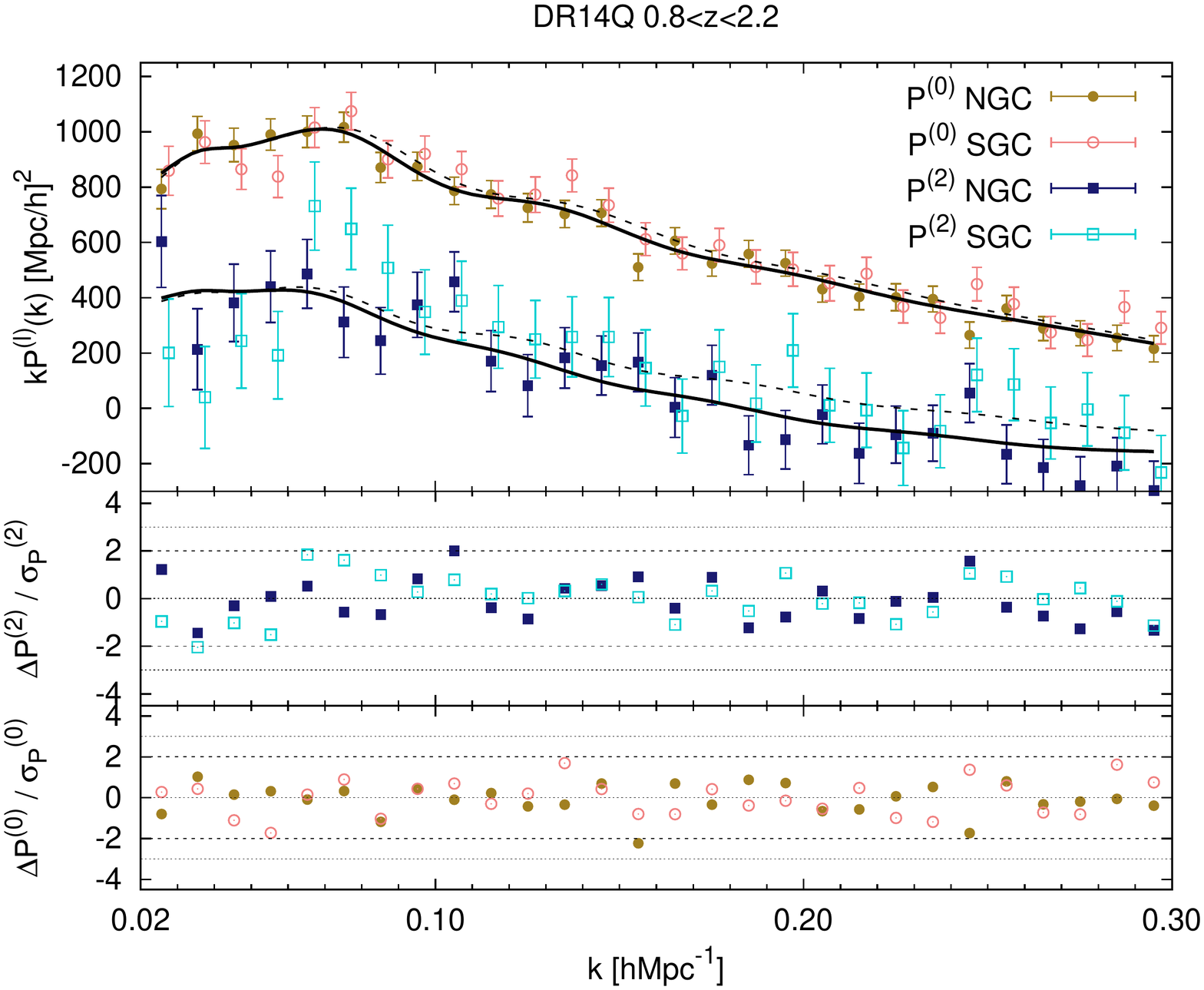}

\includegraphics[scale=0.3]{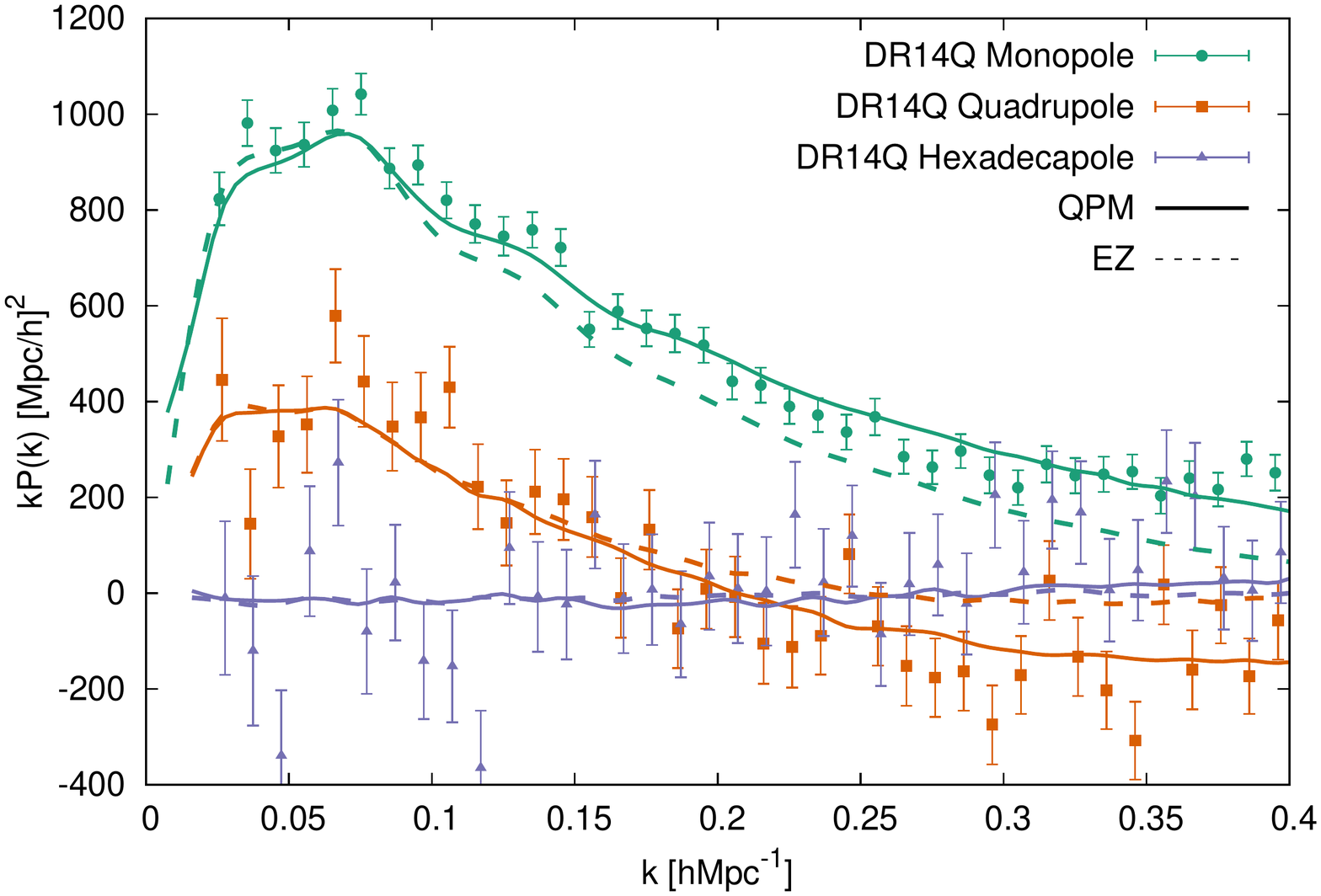}

\caption{{\it Top panel}: DR14 quasar sample measurement for the power spectrum monopole (circle symbols) and quadrupole (square symbols), for the NGC region (filled symbols) and SGC region (empty symbols) using the full redshift range, $0.8\leq z\leq 2.2$. The solid and dashed lines indicate the best-fitting model for the NGC and SGC, respectively. The error-bars display the {\it rms} from the \textsc{ez} mocks. The middle and bottom sub-panels display the difference between the model and the measurement divided by the errors, for the quadrupole and monopole, respectively. The horizontal dashed and dotted lines represent the 2 and $3\sigma$ confidence levels.   {\it Bottom panel}: The measured DR14Q data power spectrum monopole, quadrupole and hexadecapole are shown for the NGC+SGC region in green, orange and purple, respectively. The solid and dotted curves are the measured multipoles from the mean of the \textsc{qpm} (400 realisations) and \textsc{ez} mocks (1000 realisations), respectively, using the same colour notation. Since the error-bar for the mean of the mocks is small it is not plotted. At large scales both mocks and data show a good agreement, but at small scales both \textsc{qpm} and \textsc{ez} mocks fail to accurately reproduce the data. However, this behaviour can be partially fixed in the monopole by modifying the shot noise value of the mocks, as it demonstrated in fig. 6 of \citet{Ataetal:2017}. }
\label{plot:measurements2}
\end{figure}

The bottom panel of Fig.~\ref{plot:measurements2} displays the performance of the mean of the 1000 realisations of the \textsc{ez} mocks (dashed lines) and the mean of the 400 realisations of the \textsc{qpm} mocks (solid lines) along with the DR14Q measurements.  For the monopole the \textsc{ez} and \textsc{qpm} underestimate and overestimate, respectively, the measurement from the data. This behaviour was also reported the fig. 6 of \cite{Ataetal:2017}. When a constant value is added to the mocks, the agreement improves significantly.  We also observe differences in the behaviour of the quadrupole at small scales, $k>0.20\,h{\rm Mpc}^{-1}$, where the \textsc{ez} mocks tend to over-estimate the DR14Q measurements. The hexadecapole measurements are similar for both \textsc{ez} and \textsc{qpm} mocks, and consistent along with the data. In section in \S~\ref{sec:sysdata}, we quantify the impact of these two covariance matrices in the cosmological parameters of interest. 

\subsection{Multiple redshift bins}\label{sec:redshiftbins}

\begin{figure*}
\centering
\includegraphics[trim={120 20 10 0},clip=false,scale=0.245]{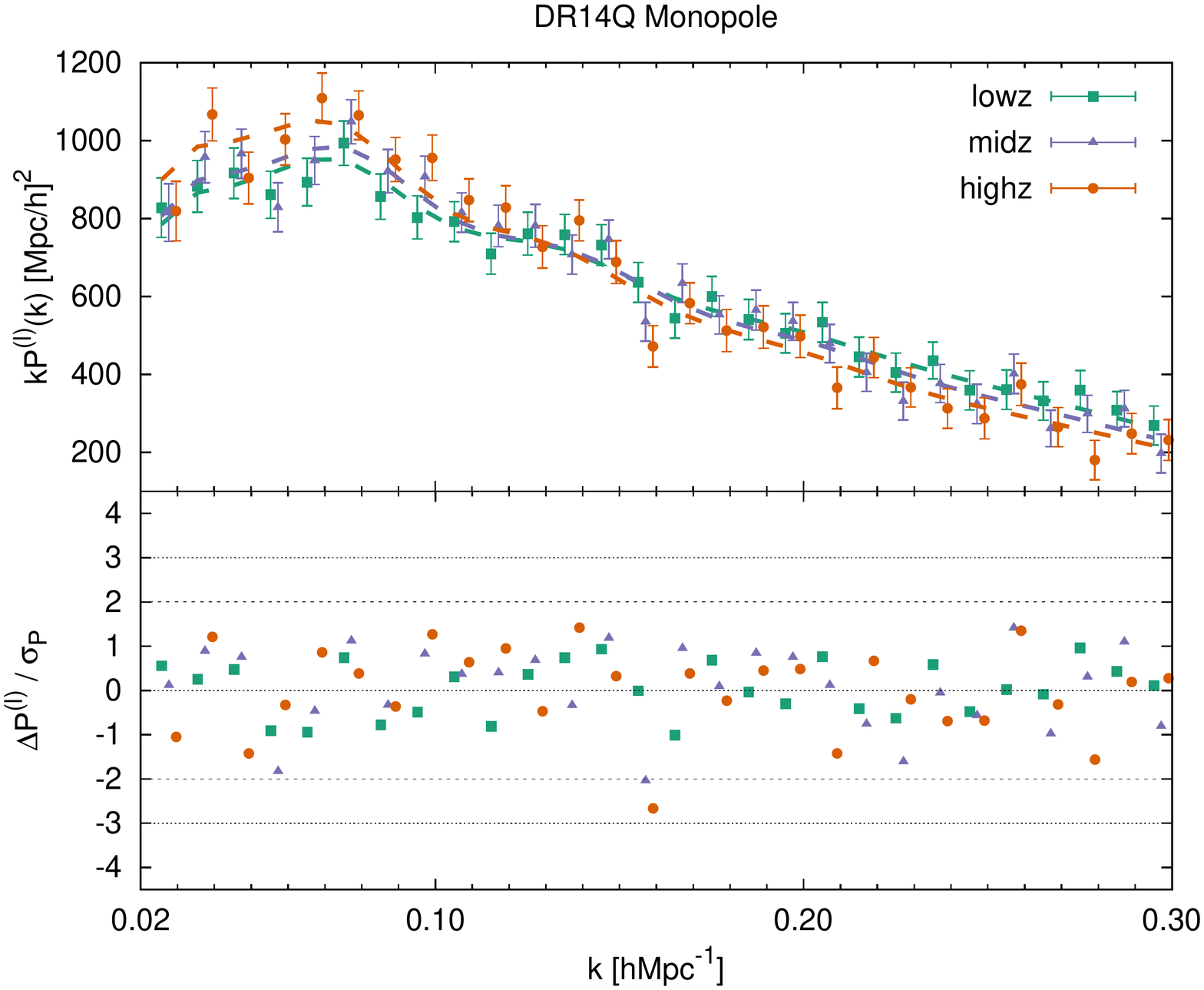}
\includegraphics[trim={120 20 10 0},clip=false,scale=0.245]{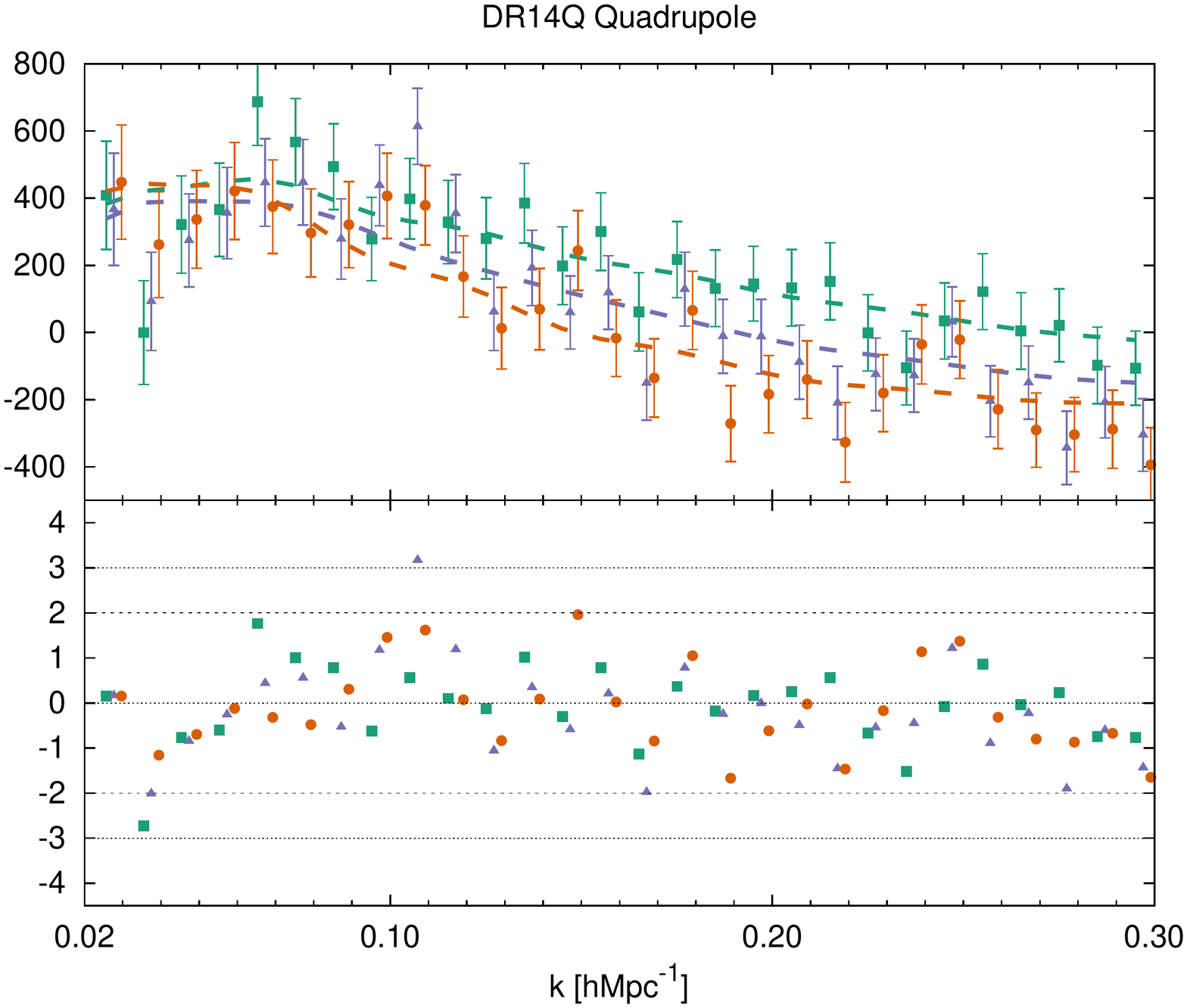}
\includegraphics[trim={120 20 100 0},clip=false,scale=0.245]{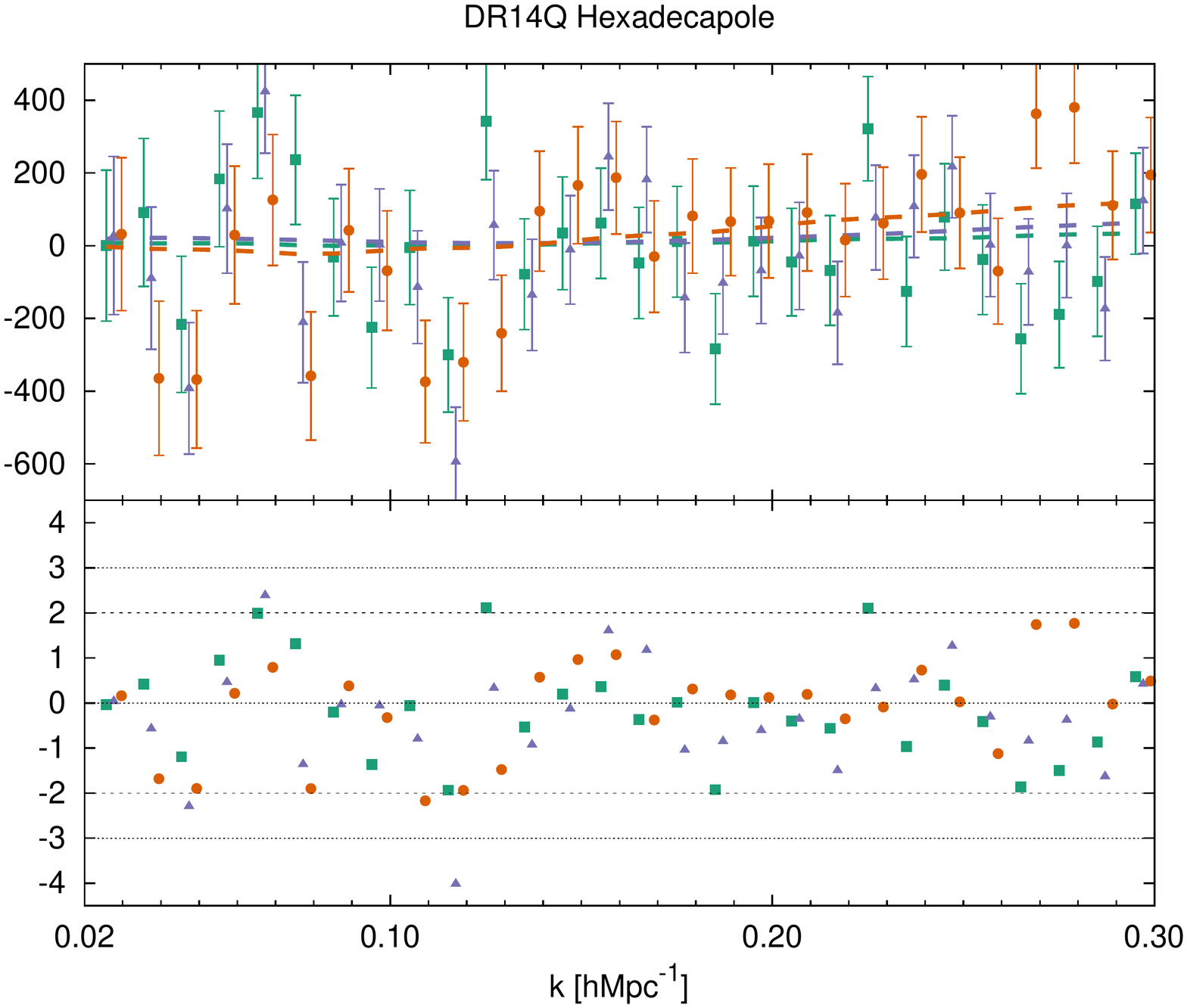}
\caption{Power Spectrum monopole (left panel), quadrupole (middle panel), and hexadecapole (right panel), for different $z$-bins: {\it lowz} $0.8\leq z\leq1.5$ (green), {\it midz} $1.2\leq z\leq1.8$ (purple), {\it highz} $1.5\leq z\leq 2.2$ (orange). The coloured symbols display the measurements from the NGC+SGC DR14Q sample, whereas the dashed lines indicate the best-fitting model. The lower sub-panels show the differences between the model and the data in terms of $1\sigma$ confidence levels. }
\label{Pkz:plot}
\end{figure*}

In the previous section we have presented the measured power spectrum multipoles for the entire redshift bin, $0.8\leq z\leq2.2$, with an effective redshift of $z_{\rm eff}=1.52$. However, the size of this redshift bin is large, which covers a wide range of epochs. During these epochs we expect that the cosmological parameters, such as $f\sigma_8$ and $b_1\sigma_8$, will significantly evolve with redshift. Constraining the evolution of these parameters with redshift, will better constrain potential departures from the standard cosmological model than just the average measurements of the whole redshift bin. 

Following this approach, we divide the DR14Q NGC+SGC sample in three overlapping redshift bins with similar effective volumes:  {\it lowz}, which covers $0.8\leq z\leq 1.5$ and whose effective volume is $V_{\rm eff}^{\rm lowz}=0.126\,{\rm Gpc}^3$; {\it midz}, which covers $1.2\leq z\leq 1.8$ and whose effective volume is $V_{\rm eff}^{\rm midz}=0.131\,{\rm Gpc}^3$;  and {\it highz}, which covers $1.5\leq z\leq 2.2$ and whose effective volume is $V_{\rm eff}^{\rm highz}=0.119\,{\rm Gpc}^3$. Since the {\it midz} range overlaps with both {\it lowz} and {\it highz}, we expect a significant correlation among these measurements, and their derived cosmological parameters. Using the \textsc{ez}-mocks, which contain an intrinsic evolution of the bias and cosmological parameters with redshift, we can compute the cross correlation coefficients among the parameters of the different redshift bins. 

The three panels of Fig.~\ref{Pkz:plot} display the power spectrum monopole (left panel), quadrupole (middle panel), and hexadecapole (right panel) for the DR14Q, for the different redshift bins, as indicated, in the coloured symbols. The coloured dashed lines indicate the best-fitting model, with the same colour notation.

For the power spectrum monopole, at large scales the amplitude of the power spectrum increases with redshift. The Kaiser boost factor for the monopole is, $([b_1\sigma_8]^2+2/3[f\sigma_8][b_1\sigma_8]+1/5[f\sigma_8]^2)$. The quantity $b_1\sigma_8(z)$ is a slightly increasing function with redshift in $0.8\leq z\leq 2.2$ ($b_1$ increases with $z$ and $\sigma_8$ decreases),  whereas, $f\sigma_8(z)$ is a decreasing function with redshift. However, the bias has a dominant effect over the logarithmic growth factor, and the overall effect is an increase of the amplitude, as observed on the data. At small scales we observe the opposite effect: the {\it highz} redshift bin presents a more important damping factor than the {\it lowz} bin. This behaviour is also expected, as the parameter $\sigma_P$ in our model accounts for not only the damping caused by the intra-halo velocity dispersion of the satellite quasars, but also for the effect of spectroscopic redshift errors. It is expected that these errors will increase with redshift, as the more distant objects tend to have lower signal-to-noise ratio spectra, which can lead to signfificantl errors in the measurement of their radial distance (see the Gaussian smearing redshift error model at the end of \S~\ref{sec:qpmmocks}). 

For the power spectrum quadrupole, we observe the opposite behaviour at large scales. In this case, the Kaiser boost factor becomes, $(4/3[b_1\sigma_8] [f\sigma_8] + 4/7 [f\sigma_8]^2)$, where the dominant component is $f\sigma_8(z)$, which drives the whole factor to decrease with redshift, as seen in the data and best-fitting model. Indeed, the importance of the bias parameters decreases for high order multipoles, which causes the Kaiser boost factor to be dominated by the $z$-evolution of the $f\sigma_8(z)$ parameter. At small scales we observe the same behaviour in the monopole. The redshift failure effects that produce that the damping factor strongly increase with redshift. 

Finally, in the hexadecapole the Kaiser boost factor is $\propto [f\sigma_8]^2$, with no bias contribution at large scales. Because of the large statistical errors we do not observe any particular trend of the hexadecapole as a function of the redshift bin. In addition, in the {\it midz} redshift bin, the hexadecapole at $k\sim0.11\,h{\rm Mpc}^{-1}$ is a $4\sigma$ outlier with respect to the expected model. This tension is reduced to the $3\sigma$ discrepancy when the full redshift range is considered, as shown in the top panel of Fig.~\ref{plot:measurements}. In \S\ref{sec:results_zbins}, we will discuss the impact of this frequency in the total $\chi^2$ and in the cosmological parameters. 

\section{Systematic Tests}\label{sec:systematictests}

We aim to identify potential systematic errors of our model, as well as potential systematics on the data, and quantify their effect on the measurements of cosmological interest. We start by using the \textsc{ez} and \textsc{qpm} mocks to recover the expected cosmological parameters. Although these mocks are not a proper N-body simulation, they can provide a first approximation on the performance of how the model works at these redshifts, and, more importantly, can test the effect of the potential systematics introduced by the spectroscopic weights.  We will later use the \textsc{OuterRim} N-body simulations to test the performance of the model using different prescriptions for the fraction of satellite quasars.

\subsection{Isotropic fits on mocks}\label{sec:resultsiso}

Table~\ref{table:mocksparameters} displays the shifts between the measured $\alpha_{\rm iso}$ and $f\sigma_8$ parameters, and the expected value from the known cosmology of the mocks and \textsc{or}-skycut,  when $\epsilon$ is fixed to 0. The expected values for both \textsc{qpm} and \textsc{ez} mocks can be found in Table~\ref{table:mocksalphas}. For \textsc{or},  the expected $\alpha_{\rm iso}$ is 1 as explained in \S~\ref{sec:OR}.  The $\langle x \rangle_i$ rows contain those quantities obtained by fitting the mean of all available realisations. In this case the errors represent the errors of the mean, where all the elements of the covariance matrix have been re-scaled by the inverse of the total number of realisations, 1000, 400, and 20 for the \textsc{ez}-, \textsc{qpm}-mocks, and \textsc{or}-skycut, respectively.  The $\langle x_i\rangle$ rows are the average of the best-fitting parameters individually on each realisation. In this case the errors represent the average of the errors of each individual fit. Additionally, the $S$ columns display the {\it rms} among best-fitting values of $\alpha_{\rm iso}$ and $f\sigma_8$ of each realisation. The $N_{\rm det}$ column is the number of mocks whose best-fitting values for ${\alpha_{\rm iso}}$ lie between $0.8$ and $1.2$. We consider those realisations as mocks with detection of $\alpha_{\rm iso}$ (this is the same definition in table 4 of \citealt{Ataetal:2017}). The average quantities $\langle {f\sigma_8}_i\rangle$ and $\langle{\alpha_{\rm iso}}_i\rangle$ (as well as the respective $S_i$) are computed only using these `detection' realisations, and discarding the rest. The $\Delta\alpha_{\rm iso}$, $\Delta f\sigma_8$, and $S$ values are expressed in terms of $10^{-2}$ units. Thus, $\Delta x=1$ corresponds, for instance, to a shift of $0.01$ with respect to the true expected value.  

For the \textsc{ez} and \textsc{qpm} mocks cases, the rows labeled with ``{\it raw}'' correspond to those results obtained when no spectroscopic effects, such redshift failures and fibre collision, are added (and therefore there is no need for these corrections). The rows labeled with $z_f$ correspond to those mocks with the redshift failure effect applied and corrected using the $w_{\rm foc}$ weight (according to Eq. \ref{eq:wfoc}). Those rows labeled as $w_{\rm foc}w_{\rm cp}$ refer to mocks where both fibre collisions and redshift failures are applied and corrected following the prescription described in \S~\ref{sec:spectroweights}. The rows labeled as $w_{\rm noz}w_{\rm cp}$ also contain both fibre collisions and redshift failures, but in this case the redshift failures have been corrected using the near-neighbour technique. Those rows noted as $+P^{(4)}$ represent the analysis including the hexadecapole signal. The \textsc{ez} and \textsc{qpm} mocks are analysed using their own covariance. However,  the \textsc{ez} mocks also include the case where the \textsc{qpm}-derived covariance is used in order to test its impact.

\begin{table*}
\caption{Shifts on the cosmological parameters, $x$, with respect to their expected value, $x_{\rm exp}$, $\Delta x\equiv x-x_{\rm exp}$; for $x=\alpha_{\rm iso}$ and $x=f\sigma_8$ on \textsc{ez} , \textsc{qpm} mocks, in the redshift range $0.8\leq z\leq2.2$ and on the \textsc{or}-skycut mock in $0.8\leq z\leq2.0$. $\langle x_i\rangle$ quantities are the average of the best-fitting values measured in individual realisations, whereas $\langle x \rangle_i$ are the quantities obtained by fitting the average of all the realisations. The errors correspond to the average value of the errors of individual fits and the error of the mean, in each case. For the $\langle x_i\rangle$ rows, the average is only performed among those realisations whose best-fitting value for $\alpha_{\rm iso}$ are between $0.8$ and $1.2$, for the $\alpha_{\rm iso}$ and $f\sigma_8$ columns. We call such fits a detection. The number of detection realisations is presented by the column $N_{\rm det}$. The $S_i$ columns display the {\it rms} of $\alpha_{\rm iso}$ and $f\sigma_8$, as indicated by the sub-index $i$. The units of $\Delta x$ and $S_x$ are $10^{-2}$, such that $\Delta x=1$ corresponds to a shift of 0.01. All results use the $0.02\leq k\,[h{\rm Mpc}^{-1}]\leq0.30$ range for the fits. For both \textsc{qpm} and \textsc{ez}-mocks, the covariance elements have been computed using their own covariance, unless the contrary is explicitly stated. For the \textsc{or}-skycut mocks the NGC \textsc{ez} covariance is used and is re-scaled to match the {\it rms} of the diagonal elements of the 20 \textsc{or}-skymocks realisations. Additionally, we also present the effect of correcting the spectroscopic effects of fibre collision and redshift failures on the \textsc{ez} and \textsc{qpm} mocks (see text for full description and notation), where the rows labeled ``{\it raw}'' are those corresponding to the mocks with no observational effects applied. For the \textsc{or}-skycut mocks no observational effects, other than a selection function, is applied. }
\begin{center}
\begin{tabular}{cccccc}
\hline
\hline
 & $\Delta\alpha_{\rm iso}$ & $S_{\alpha}$ & $\Delta f \sigma_8$ & $S_{f\sigma_8}$ & $N_{\rm det}$\\
\hline
{\bf \textsc{ez} mocks} & & & & & \\
 $\langle x \rangle_i$ raw & $-1.64\pm0.13$ & $-$ & $-1.14\pm0.16$ & $-$ & $-$ \\
 $\langle x \rangle_i$ $w_{\rm noz}w_{\rm cp}$ & $-1.90\pm0.13$ & $-$ & $2.53\pm0.17$ & $-$ & $-$ \\
 $\langle x \rangle_i$ $z_f$ & $-1.73\pm0.13$ & $-$ & $-1.30\pm0.16$ & $-$ & $-$ \\
 $\langle x \rangle_i$ $w_{\rm foc}w_{\rm cp}$ & $-1.80\pm0.13$ & $-$ & $-0.19\pm0.16$ & $-$ & $-$ \\
  $\langle x \rangle_i$ $w_{\rm foc}w_{\rm cp}$ \textsc{qpm}-Cov & $-1.90\pm0.12$ & $-$ & $-0.06\pm0.16$ & $-$ & $-$ \\
 $\langle x \rangle_i$ $w_{\rm foc}w_{\rm cp}+P^{(4)}$ & $-1.64\pm0.13$  & $-$ & $-0.10\pm0.17$ & $-$ & $-$ \\
 \hline
 $\langle x_i \rangle$ raw & $-1.65\pm3.91$ & $4.41$ & $-1.09\pm4.89$ & $4.91$ & $979$ \\
 $\langle x_i \rangle$ $w_{\rm foc}w_{\rm cp}$ & $-1.63\pm4.16$ & $4.47$ & $-0.03\pm5.24$ & $5.12$ & $973$ \\
 $\langle x_i \rangle$ $w_{\rm foc}w_{\rm cp}$ \textsc{qpm}-Cov & $-1.69\pm3.94$ & $4.83$ & $+0.12\pm 5.12$ & $5.69$ & $945$ \\
 $\langle x_i \rangle$ $w_{\rm foc}w_{\rm cp}+P^{(4)}$ & $-1.49\pm4.16$ & $4.43$ & $-0.03\pm5.30$ & $5.10$ & $970$ \\
\hline
{\bf \textsc{qpm} mocks} & & & & & \\
 $\langle x \rangle_i$ raw & $0.28\pm0.22$ & $-$ & $0.51\pm0.21$ & $-$ & $-$ \\
 $\langle x \rangle_i$ $w_{\rm foc}w_{\rm cp}$ & $0.15\pm0.21$ & $-$ & $1.05\pm0.24$ & $-$ & $-$ \\
 \hline
  $\langle x_i \rangle$ raw & $0.37\pm4.1$ & $4.48$ & $0.84\pm4.4$ & $4.21$ & $397$ \\
 $\langle x_i \rangle$ $w_{\rm foc}w_{\rm cp}$ & $-0.28\pm4.4$ & $4.84$ & $1.57\pm5.0$ & $4.80$ & $396$ \\
\hline
\hline
{\bf \textsc{or}-skymock} {\it w/o smearing} & & & & \\
no-sat $\langle x \rangle_i$ & $0.46\pm 0.79$ & $-$ & $-1.25\pm0.95$ & $-$ & $-$ \\
std $\langle x \rangle_i$ & $-1.80\pm 0.75$ & $-$ & $-1.77\pm 0.86$ & $-$  & $-$\\
high $\langle x \rangle_i$ & $-2.33\pm 0.64$ & $-$ & $-1.02\pm 0.80$ & $-$ & $-$ \\
\hline
{\bf \textsc{or}-skymock} {\it w/ smearing} & & & & \\
no-sat $\langle x \rangle_i$ & $0.82\pm0.84$ & $-$ & $-0.63\pm 1.02$ & $-$ & $-$ \\
std $\langle x \rangle_i$ & $-0.86\pm 0.75$ & $-$ & $-0.60\pm 0.94$ & $-$ & $-$ \\
high $\langle x \rangle_i$ & $-2.05\pm 0.66$ & $-$ & $0.37\pm 0.88$ & $-$ & $-$ \\
\hline
\hline
\end{tabular}
\end{center}
\label{table:mocksparameters}
\end{table*}%

In general there is a concordance between the shifts observed for the $\langle x_i\rangle$ and $\langle x \rangle_i$ variables. The $\alpha_{\rm iso}$ variable is robust under the different analysis methods, and we observe a consistent $1\%-2\%$ shift for the \textsc{ez} mocks to systematically lower values than expected, and $<1\%$ shift for the \textsc{qpm} mocks. The systematic $1\%-2\%$ shift observed on the \textsc{ez} mocks (and not observed in the \textsc{qpm}) may be due to either some systematic of the model or an intrinsic systematic of the mocks. 

The $f\sigma_8$ variable is more sensitive to the spectroscopic weights, in particular to the redshift failures when they are corrected through the $w_{\rm noz}$ prescription. In this case the systematic shifts on $f\sigma_8$ can reach $\sim6\%$, whereas the correction through $w_{\rm foc}$ does not produce any measurable systematic shift. Moreover, the impact of fibre collisions through the $w_{\rm cp}$ weight correction is $<2\%$. Adding the hexadecapole does not produce any significant change on the fits. This is because we are performing fits with $\epsilon$ set to 0, whereas the main extra information of the hexadecapole comes for breaking the degeneracy among $\alpha_\parallel$ and $\alpha_\perp$. By adding the spectroscopic weights, the errors and the {\it rms} values increase because of the increase of the shot noise component in the covariance matrix, which is the dominant term. Finally, the fits on the \textsc{ez} mocks with either the \textsc{ez}-derived covariance matrix or the \textsc{qpm}-derived covariance matrix do not produce any significant shift. 

The \textsc{or}-skycut mock results are also displayed in the lower columns of Table~\ref{table:mocksparameters} for three different satellite fraction, $f_{\rm no-sat}=0$, $f_{\rm std}=0.13$ and $f_{\rm high}=0.22$, and with and without a redshift smearing effect, which would mimic the expected uncertainty in the model determination. The \textsc{qpm} mocks, which are also based on a HOD technique, have a satellite fraction similar to $f_{\rm std}$. For the \textsc{or}-skycut mocks analyses no fibre collision or redshift failure effect has been included.
The \textsc{or}-skycut mock results show that neither the satellite fraction nor the redshift error smearing has an important impact on $f\sigma_8$; the systematic shifts are below $\sim 4\%$ (shift of $\leq0.02$). Similarly, the value of the $\alpha_{\rm iso}$ parameter does not produce shifts higher than $2\%$, although the systematic shift increases as the satellite fraction increases. 

\begin{figure}
\centering
\includegraphics[scale=0.3]{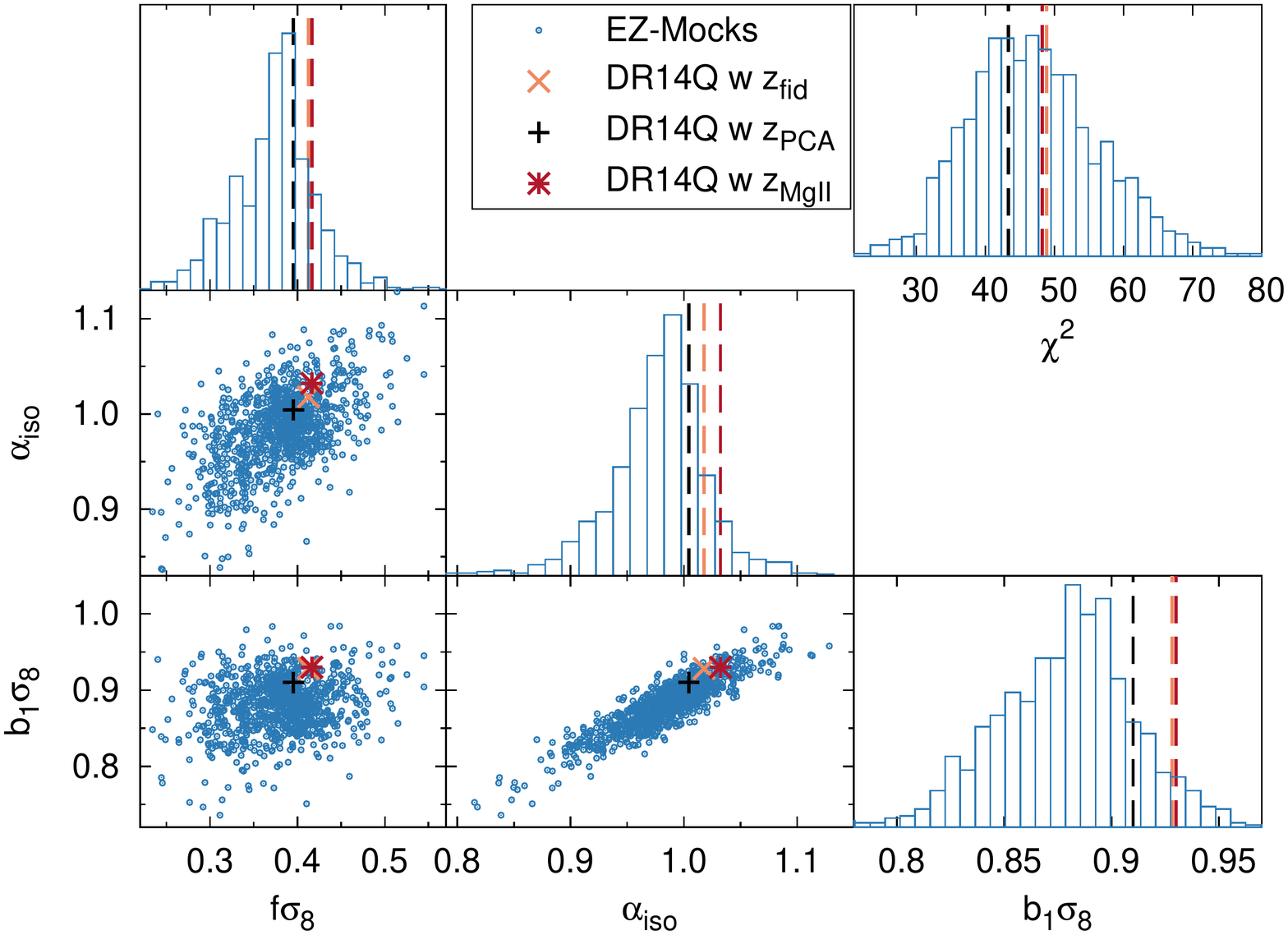}

\includegraphics[scale=0.3]{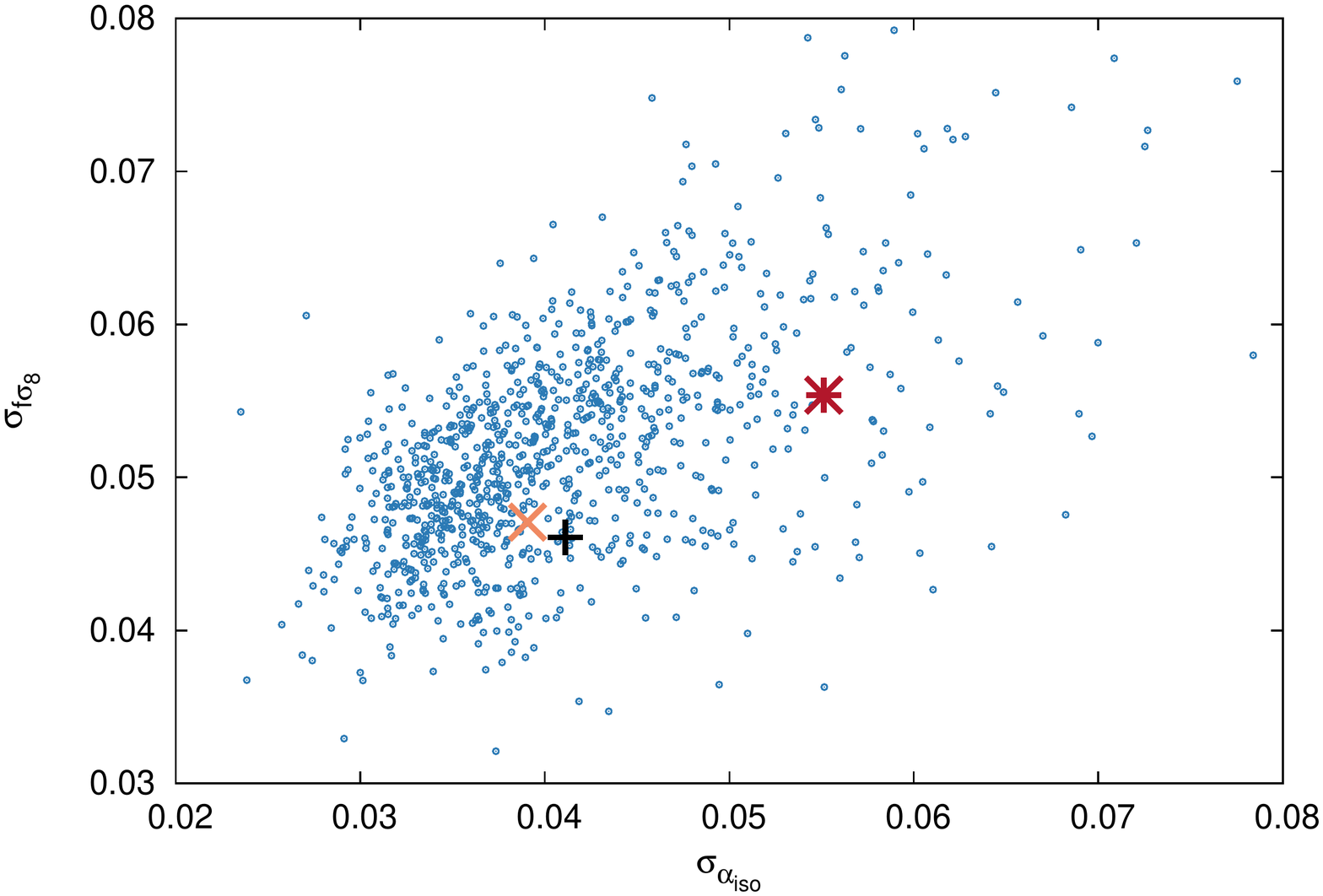}
\caption{{\it Top panel}: Best-fitting solution for the 1000 realisations of the \textsc{ez} mocks (in blue circles) and DR14Q dataset: orange cross for $z_{\rm fid}$, red for $z_{\rm MgII}$, and black for $z_{\rm PCA}$, under the constraint $\epsilon=0$ for the parameters $b_1\sigma_8$, $f\sigma_8$, $\alpha_{\rm iso}$ when the power spectrum monopole and quadrupole are used. The histograms show the distribution of parameters, as well as the $\chi^2$ values. The panels display the correlation among parameters. {\it Bottom panel}: Distribution of the $1\sigma$ errors of the $f\sigma_8$ and $\alpha_{\rm iso}$ parameters, for the \textsc{ez} mocks in blue circles, and DR14Q data using the same colour notation. The uncertainty of $f\sigma_8$ and $\alpha_{\rm iso}$ parameters have a positive correlation. In both panels, the results from the DR14Q data, when the $z_{\rm fid}$ and $z_{\rm PCA}$ estimates are used, are a typical individual case of the mocks. The errors obtained from the $z_{\rm MgII}$ redshift estimates are $\sim2\sigma$ larger than those expected from the mocks.}
\label{scatter:ezmocks}
\end{figure}

The upper panel of Fig.~\ref{scatter:ezmocks} displays the distribution of the best-fitting solution corresponding to the 1000 realisations of  the \textsc{ez} mocks with the $w_{\rm foc}w_{\rm cp}$ weights applied (blue circles) and DR14Q data (for the same weighting scheme) for the different redshift estimates: $z_{\rm fid}$ (orange cross), $z_{\rm Mg II}$ (red cross) and $z_{\rm PCA}$ (black cross) for the parameters, $\alpha_{\rm iso}$, $f\sigma_8$ and $b_1\sigma_8$; as well as the corresponding histograms including the $\chi^2$ distribution. The figure shows the degeneracies among parameters, in particular the strong correlation between $\alpha_{\rm iso}$ and $b_1\sigma_8$, as both parameters are sensitive to the amplitude of the power spectrum at all scales. Given these results, the data accordingly fit into a typical realisation of the mocks, and all the redshift estimates provide similar results on the studied parameters. The bottom panel of Fig.~\ref{scatter:ezmocks} displays the distribution of the $1\sigma$ error for $\alpha_{\rm iso}$ in the $x$-axis and $f\sigma_8$ in the $y$-axis. The mocks show a positive correlation between the error of $f\sigma_8$ and $\alpha_{\rm iso}$, as expected, since $\alpha_{\rm iso}$ and $f\sigma_8$ are correlated as well, i.e., $\alpha_{\rm iso}$ is well determined in a particular realisation, the probability that $f\sigma_8$ is well determined in that particular realisation is high. The errors measured from the data when $z_{\rm fid}$ and $z_{\rm PCA}$ are used are consistent with the mocks. The measured errors on the data with redshift estimate $z_{\rm MgII}$ are $\sim2\sigma$ larger than the distribution of the mocks. This behaviour may be caused by the broadening on spectroscopic redshift errors when estimating the redshifts using the MgII line with respect to the other two methodologies. 

\subsection{Anisotropic fits on mocks}\label{sec:anisofitsonmocks}

In this section we extend the above tests by relaxing the $\epsilon=0$ prior. We refer to the fit under these conditions as `full-AP'' fit.  The results are shown in  Table~\ref{table:mocksanis} for the \textsc{ez} mocks, \textsc{qpm} mocks, and \textsc{or}-skycut mocks, similarly as it was presented in the previous section. 

We start by describing the results on the \textsc{ez} and \textsc{qpm} mocks. When the power spectrum monopole and quadrupole are used, the results on the ``raw'' mocks present less than $2-3\%$ systematic shifts, similar to what was observed in the isotropic case. By including the hexadecapole on the \textsc{ez} mocks there is a slight increase on the systematics of the $\alpha_\parallel$ parameter, which reaches a $2.5\%$ shift, but it does not have any effect on the rest of parameters or on the \textsc{qpm} mocks, other than reducing the statistical uncertainty by $\sim20\%$. 

We compare the observed shifts on the ``raw'' mocks with the rest of the mocks using different weigh-correction prescriptions. In the $z_f$ case, the $w_{\rm foc}$ weight perfectly accounts for the redshift failure effects (see also in Appendix~\ref{ap:weights} the effect on the signal from the power spectrum multipoles), even when the hexadecapole signal is considered. For the  $w_{\rm foc}w_{\rm cp}$ case,  when only the power spectrum monopole and quadrupole are considered the inferred cosmological parameters shift slighly more than $1\%$, both for \textsc{ez} and \textsc{qpm} mocks. When the hexadecapole is added we observe a systematic $2\%$ for \textsc{ez} mocks and for \textsc{qpm} mocks toward lower values of $\alpha_\parallel$; $\lesssim1\%$ shift on $\alpha_\perp$ towards higher values; and $\sim0.02$ shift toward higher values of $f\sigma_8$, which represent a $\sim5\%$ shift. We interpret these changes as uncorrected systematics caused by the LOS distortion by the nearest-neighbour correction for the fibre collisions.

Similarly to Fig.~\ref{scatter:ezmocks} for the isotropic case, Fig.~\ref{plot:anis_hexa_scatter} displays the scatter of the \textsc{ez} mocks with respect to the DR14Q for the three different redshift estimates studied in this paper. As for the isotropic case, the main values drawn from these three redshifts are similar and are consistent with typical realisation of the mocks.  Similarly, the errors associated to the DR14Q for $z_{\rm MgII}$ and $z_{\rm PCA}$ estimates are larger than those corresponding to $z_{\rm fid}$ for $\alpha_\parallel$, but in any case within the expected range from the mocks. 

\begin{figure}
\centering
\includegraphics[scale=0.3]{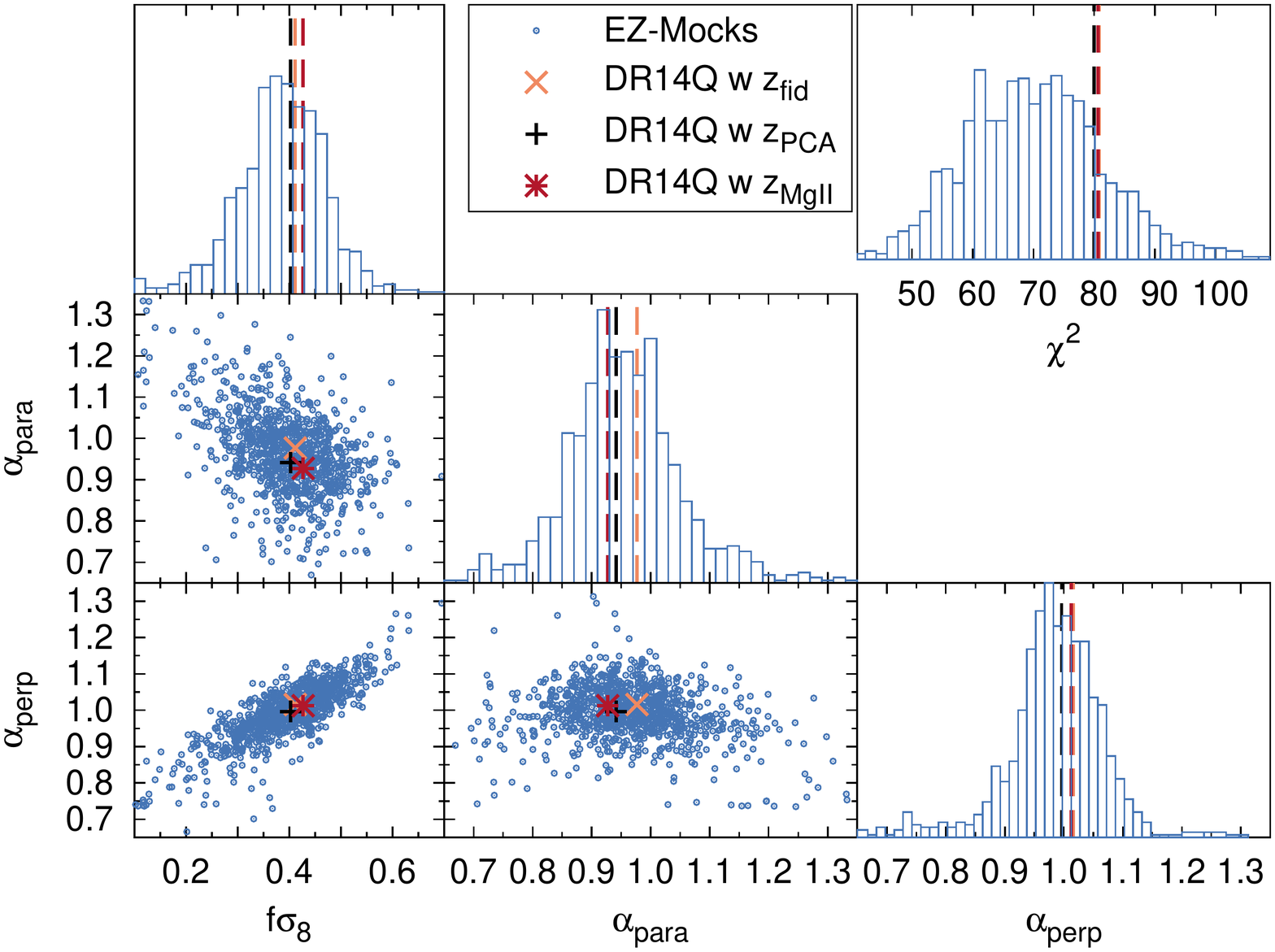}
\includegraphics[scale=0.3]{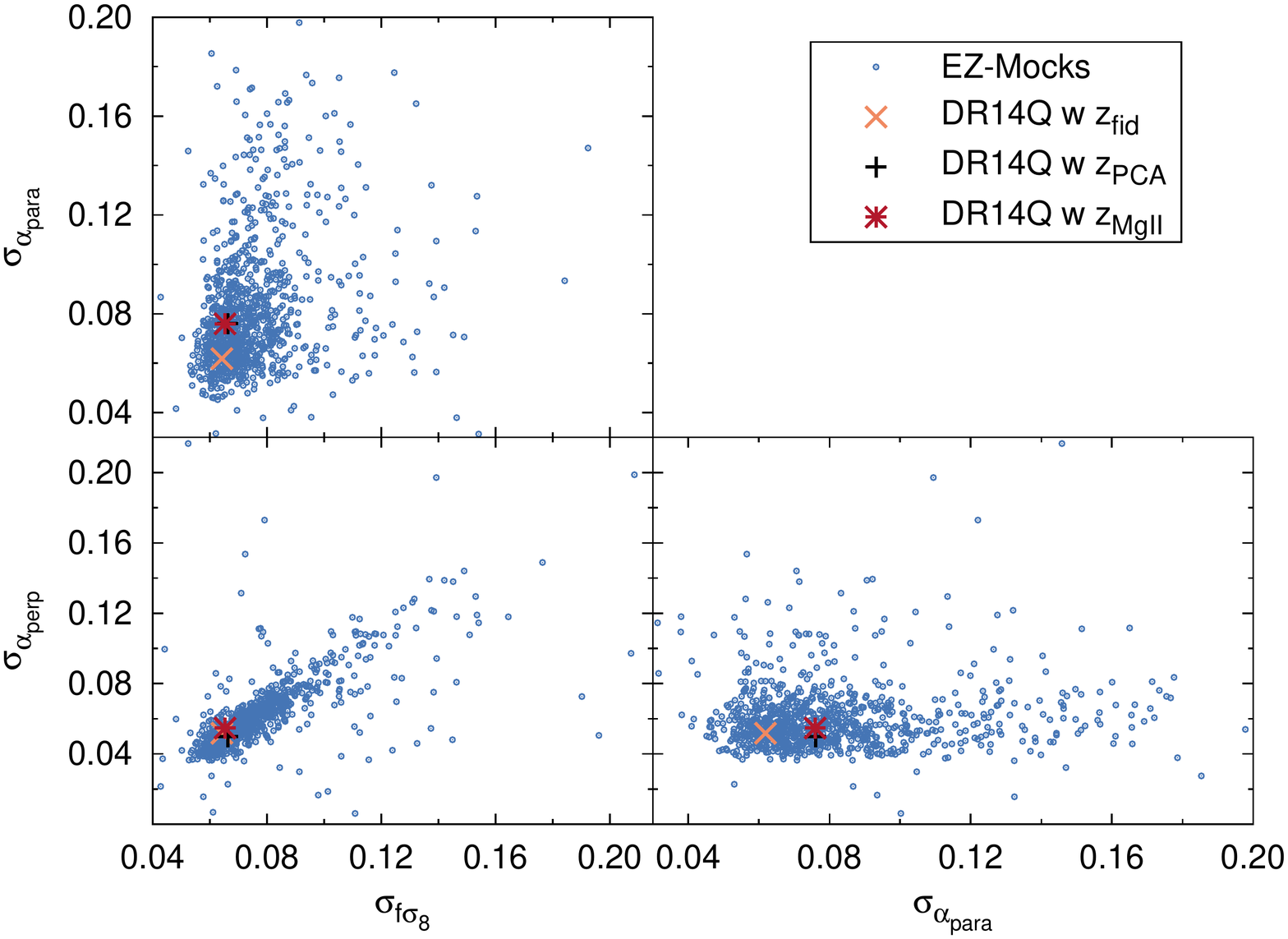}
\caption{Same format as in Fig.~\ref{scatter:ezmocks}, but for the full-AP fits, for the power spectrum monopole, quadrupole, and hexadecapole. Unlike for the $\epsilon=0$ case of Fig.~\ref{scatter:ezmocks}, for the full-AP case with the monopole, quadrupole, and hexadecapole are considered. The main values drawn from the different redshift estimates of the DR14Q dataset (top panels) agree very well among themselves and represent a typical single example from the mocks. On the error panels (bottom panels), the datasets corresponding to the three redshift estimates present a typical behaviour with respect to the mocks.}
\label{plot:anis_hexa_scatter}
\end{figure}

As a summary, the observed shifts due to redshift failures are totally negligible both for the cases $P^{(0)}+P^{(2)}$ and $P^{(0)}+P^{(2)}+P^{(4)}$ . The systematic shifts produced by fibre collisions on the  $P^{(0)}+P^{(2)}$ case are $\lesssim0.01$ for $\alpha_\parallel$, $\alpha_\perp$ and $f\sigma_8$.  Finally, the systematic shifts produced by fibre collisions when the hexadecapole is added are $\sim0.02$ for $\alpha_\parallel$ toward lower values, $\sim0.01$ for $\alpha_\perp$ towards higher values; and $0.027$ on $f\sigma_8$ towards higher values. These systematic shifts are below the statistical budget we expect from the data, given in the errors of the $\langle x_i\rangle$ rows. However, we will add in quadrature these errors in \S\ref{sec:results} and \ref{sec:cosmo} when the covariances and cosmological results are presented. 

We are also interested in testing the potential systematics of the theoretical modelling using full $N$-body mocks. We employ the same \textsc{or}-skymocks used in the section above, but now we perform a full-AP analysis on all three sets of mocks including the hexadecapole signal. The results are presented in Table~\ref{table:mocksanis}, for the different satellite fractions, with and without the redshift smearing effect. When the fraction of satellite quasars is kept to 0, the $f_{\rm no-sat}$ case, the systematic shifts observed for all the parameters are $\lesssim0.01$, for both with and without redshift smearing, and the model is able reproduce the expected signal. Introducing a satellite fraction produces some systematic shifts in some parameters. When the satellite fraction is kept to a value close to that used on the \textsc{qpm} mocks, $f_{\rm std}=0.13$, we observe shifts above $0.01$ on all the parameters, which reaches deviations between $0.02$ and $0.03$  for all variables. When the fraction of satellites is ${\it high}$ and no smearing is applied, the shifts on $\alpha_\parallel$ are $\sim0.03$, whereas on the rest of the cases are $\lesssim0.02$. We believe that these shifts are entirely due to a limitation of the theoretical model used to describe the clustering of quasars, either by the adopted bias model, or by the redshift space distortions. In this work we do not investigate further the origin of these discrepancies, which in all cases are $\lesssim 1/3$ of the expected statistical errors. 

The top panel of Fig.~\ref{fig:systematics} summarises the systematic shifts of Table~\ref{table:mocksanis} along with the expected statistical errors (horizontal dashed lines) computed from the average errors of the $\langle x\rangle_i\,w_{\rm cp}w_{\rm foc}$ row of \textsc{ez} mocks. For all the cases the shifts are computed using the $\langle x_i \rangle$ rows. The shifts obtained with the \textsc{ez} (dark blue) and \textsc{qpm} (light blue) mocks are driven by our treatment of observational inefficiencies with respect to their corresponding measured raw value. Therefore, these shifts display only the effects of fibre collisions and redshift failures on the measurements (any other systematic cancels). The effects are consistent among \textsc{ez} and \textsc{qpm} mocks, as expected, since they contain the same spectroscopic systematic effects. Examining the highest deviation either on \textsc{ez} or \textsc{qpm} mocks reveals that, {\it i)} $f\sigma_8$ the observational weights tend to produce a systematic shift towards higher values of $0.027$; {\it ii)} for $\alpha_\parallel$ these weights tend to reduce the measured quantity by $0.018$; {\it iii)} for $\alpha_\perp$ the observational systematics tend to increase the measured quantity by $0.013$. These values are summarised in Table~\ref{table:systematics} in the systematic observational budget column, $\sigma_{\rm obs}$. The shifts obtained with the \textsc{or}-skycuts mocks and various satellite fractions, however, indicate modelling errors. We adopt as a systematic error the largest shift obtained with among the different satellite fractions and smearing effects, with respect to the true underlaying value. Orange-circle, red-square and green-triangle symbols represent the results for $f_{\rm no-sat}$, $f_{\rm std}$ and $f_{\rm high}$, respectively. The information in Fig.\ref{fig:systematics} does not display a clear correlation between the modelling systematic shift and the satellite fraction value. Although the true satellite fraction of the data is unknown, we expect the upper limit to be less than $f_{\rm high}$, so we adopt the highest deviation among the six possible combinations as the modelling systematic contribution. These shifts are summarised in Table~\ref{table:systematics} under the modelling systematic column, $\sigma_{\rm mod}$. As $\sigma_{\rm obs}$ and $\sigma_{\rm mod}$ are different sources of uncertainties, we add them in quadrature to our uncertainty budget $\sigma^2_{\rm systot}\equiv \sigma_{\rm obs}^2+\sigma_{\rm mod}^2$. As both shifts tend to be of opposite signs and its origin is uncertain, we do not apply any shift to the best-fit values of the measured quantities.

In addition, the bottom panel of Fig.~\ref{fig:systematics} display the same quantities of the top panel as a function of the maximum wave-number used for the fit, $k_{\rm max}$. For simplicity we only show the results for the \textsc{ez} mocks and the \textsc{or}-skycut mocks with smearing. The systematic shifts reported in the top panel and Table~\ref{table:systematics} do not present any significant dependence with $k_{\rm max}$, except for the extreme case of $k_{\rm max}=0.37\,h{\rm Mpc}^{-1}$, which is outside the range of validity of the theoretical modelling at this redshift.  

\begin{figure}
\centering
\includegraphics[scale=0.3]{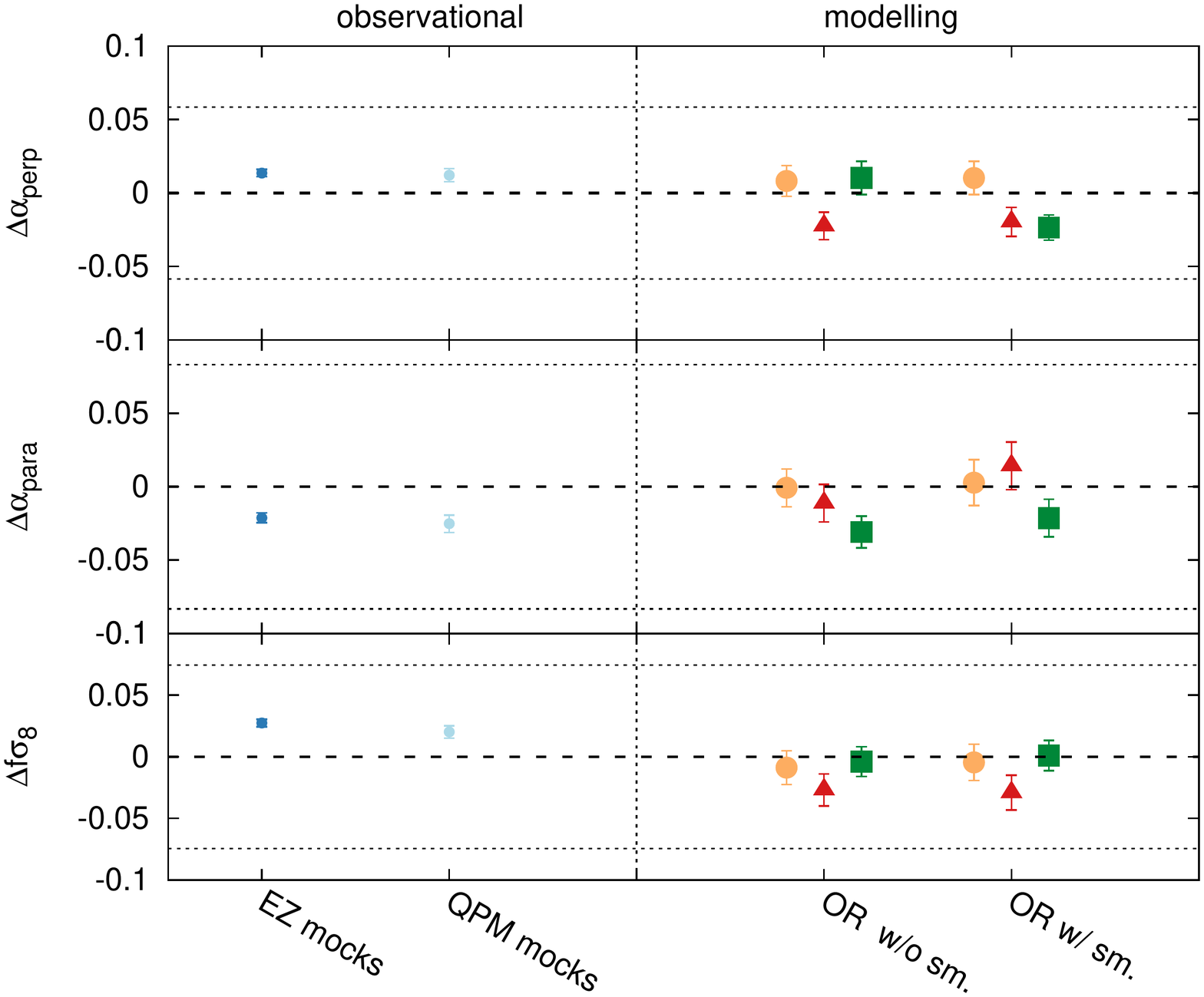}
\includegraphics[scale=0.3]{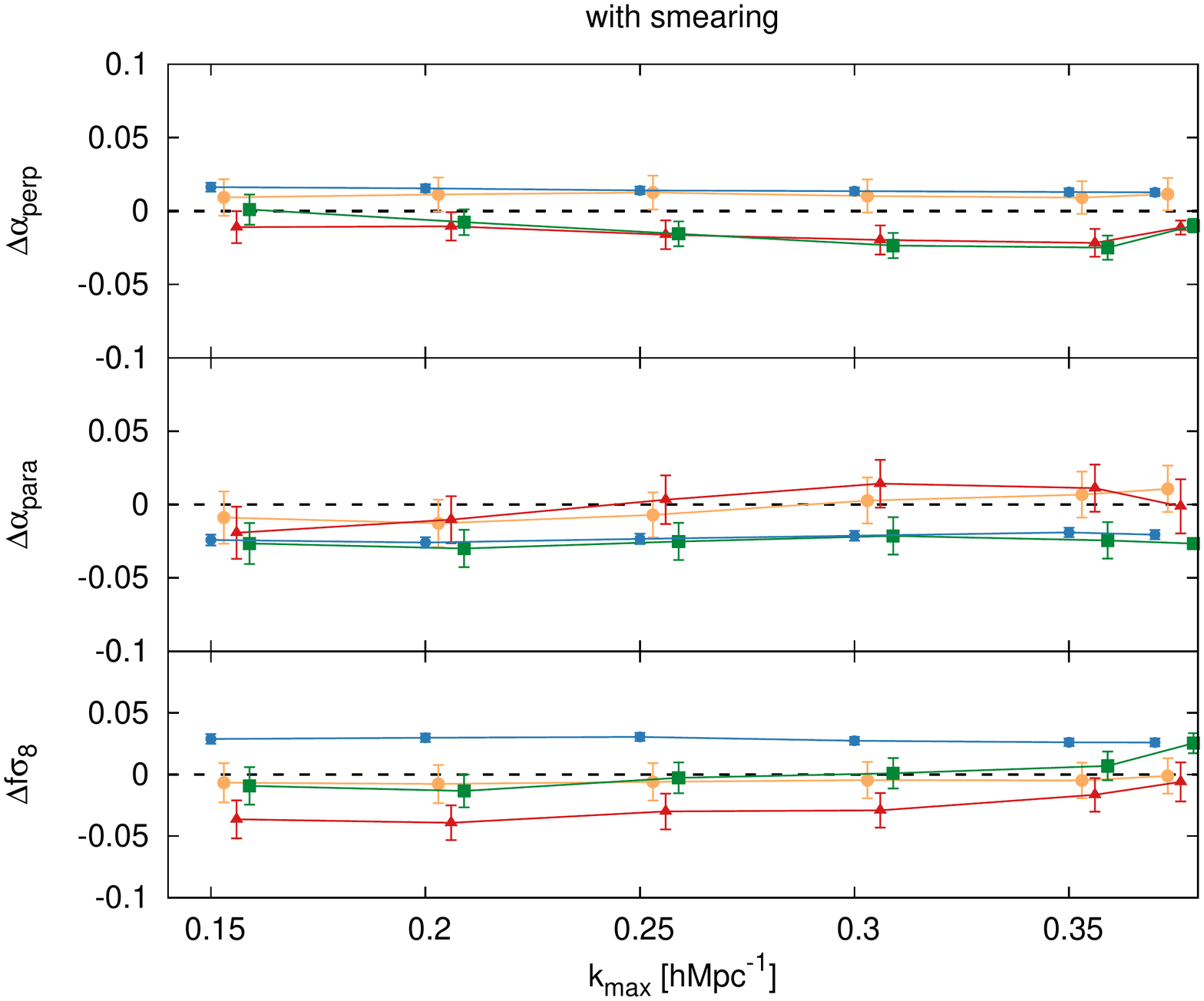}

\caption{{\it Top panel}: Systematic error shifts for $f\sigma_8$, $\alpha_\parallel$, and $\alpha_\perp$ computed from the results presented in Table~\ref{table:mocksanis}. The observational systematics represent the residual effects of the fibre collision + redshift failures. These shifts are computed as, $[\Delta x]_{\rm obs}\equiv  \langle x_i\rangle\,w_{\rm cp}w_{\rm foc} - \langle x_i\rangle\,{\rm raw}$ on both the \textsc{ez} and \textsc{qpm} mocks, as indicated; both sets have consistent systematic offsets. The modelling systematic results are computed only from the \textsc{or}-skycut mocks and they represent the uncertainties when recovering the true cosmological parameters due only to modelling limitations, $[\Delta x]_{\rm mod}\equiv  \langle x_i\rangle- x_{\rm true}$. The different colours and symbols represent the satellite fractions, $f_{\rm no-sat}$, $f_{\rm std}$, and $f_{\rm high}$ for orange-circles, red-triangles, and green-squares, respectively. The dashed black horizontal lines represent the expected statistical error for the data, computed as the average errors over the \textsc{ez} mocks. {\it Bottom panel}: The quantities of the top panel plotted as a function of the maximum $k$-wave number chosen to perform the fit. For simplicity we only show the \textsc{or}-skycut mocks with smearing and we do not display the \textsc{qpm} mocks results. The same colour code displayed in the top panel is used. Note that for the \textsc{or}-skycut mocks we display the deviation with respect to their expected value whereas for the \textsc{ez} mocks the deviation is plotted with respect to the `raw' case. }
\label{fig:systematics}
\end{figure}

\begin{table*}
\caption{Shifts of $\alpha_\parallel$, $\alpha_\perp$ and $f\sigma_8$ with respect to the fiducial values when the full-AP test is performed. The notation is the same that the one used in Table~\ref{table:mocksparameters}, with the difference that the definition of detection has been relaxed to be $0.7<\alpha<1.3$. Fig.~\ref{fig:systematics} summarises the systematics shifts observed in this table.}
\begin{center}
\begin{tabular}{cccccccc}
\hline
\hline
 & $\Delta\alpha_{\rm \parallel}$ & $S_{\alpha}$ & $\Delta\alpha_{\rm \perp}$ & $S_{\alpha}$ & $\Delta f \sigma_8$ & $S_{f\sigma_8}$ & $N_{\rm det}$\\
\hline
{\bf \textsc{ez} mocks} & & & & & & & \\
 $\langle x \rangle_i$ raw & $-1.58\pm 0.30$  & $-$ & $-1.71\pm0.20$  & $-$ & $-1.29\pm 0.25$  & $-$ & $-$ \\
 $\langle x \rangle_i$ $z_f$ & $-1.67\pm0.31$ & $-$ & $-1.75\pm0.21$   & $-$ & $-1.40\pm0.26$ & $-$ & $-$ \\
 $\langle x \rangle_i$ $w_{\rm foc}w_{\rm cp}$ & $-2.02\pm0.31$  & $-$ & $-1.69\pm0.22$  & $-$ & $-0.13\pm0.27$  & $-$ & $-$ \\
 $\langle x \rangle_i$  $ {\rm raw} +P^{(4)}$ & $-2.24\pm 0.23$  & $-$ & $-1.33\pm 0.17$  & $-$ & $-0.82\pm 0.21$  & $-$ & $-$ \\
 $\langle x \rangle_i$ $w_{\rm foc}w_{\rm cp}+P^{(4)}$ & $-4.36\pm0.24$ & $-$ & $0.03\pm0.19$  & $-$ & $1.92\pm0.23$ & $-$ & $-$ \\
  $\langle x \rangle_i$ $z_f$ $+P^{(4)}$ & $-2.28\pm0.24$  & $-$  & $-1.38\pm0.18$  & $-$ & $-0.94\pm0.22$ & $-$ & $-$ \\
 \hline
 $\langle x_i \rangle$ raw & $-1.61\pm 9.64$ & $8.11$ & $-1.69\pm 6.13$ & $6.80$ & $-1.18\pm 8.06$ & $8.20$ & $809$ \\
 $\langle x_i \rangle$  $ {\rm raw} +P^{(4)}$ & $-2.50\pm 7.72$ & $7.53$ & $-1.44\pm 5.38$ & $6.34$ & $-0.56 \pm 6.76$ & $7.18$ & $951$ \\
  $\langle x_i \rangle$ $w_{\rm foc}w_{\rm cp}+P^{(4)}$ & $-4.36\pm 8.32$ & $7.70$ & $0.10\pm 5.85$ & $6.26$ & $2.31\pm 7.44$ & $7.39$ & $902$ \\
 
\hline
{\bf \textsc{qpm} mocks} & & & & & & & \\
 $\langle x \rangle_i$ raw  &  $1.12\pm 0.47$ & $-$ & $0.25\pm0.35$  & $-$ & $0.19 \pm 0.41$  & $-$ & $-$ \\
 $\langle x \rangle_i$ $w_{\rm foc}w_{\rm cp}$ & $0.22\pm 0.53$  & $-$ & $0.13\pm0.36$  & $-$ & $0.87\pm 0.43$  & $-$ & $-$ \\
 $\langle x \rangle_i$  $ {\rm raw} +P^{(4)}$ & $1.18\pm 0.39$  & $-$ & $0.39\pm 0.30$  & $-$ & $-0.14\pm 0.33$  & $-$ & $-$ \\
 $\langle x \rangle_i$ $w_{\rm foc}w_{\rm cp}+P^{(4)}$ & $-1.34\pm 0.44$ & $-$ & $1.61\pm 0.33$  & $-$ & $1.88\pm 0.37$ & $-$ & $-$ \\
\hline
 $\langle x_i \rangle$ raw & $0.94\pm 11.03$ & $8.03$ & $0.70\pm 6.27$ & $7.24$ & $1.26\pm  8.03$ & $7.64$ & $307$ \\
 $\langle x_i \rangle$  $ {\rm raw} +P^{(4)}$ & $0.63\pm 8.87$ & $8.09$ & $0.27\pm  4.92$ & $7.37$ & $0.93 \pm 6.48$ & $7.57$ & $345$ \\
  $\langle x_i \rangle$ $w_{\rm foc}w_{\rm cp}+P^{(4)}$ & $-2.43\pm 8.15$ & $8.77$ & $0.63\pm 4.89$ & $8.43$ & $2.61\pm 6.53$ & $9.46$ & $326$ \\
\hline
\hline
{\bf \textsc{or}-skycut} {\it w/o smearing} & & & & & & & \\
$\langle x \rangle_i$  no-sat $+P^{(4)}$   & $-0.07\pm1.28$ & $-$ &  $0.83\pm1.04$ & $-$ & $-0.89\pm1.38$ & $-$ & $-$ \\
$\langle x \rangle_i$  std $+P^{(4)}$  & $-1.12\pm1.27$ & $-$ & $-2.23\pm0.94$ & $-$ & $-2.69\pm1.30$ & $-$ & $-$ \\
$\langle x \rangle_i$  high $+P^{(4)}$  & $-3.09\pm1.08$ & $-$ &$-2.11\pm0.84$& $-$ & $-0.39\pm1.20$ & $-$ & $-$ \\
\hline
{\bf \textsc{or}-skycut} {\it w/ smearing} & & & & & & & \\
$\langle x \rangle_i$  no-sat $+P^{(4)}$   & $0.28\pm1.56$ & $-$ & $1.03\pm1.13$ & $-$ & $-0.45\pm1.48$ & $-$ & $-$ \\
$\langle x \rangle_i$  std $+P^{(4)}$  & $1.50\pm1.63$ & $-$ & $-1.82\pm  0.99$ & $-$ & $-2.91\pm1.41$ & $-$ & $-$ \\
$\langle x \rangle_i$  high $+P^{(4)}$  & $-2.13\pm1.28$ & $-$ & $-2.35\pm0.86$ & $-$ & $0.09\pm1.24$ & $-$ & $-$ \\
\hline
\hline
\end{tabular}
\end{center}
\label{table:mocksanis}
\end{table*}%

\begin{table}
\caption{Total systematic error budget associated to the cosmological parameters, $f\sigma_8$, $\alpha_\parallel$, and $\alpha_\perp$ computed from the results of Table~\ref{table:mocksanis} and Fig.~\ref{fig:systematics}. The sign and magnitude represent the over- or under-estimated shift of the highest observed case, \textsc{ez} and \textsc{qpm} mocks for the observational systematics, and satellite fraction with and without smearing for the modelling systematics. Both contributions are added in quadrature, $\sigma_{\rm systot}^2\equiv \sigma_{\rm obs}^2+\sigma_{\rm mod}^2$, in order to produce a total systematic contribution which is added to the diagonal component of the covariance matrix of the data.}
\begin{center}
\begin{tabular}{cccc}
\hline
\hline
& $\sigma_{\rm obs}\times10^2$ & $\sigma_{\rm mod}\times10^2$ & $\sigma_{\rm sys tot}^2\times10^{3}$ \\
\hline
$f\sigma_8$ & $+2.74$ & $-2.91$ & $1.598$ \\
$\alpha_\parallel$ & $-2.12$ & $-3.09$ & $1.404$ \\
$\alpha_\perp$ & $+1.36$ & $-2.35$ & $0.737$ \\
\hline
\hline
\end{tabular}
\end{center}
\label{table:systematics}
\end{table}

\subsection{Isotropic fits on data}\label{sec:sysdata}

As an additional sanity check on our analysis procedure we perform a set of fits on the actual DR14Q dataset changing different specifications in order to examine how strong the results are with respect to the data parametrisation. For simplicity we only perform these fits fixing $\epsilon$ to 0. We call the ``standard'' (std) choice of specifications: {\it i)} $z_{\rm fid}$ as redshift estimates, {\it ii)} the power spectrum monopole and quadrupole as observables,  {\it iii)}  of both NGC and SGC regions, {\it iv)}  sampled with a linear binning of $\Delta k=0.01$ with $k$-centres at $k_i=(i+0.5)\Delta k$, {\it v)} scale range considered between $0.02\,\leq k\,[h{\rm Mpc}^{-1}]\leq 0.30$, {\it vi)}  where the fibre collisions and redshift failures are corrected through $w_{\rm cp}w_{\rm foc}$, and {\it vii)}  where the covariance matrix of the power spectrum multipoles is inferred from 1000 \textsc{ez}-mocks. 

Table~\ref{table:results_models} and Fig.~\ref{fig:dataoptions} display the measured $f\sigma_8$, $\alpha_{\rm iso}$ and $b_1\sigma_8$ measurements when some of these seven conditions are relaxed or modified. The $+1/4$, $+2/4$, and $+3/4$ rows correspond to the effect of shifting the centres of the $k$-bins by the respective fraction (with respect to the standard $k$-bin centre positions), and the `comb' rows to the result of combining these four results into a single measurement\footnote{We combine the results of the four different $k$-bin centres by averaging their normalised likelihood for each individual parameter.}. Hereafter, we will refer to the `standard' specification with the 4 $k$-bin centres combined as `comb' or `combined'. Likewise, we refer to $z_{\rm MgII}$-comb and $z_{\rm PCA}$-comb as the `combined' specification with the redshift estimate changed from the fiducial case to the $z_{\rm MgII}$ and $z_{\rm PCA}$ cases, respectively. The horizontal dashed lines in Fig.~\ref{fig:dataoptions} correspond to the $\pm1\sigma$ errors of the ``comb'' case. Along with the fiducial combined case we present the combined measurements for the two other redshift estimates, $z_{\rm MgII}$ and $z_{\rm PCA}$. The NGC and SGC rows correspond to the measurements when only one Galactic Cap is used. Those rows labeled with $\log k$ correspond to changing the linear binning to logarithmic binning in the same $k$-range and with similar number of $k$-bins. Consequently, the $\log k$ cases have more $k$-measurements at large scales (low $k$ values) with respect the linear binning (and the covariance accounts for an extra statistical correlation because of this effect). The \textsc{qpm} row displays the result when the covariance matrix is changed by that inferred by the 400 realisations of the \textsc{qpm} mocks. For comparison, the results inferred from only 400 realisations of the \textsc{ez} mocks is also displayed. Finally, the effect of changing or turning off the imaging and spectroscopic weights is also presented. Bear in mind that all the results are highly correlated as they are based on an identical dataset, with the exception of those NGC and SGC cases, which can be considered as totally uncorrelated. 

We observe a $\sim1\sigma$ deviation between NGC and SGC on the $\alpha_{\rm iso}$ and $b_1\sigma_8$ (both parameters are strongly correlated as is shown in Fig.~\ref{scatter:ezmocks}), but since these two regions are statistically independent a $1\sigma$ shift is expected. The higher value of $b_1\sigma_8$ in the SGC is related to the observed excess of power in the power spectrum monopole that is visible on top panel of Fig.~\ref{plot:measurements2}. As stated previously, we conclude that this shift is not statistically significant. 

For the rest of parameters studied, none have a strong effect on the cosmological parameters, with the exception of the redshift estimates (which we have already commented above) and the \textsc{qpm} covariance matrix, which produce a $\sim1\sigma$ shift on $b_1\sigma_8$ and $\alpha_{\rm iso}$, but no significant effect on $f\sigma_8$. However, this $1\sigma$ shift appears only when we compare the `comb' results with the \textsc{qpm}-cov results, which is drawn from the same $k$-bin centre condition as in `std' case. Therefore, a fair comparison between `std' and \textsc{qpm}-cov case, yields only a $0.67\sigma$ offset. Further investigation on the potential impact of the choice of the covariance matrix suggests that this $0.67\sigma$ shift is not caused by the limited amount of realisations (400 on \textsc{qpm} vs. 1000 on the \textsc{ez} mocks), nor by the differences in their off-diagonal elements which are sub-dominant in the total $\chi^2$ contribution. Therefore, the origin of this shift is located in the diagonal elements of the covariance, which present variations from 5 to $10\%$. 

Fig.~\ref{plot:covariance2} shows the ratio between the diagonal errors of \textsc{ez} and \textsc{qpm} mocks. In general \textsc{qpm}-derived errors are $5\%$ larger than \textsc{ez}-derived errors, but for the monopole at large scales this tendency is inverted. Examining at the shifts on the \textsc{ez} mocks results presented in Table~\ref{table:mocksparameters} when the \textsc{ez}- and \textsc{qpm}-derived covariances were applied, we find that $134/962$\footnote{We account for those realisations with double detection of $\alpha_{\rm iso}$ for both \textsc{ez}- and \textsc{qpm}-derived covariances, in total 962 out of 1000 mocks.} of the mocks have shifts $>0.67\sigma$ on $\alpha_{\rm iso}$ when the covariance matrix is changed, 75 towards lower values and 59 towards higher values, and consequently the average mean value of the inferred $\alpha_{\rm iso}$ is not significantly affected by the choice of the covariance (see Table~\ref{table:mocksparameters}). We conclude that the observed behaviour on the data is consistent with the behaviour of the mocks (occurs 14\% of times on the mocks), and that the origin of this effect are the 5 to $10\%$ differences in the diagonal terms of the two covariances, which in combination with the intrinsic statistical noise of the data can produce the observed $\sim0.67\sigma$ fluctuation. In any case, we believe that the \textsc{ez}-mocks are a better representation of the actual DR14Q dataset, as they have been produced using different epoch snapshots, whereas the \textsc{qpm} is generated from a single one. Therefore we assign a higher level of likelihood to those results derived from the \textsc{ez}-mocks covariance, over those from the \textsc{qpm}.

We conclude that none of the studied specifications produce major systematic shifts in the studied parameters, and we consider that our results are robust under the change of the specifications presented in this section. 

\begin{table*}
\caption{Impact of different parameters (see text for definitions) on the values of the DR14Q dataset measurements: $b_1\sigma_8$, $f\sigma_8$ and $\alpha_{\rm iso}$, where the fits have been performed keeping $\epsilon=0$. The parameter which has the largest impact is the choice of covariance. The \textsc{qpm} covariance matrix shifts by $0.67\sigma$ the values of $b_1\sigma_8$ and $\alpha_{\rm iso}$ with respect to the ``std'' case. However, further studies demonstrate that $\sim14\%$ of the mocks present such behaviour when the covariance matrix is changed from \textsc{qpm}- to \textsc{ez}-derived one. The SGC row also contains a $1\sigma$ shift, but in this case the information content is significantly independent, as the std case contains more area than SGC, and therefore such $\sim1\sigma$ shift is expected. For all cases the mean value between the $\pm1\sigma$ edges is reported. The reported errors only represent the statistical error budget. Fig.~\ref{fig:dataoptions} displays the values of this table.}
\begin{center}
\begin{tabular}{|c|c|c|c|c|}
\hline
\hline
case & $b_1\sigma_8$ & $f\sigma_8$ & $\alpha_{\rm iso}$ & $\chi^2/{\rm d.o.f}$ \\
\hline
std & $0.928\pm 0.037$ & $0.411\pm 0.047$ & $1.017\pm 0.039$ & $49/(56-6)$ \\
std +1/4 & $0.907\pm0.034$  & $0.411\pm0.046$  & $1.000\pm0.035$  & $51/(56-6)$  \\
std +2/4 & $0.908\pm0.034$  & $0.390\pm0.045$  &$0.994\pm0.034$  & $43/(56-6)$  \\
std +3/4 &  $0.934\pm0.035$ & $0.392\pm0.045$ & $1.023\pm0.035$  & $47/(56-6)$  \\
std comb. & $0.918\pm0.035$  & $0.401\pm0.046$  & $1.006\pm0.036$  & $47/(56-6)$  \\
$z_{\rm MgII}$ comb. & $0.908\pm0.045$  & $0.404\pm0.052$  & $1.009\pm0.050$ & $48/(56-6)$   \\
$z_{\rm PCA}$ comb. & $0.896\pm0.037$  & $0.390\pm0.045$  & $0.989\pm0.038$  & $45/(56-6)$ \\
NGC & $0.913\pm 0.044$ &  $0.408\pm 0.062$ & $1.005\pm 0.046$ & $41/(56-6)$ \\
SGC & $0.966\pm 0.055$ & $0.421\pm 0.076$  & $1.051\pm 0.056$ & $42/(56-6)$ \\
log$k$ & $0.918\pm0.033$ & $0.403\pm0.044$ & $1.002\pm0.033$ & $49/(54-6)$ \\
log$k$ NGC & $0.915\pm0.041$ & $0.395\pm0.060$ & $1.002\pm0.041$ & $37/(54-6)$ \\
log$k$ SGC & $0.948\pm0.054$ & $0.402\pm0.073$ & $1.026\pm0.053$ & $53/(54-6)$ \\
$\alpha_{\rm iso}=1$ & $0.914\pm0.018$ & $0.402\pm0.041$ & 1 & $49/(56-5)$ \\
$k_{\rm max}=0.20\,h{\rm Mpc}^{-1}$ & $0.931\pm0.045$ & $0.387\pm0.049$ & $1.011\pm0.040$ & $34/(36-6)$ \\
\textsc{qpm} cov & $0.951\pm0.037$ & $0.413\pm0.047$ & $1.043\pm0.039$ & $46/(54-6)$ \\
\textsc{ez} cov 400 real. & $0.921\pm0.039$ & $0.385 \pm 0.045$ & $1.005 \pm 0.040$ & $43/(54-6)$ \\
no $w_{\rm sys}$ & $0.936\pm0.039$ & $0.401\pm0.047$ & $1.017\pm0.041$ & $46/(56-6)$ \\
no $w_{\rm foc}$ & $0.929\pm0.037$ & $0.411\pm0.047$ & $1.017\pm0.039$ & $49/(56-6)$ \\
no $w_{\rm cp}$ & $0.928\pm0.036$ & $0.401\pm0.046$ & $1.012\pm0.037$ & $48/(56-6)$ \\
$w_{\rm noz}$ & $0.924\pm0.038$ & $0.400\pm0.046$ & $1.013\pm0.039$ & $43/(56-6)$ \\
${\rm fid}+P^{(4)}$ & $0.926\pm0.038$ & $0.399\pm0.045$ & $1.011\pm0.039$  & $84/(84-6)$ \\
\hline
\hline
\end{tabular}
\end{center}
\label{table:results_models}
\end{table*}%

\begin{figure}
\centering
\includegraphics[scale=0.3]{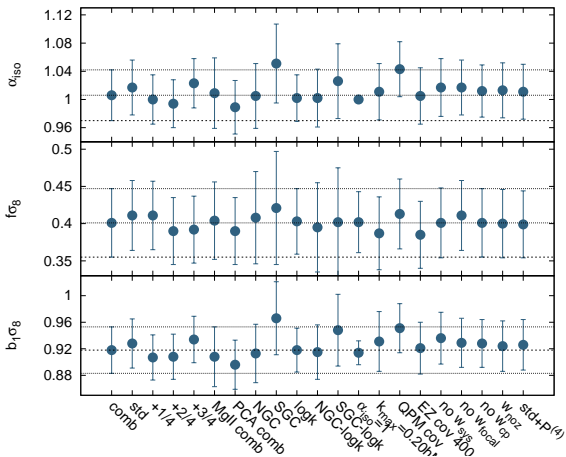}
\caption{Impact of different parameters (see text for definitions) on the measured $b_1\sigma_8$, $f\sigma_8$ and $\alpha_{\rm iso}$ quantities, corresponding to those results presented in Fig.~\ref{table:results_models}. For reference, the horizontal dashed lines correspond to the fiducial choice of parameters when the different redshift bin centres results have been combined (comb option), and the horizontal dotted lines to the $1\sigma$ deviation for the same case.}
\label{fig:dataoptions}
\end{figure}

\section{Results}\label{sec:results}
In this section we describe the results on the DR14Q data sample. We present the results for both isotropic fits (keeping $\epsilon=0$) and full-AP fits. Additionally, we perform two parallel analyses, {\it i)} we consider the full $0.8\leq z \leq 2.2$ redshift range as a single redshift bin, {\it ii)} we divide this redshift range into three overlapping redshift bins, as described above in \S\ref{sec:redshiftbins}. 

\subsection{Isotropic fits}\label{sec:isotropicfits}

We start by performing an isotropic fit to the power spectrum monopole and quadrupole of the DR14Q dataset, keeping $\epsilon=0$. Here we do not consider to use the hexadecapole because, as demonstrated in \S~\ref{sec:resultsiso}, it does not add any extra information when $\epsilon$ is set to a constant value.  
Table~\ref{table:full_results_iso_data} displays the mean value between the $1\sigma$ errors, $x_{\rm mean}\equiv [(x_{\rm bf}+\sigma^+) + (x_{\rm bf}-\sigma^-)]/2$\footnote{$\sigma^+$ and $\sigma^-$ are defined in such a way that $x_{\rm bf}\pm\sigma^\pm$ corresponds to $\chi_{\rm min}^2+1$.} for the free parameters of the model. 
The different columns display the results using the fiducial redshift estimate, $z_{\rm fid}$, applied to the individual NGC and SGC, along with the combination of both. In addition, the results for the two extra redshift estimates,  $z_{\rm MgII}$ and $z_{\rm PCA}$ are presented. For all the cases, the rest of the specifications correspond to the  `standard'  case defined in \S~\ref{sec:sysdata}, unless otherwise stated. Fig.~\ref{scatter:ezmocks} has previously shown the results on the three redshift estimates corresponding to the entire NGC+SGC sample for the parameters  $b_1\sigma_8$, $f\sigma_8$ and $\alpha_{\rm iso}$, along with the \textsc{ez} mocks. As described in \S\ref{sec:sysdata}, the results of the NGC and SGC are consistent within $1\sigma$ for all the parameters. 
The results on the different redshift estimates are also consistent and there is no observed evident tension in any of the parameters. The errors associated to the parameters inferred from the $z_{\rm MgII}$ redshifts are in general higher, as already discussed in \S~\ref{sec:resultsiso}. 

\begin{table*}
\caption{Inferred parameters of the model for the isotropic fit ($\epsilon=0)$ when the power spectrum monopole and quadrupole are used. We report the results for the DR14Q dataset for a fiducial redshift estimate, $z_{\rm fid}$, along with two additional redshift estimates, $z_{\rm  MgII}$ and $z_{\rm PCA}$. For the  $z_{\rm fid}$ case we present the results of fitting the NGC and SGC independently. For all cases the \textsc{ez}-derived covariance is used. The rest of specifications are fixed to the `standard' choice (see definition in \S\ref{sec:sysdata}). The units of $\sigma_P$ are $[h^{-1}{\rm Mpc}]$. The errors shown only represent the statistical error budget. For all parameters, we report the mean value between the $1\sigma$ errors, $x_{\rm mean}\equiv [(x_{\rm bf}+\sigma^+) + (x_{\rm bf}-\sigma^-)]/2$.}
\begin{center}
\begin{tabular}{|c|c|c|c|c|c}
\hline
\hline
  & $z_{\rm fid}$ &  $z_{\rm fid}$ (NGC) & $z_{\rm fid}$ (SGC) & $z_{\rm MgII}$ & $z_{\rm PCA}$ \\
  \hline
 $b_1\sigma_8$& $0.928\pm 0.037$ & $0.913\pm 0.044$ & $0.966\pm 0.055$ & $0.933\pm0.049$ & $0.909 \pm 0.039$ \\
 $f\sigma_8$   & $0.411\pm 0.047$ & $0.408\pm 0.062$ & $0.421\pm 0.076$ & $0.414\pm 0.055$ & $0.393 \pm 0.046$ \\
 $\alpha_{\rm iso}$ & $1.017\pm 0.039$ & $1.005\pm 0.046$ & $1.051\pm 0.056$ & $1.034 \pm 0.055$ & $1.003 \pm 0.041$ \\
 $b_2\sigma_8$  & $0.60\pm 0.57$  & $0.75\pm 0.63$ & $-0.24\pm 1.13$ & $0.20 \pm 0.72$ & $0.58 \pm 0.53$ \\
 $\sigma_P$  & $5.20\pm0.42$  & $5.54\pm0.58$ & $4.75\pm0.77$ & $5.30 \pm 0.52$ & $4.98 \pm 0.42$ \\
 $A_{\rm noise}$ & $7.2\pm2.3$ & $6.0\pm3.3$ & $8.1\pm7.3$ & $5.4\pm 3.8$ & $4.6 \pm 2.6$ \\ 
 \hline
 $\chi^2/{\rm d.o.f}$ & $49/(56-6)$  & $41/(56-6)$ & $42/(56-6)$ & $48/(56-6)$ & $43/(56-6)$ \\
 \hline
 \hline
\end{tabular}
\end{center}
\label{table:full_results_iso_data}
\end{table*}%
We opt to present the cosmological parameters derived from Table~\ref{table:full_results_iso_data} in a form of covariance matrix. We define the data-vector containing the cosmology parameters of interest $f(z)\sigma_8(z)$ and $D_V(z)/r_s(z_d)$ as, 
\begin{equation}
 \label{datanoAP}
D_{\rm data}(z) = 
 \begin{pmatrix}
  f\sigma_8(z)  \\
  D_V(z)/r_s(z_d)
 \end{pmatrix}.
 \end{equation}
 We fill in the data-vector  from the {\it rms} of the \textsc{mcmc}-chain steps within $3\sigma$ confident levels around the minimum, which for the two-parameter vector corresponds to those steps with $\chi^2<\chi^2_{\rm min}+12$. For the redshift estimate $z_{\rm fid}$,
 \begin{equation}
 \label{data_iso}
D_{\rm data}(z_{\rm eff}) = 
 \begin{pmatrix}
   0.426878 \\
  27.041179 
   \end{pmatrix}.
 \end{equation}
We compute the corresponding covariance matrix of this data-vector from 20 \textsc{mcmc} chains, with half a million steps in each chain using a convergence $R-1$ factor $\lesssim10^{-3}$ and retaining those steps within the $3\sigma$ confident level. In addition, we add a contribution on the diagonal elements corresponding to the modelling systematics derived from the \textsc{OR}-skymocks and previously described in \S~\ref{sec:resultsiso}. Recall that for the isotropic case these systematic shifts are of $\lesssim0.02$ for $f\sigma_8$ and $\alpha_{\rm iso}$. We do not consider any additional source of systematic error because, as presented in Table~\ref{table:mocksparameters}, the potential observational systematics caused by fibre collisions are below 0.01. With all these contributions the covariance matrix is, 
 \begin{equation}
\label{cov_iso}
C = 10^{-3}
 \begin{pmatrix}
2.188+0.4 & 22.93  \\
 - &  993.0+274.5 \\
 \end{pmatrix},
 \end{equation}
where the sums in the diagonal elements correspond to the systematics. The marginalised errors for the cosmological parameters which contain the full error budget are $f\sigma_8(1.52)=0.427\pm0.051$ and $D_V(1.52)/r_s(z_d)= 27.0 \pm 1.1$, with a correlation coefficient of $\rho_{[f\sigma_8-D_V/r_s]}=0.49$. 

These results are affected by the prior condition of $\epsilon=0$: These results imply that $H(z)D_A(z)$ is equal to $H^{\rm fid}(z)D_A^{\rm fid}(z)$, which is similar to the best-fitting Planck cosmology. Therefore, the isotropic derived results {\it are not independent} from Planck results, and thus, Planck-CMB data should not be added to these results as an extra uncorrelated dataset. 

\subsection{Anisotropic fits}\label{sec:anisotropic_results}
We consider the power spectrum monopole, quadrupole, and hexadecapole, and perform a full-AP fit, relaxing the $\epsilon=0$ condition of the previous section. Thus, $\alpha_\parallel$ and $\alpha_\perp$ can freely vary, which enable constraints to be placed on $H(z)r_s(z_d)$ and $D_A(z)/r_s(z_d)$ independent of the CMB data\footnote{BAO or RSD analyses are not able to measure $D_A(z)$ or $H(z)$ independently from the  sound horizon scale at the baryon drag epoch, $r_s(z_d)$, which is usually taken from CMB measurements. However, this scale can be computed from models of Big Bang Nucleosynthesis, and therefore one can consider RSD and BAO derived quantities independent from the CMB data }.  
Table  \ref{table:full_results_data} displays these results using the same notation as in Table~\ref{table:full_results_iso_data}. The first three columns show the results for the different redshift estimates when only the power spectrum monopole and quadrupole have been fitted. The following three columns present the results when the power spectrum hexadecapole is added to the analysis. For all the cases we fix the rest of specifications to `combined', as they are described in \S~\ref{sec:sysdata}, which consist on the `standard' specifications combining the four shifts on the $k$-bin centres.

The performance of the model corresponding to column $z_{\rm fid}+P^{(4)}$ was previously displayed in the top panel of Fig.~\ref{plot:measurements} for the full NGC+SGC dataset along with the DR14Q measurements. Likewise, Fig.~\ref{plot:anis_hexa_scatter} presents the results from the different $+P^{(4)}$ columns along with the results from the \textsc{ez} mocks for the parameters of cosmological interest, $f\sigma_8$, $\alpha_\parallel$, and $\alpha_\perp$.

\begin{table*}
\caption{Parameters of the model for the full AP-fit. We report the results for the DR14Q dataset for a fiducial redshift estimate, $z_{\rm fid}$, along with two extra redshift estimates, $z_{\rm  MgII}$ and $z_{\rm PCA}$. The first three columns correspond to the results when the power spectrum monopole and quadrupole are used; the three following columns are produced when the power spectrum hexadecapole is added to the analysis. For all cases the \textsc{ez}-derived covariance is used, and the rest of specifications correspond to `comb'  as described in \S\ref{sec:sysdata}. The units of $\sigma_P$ are $[{\rm Mpc}\,h^{-1}]$. The errors shown only represent the statistical error budget. }
\begin{center}
\begin{tabular}{|c|c|c|c|c|c|c}
\hline
\hline
  & $z_{\rm fid}$ & $z_{\rm MgII}$ & $z_{\rm PCA}$ & $z_{\rm fid}+P^{(4)}$ & $z_{\rm MgII}+P^{(4)}$ & $z_{\rm PCA}+P^{(4)}$  \\
  \hline
 $b_1\sigma_8$  & $0.930\pm0.041$  & $0.965\pm0.059$  &  $0.941 \pm0.059$ & $0.908\pm0.038$  & $0.859\pm0.049$ & $0.871\pm0.046$   \\
 $f\sigma_8$   & $0.366\pm0.072$  & $0.329\pm0.083$  & $0.314\pm0.079$ & $0.412\pm 0.064$ & $0.427\pm0.065$ & $0.404\pm0.066$   \\
  $\alpha_{\rm \parallel}$  & $1.040\pm0.089$  & $1.17\pm0.12$  &  $1.13\pm0.11$ & $0.976 \pm 0.062$ & $0.919\pm0.076$ & $0.937\pm 0.076$  \\
  $\alpha_{\rm \perp}$  & $0.977\pm0.056$  & $0.953\pm0.063$ & $0.941\pm0.054$ & $1.015\pm0.052$  & $1.011\pm0.055$ & $0.994\pm 0.050$   \\
 $b_2\sigma_8$  & $0.49\pm0.58$ & $-0.21\pm0.65$ & $0.33\pm0.55$ & $0.56\pm0.55$ &  $0.51\pm0.65$ & $0.61\pm0.52$  \\
 $\sigma_P$  & $5.18\pm0.49$ & $5.73\pm0.95$  & $5.22\pm0.65$ & $4.98\pm0.41$ & $5.05\pm0.46$ & $4.75\pm 0.40$  \\
 $A_{\rm noise}$  & $6.8\pm3.5$  & $2.2\pm4.5$  & $3.8\pm3.2$ & $6.5\pm2.7$  & $4.2\pm2.9$  &  $3.7\pm2.5$  \\
 \hline
 $\chi^2/{\rm d.o.f}$ & $47/(56-7)$ & $47/(56-7)$  & $43/(56-7)$ & $81/(84-7)$  & $80/(84-7)$&$81/(84-7)$   \\
 \hline
 \hline
\end{tabular}
\end{center}
\label{table:full_results_data}
\end{table*}%

When the power spectrum hexadecapole is added to the analysis some parameters shift their value at the $1\sigma$ level. When adding the hexadecapole signal we are introducing 28 new data-points (to the 56 already from monopole and quadrupole), which are highly independent, as indicated by the off-diagonal terms of the covariance matrices in Fig.~\ref{plot:covariance1}. We investigate the significance of those shifts using the \textsc{ez}-mocks. The top panel of  Fig.~\ref{plot:anis_hexa_scatter_comparison} displays the quantity $\Delta x\equiv x^{\rm MQ}-x^{\rm MQH}$, where $x^{\rm MQ}$ is the variable estimated from the monopole and quadrupole measurement, and $x^{\rm MQH}$ is the value when the hexadecapole is added. The bottom panel shows the same information but for the errors of the $x$-corresponding quantity. From the top panel,  the shifts presented by the results for the $z_{\rm fid}$ redshift estimate (orange symbols) are typical with respect to the observed shifts of the mocks, for the three variables of interest. For the $z_{\rm MgII}$ and $z_{\rm PCA}$ cases, however, the shifts on $\alpha_\parallel$ deviate by $2\sigma$ from the expected behaviour of the mocks, although they are along the degeneracy region among the studied parameters. Certainly, the discrepancy among the different redshift estimates is larger when only the monopole and quadrupole are considered, as the values of $\alpha_\parallel$ estimated from $z_{\rm MgII}$ and $z_{\rm PCA}$ are about $1\sigma$ from the value obtained with $z_{\rm fid}$. Adding the hexadecapole produces more consistent results among the three redshift estimators, as shown in Fig.~\ref{plot:anis_hexa_scatter}. The bottom panel of Fig.~\ref{plot:anis_hexa_scatter_comparison}  reveals that the reduction on the errors obtained by adding the hexadecapole are typical with respect to the behaviour observed by the mocks. 

\begin{figure}
\centering
\includegraphics[scale=0.3]{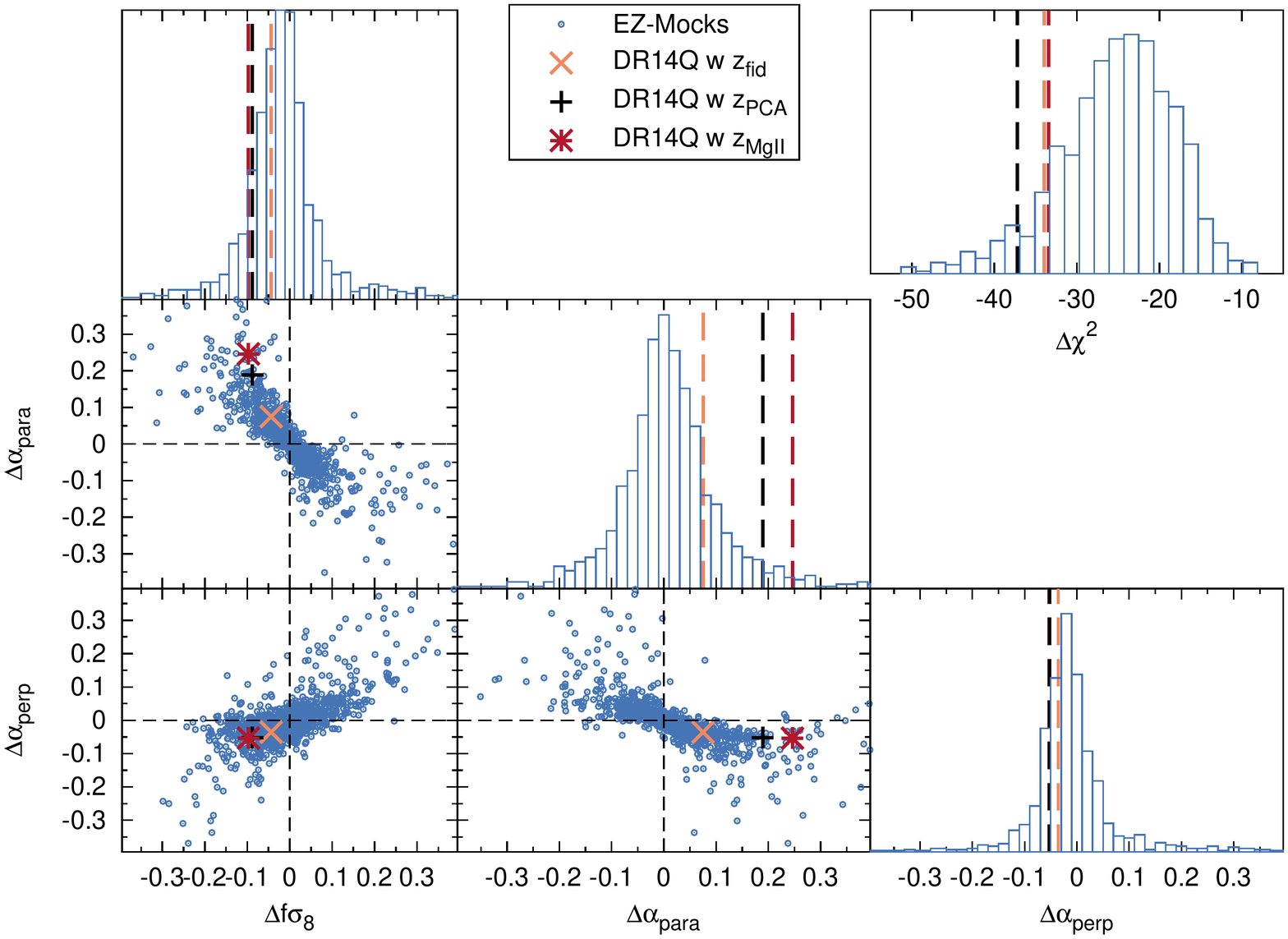}
\includegraphics[scale=0.3]{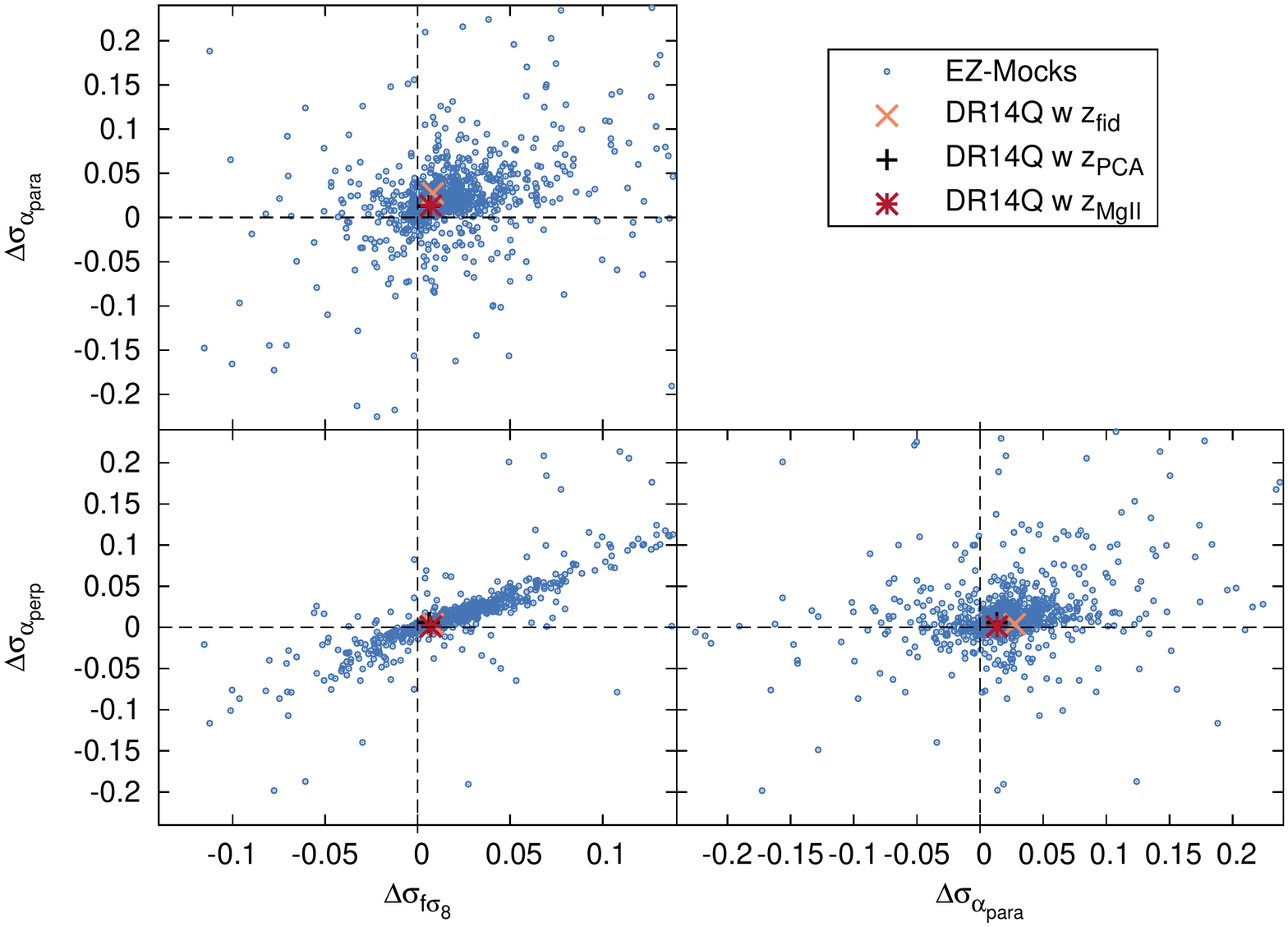}
\caption{{\it Top panel:} Difference between the best-fitting value of the mocks (blue symbols) and the DR14Q data (different colour symbols for the different redshift estimates) when only the monopole and quadrupole are used, $x^{\rm MQ}$, and when the hexadecapole is added, $x^{\rm MQH}$, for the different cosmological parameters: $\Delta x\equiv x^{\rm MQ}-x^{\rm MQH}$. {\it Bottom panel}: The same format as the top panel but for the error associated to the each parameter, $\sigma_x$, $\Delta\sigma_x\equiv \sigma_{x^{\rm MQ}}-\sigma_{x^{\rm MQH}}$. On average the expected shift on $\Delta x$ should be 0 if no extra systematic is added by the hexadecapole. However, from Table~\ref{table:mocksanis}, we know that adding the hexadecapole produces a systematic shift of $-0.02$ on $\alpha_\parallel$ and $+0.01$ on $\alpha_\perp$, which slightly shifts the centre of the measurements from the black dashed lines. We expect the hexadecapole to reduce the errors on the measured quantities (on average) and therefore $\sigma_x^{\rm MQH}<\sigma_x^{\rm MQ}$, which shifts the centre of the $\Delta\sigma_x$ distribution towards the positive quadrant for all the variables. }
\label{plot:anis_hexa_scatter_comparison}
\end{figure}

\begin{figure*}
\centering
\includegraphics[scale=0.5]{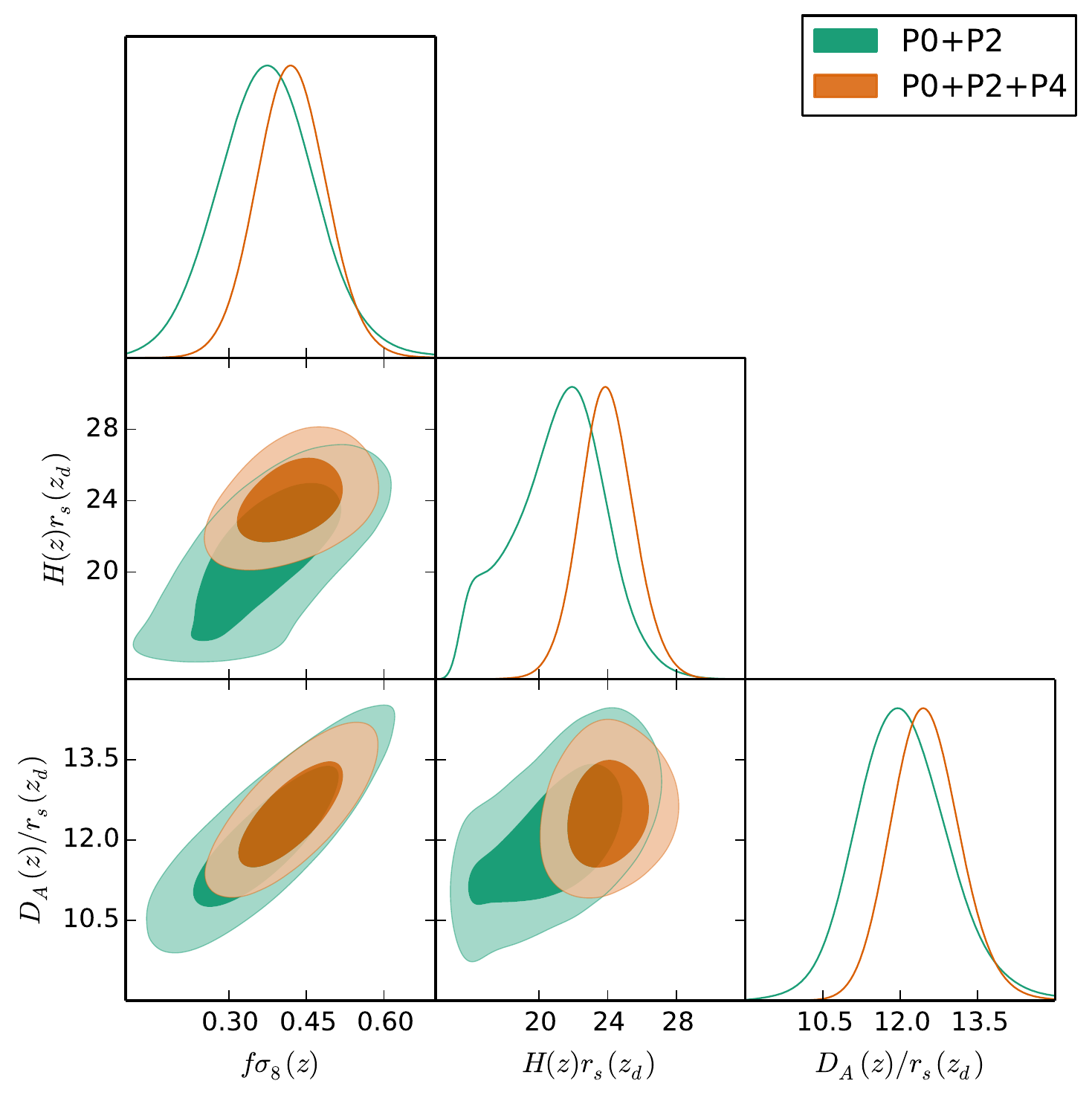}
 \includegraphics[scale=0.5]{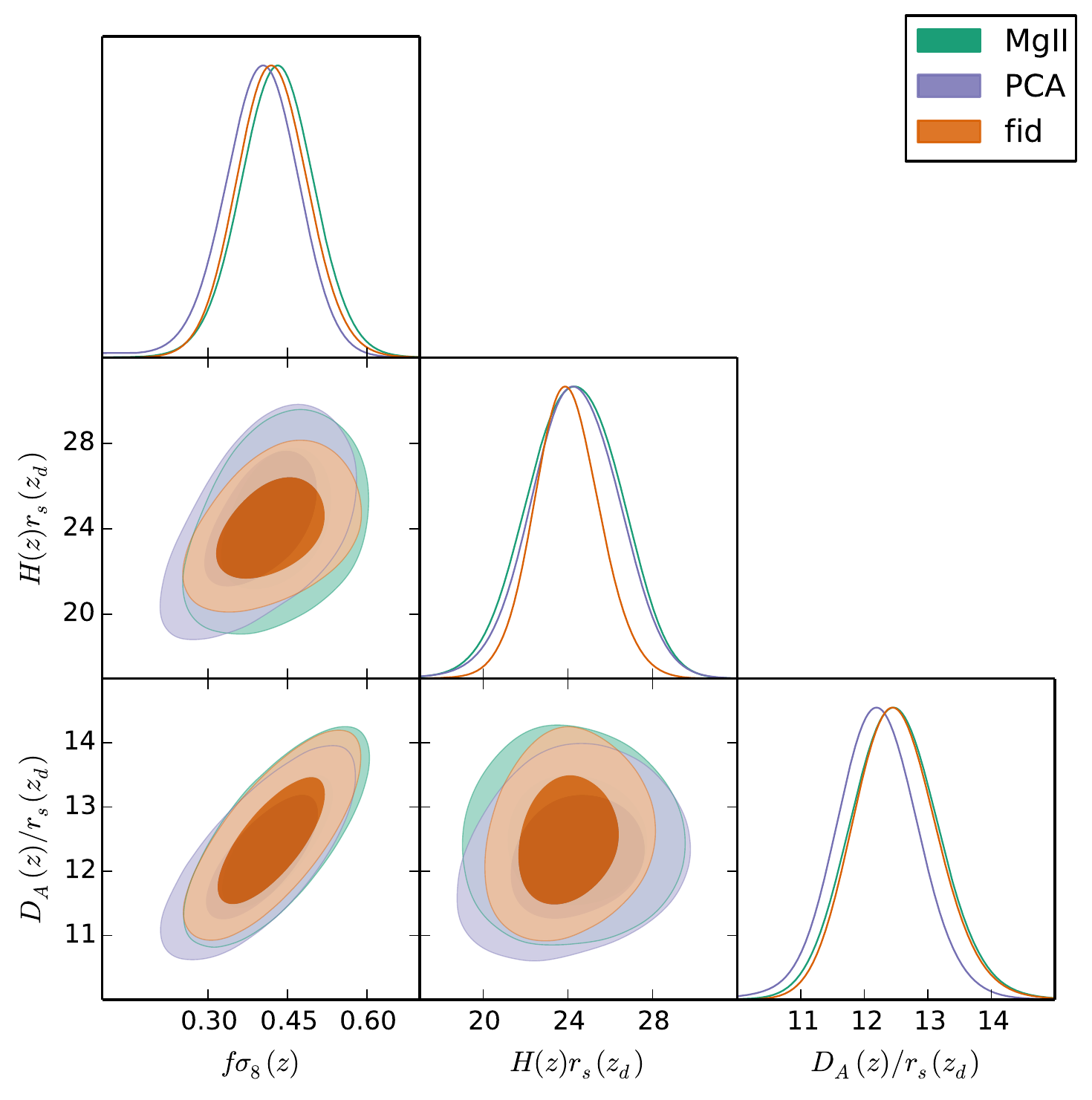}
\caption{Posterior likelihood contours from the DR14Q data in the redshift range $0.8\leq z \leq 2.2$ for the $f\sigma_8$, $D_A/r_s$, and $Hr_s$ cosmological parameters derived from the \textsc{mcmc} chains. The left panels display the results when the power spectrum monopole and quadrupole are used (green contours), and  when the hexadecapole signal is added (orange contours). The right panel shows the comparison for different redshift estimates, $z_{\rm fid}$ (in orange), $z_{\rm MgII}$ (in green) and $z_{\rm PCA}$ (in violet contours), for the power spectrum monopole, quadrupole and hexadecapole. When the power spectrum monopole and quadrupole are used alone (green contours on the left panel), we use a flat prior between 0.5 and 1.5 on the scale dilation factors, $\alpha_\parallel$ and $\alpha_\perp$, in order to improve the convergence of the chains. The units of $H(z)r_s(z_d)$ are $10^3\,{\rm km}\,s^{-1}$.}
\label{plot:contours}
\end{figure*}

Fig.~\ref{plot:contours} presents a further comparison among the different cases presented in Table~\ref{table:full_results_data} via 2D contour plots of the cosmological parameters of interest. All the measurements are made using 20 \textsc{mcmc} chains with half a million of steps each. For the case of monopole and quadrupole, a flat prior between 0.5 and 1.5 on $\alpha_\parallel$ and $\alpha_\perp$ is set to speed the convergence. Outside these ranges a $\sim x^2$ function is added to the $\chi^2$, where $x$ is the difference between the studied parameter (in this case $\alpha_\parallel$ and $\alpha_\perp$) and the limit of the prior (in this case 0.5 and 1.5). This condition correspond to flat priors within the range $15.83\leq H(z)r_s(z_d) [10^3\,{\rm km}\,s^{-1}] \leq 47.5$ and $6.07 \leq D_A(z)/r_s(z_d) \leq 18.22$. For  $H(z)r_s(z_d)<15.83\,\times10^3\,{\rm km}\,s^{-1}$, the prior is the responsible of the abrupt change in the shape of the likelihood function.

The left panel shows the comparison among $D_A(z)/r_s(z_d)$, $H(z)r_s(z_d)$ and $f\sigma_8$ for the redshift estimate, $z_{\rm fid}$ when the power spectrum monopole and quadrupole is used (green contours), and when the hexadecapole is added (orange contours). The information contained by the hexadecapole improves the constraint on $H(z)r_s(z_d)$ and consequently breaks the degeneracy between $H(z)r_s(z_d)$ and the other two parameters. The right panel we show the constraints on the same variables for the case where the three multipoles are used, but now under different redshift estimates: $z_{\rm fid}$ as before in orange contours, $z_{\rm MgII}$ in green and $z_{\rm PCA}$ in purple contours. 
 In summary, we obtain a good agreement among the different redshift estimates when the three power spectrum multipoles are used, which strongly supports the idea that the results are not significantly affected by the choice of the automated redshift classification. However, we believe that $z_{\rm fid}$ is the best procedure of obtaining the redshifts and we adopt those as the main results of this paper.

We present the results of the full-AP fits in a form of a vector and its covariance when the power spectrum monopole, quadrupole, and hexadecapole, and when the $z_{\rm fid}$ is being used as redshift estimate. The rest of the specifications are set to ``comb'', as described in \S~\ref{sec:sysdata}. We define the data vector as, 
\begin{equation}
 \label{datafullAP}
D_{\rm data} = 
 \begin{pmatrix}
  f(z_{\rm eff})\sigma_8(z_{\rm low})  \\
  H(z_{\rm eff})r_s({z_d}) 10^3\, [{\rm km\,s^{-1}}] \\
        D_A(z_{\rm eff})/r_s({z_d}) \\
 \end{pmatrix},
 \end{equation}
whose values are given by the mean of the chains steps,
\begin{equation}
 \label{datafullAPvalues}
D_{\rm data} = 
 \begin{pmatrix}
 0.420478 \\
  24.009353 \\
  12.486616 \\
 \end{pmatrix}.
 \end{equation}
 Using the \textsc{mcmc} chains described above we compute the covariance matrix. Adding the systematic budget described before in Table~\ref{table:systematics} the final covariance reads, 
  \begin{equation}
\label{covfullAPvalues}
C = 10^{-3}
 \begin{pmatrix}
4.137 +1.598 &  39.19 & 29.96  \\
 - &  2314+791.8 & 152.5  \\
  - &  -  & 397.7+108.7 \\
 \end{pmatrix},
 \end{equation}
where the sums in the diagonal elements correspond to the systematic contribution.  
As previously described in \S\ref{sec:isotropicfits}, in order to compute the data-vector and the covariance we take only those chain-steps within $3\sigma$ confidence regions around the best-fit, which for three parameters correspond to those steps whose $\chi^2$ is $\lesssim \chi^2_{\rm min}+14.16$. This choice allows a Gaussian-approximated covariance which is closer to the actual full contours (see Appendix~\ref{appendix:gaussian} for the performance of this approximation). 

The marginalised errors for the cosmological parameters which contain the full error budget are, $f\sigma_8(1.52)=0.420\pm0.076$ and $D_A(1.52)/r_s(z_d)= 12.48  \pm 0.71$ and $H(1.52)r_s(z_d)=[24.0\pm 1.8]\times 10^3\, {\rm km\,s}^{-1}$ with a correlation coefficient of $\rho_{[f\sigma_8-D_A/r_s]}=0.74$, $\rho_{[f\sigma_8-Hr_s]}=0.40$ and $\rho_{[D_A/r_s-Hr_s]}=0.16$

In \S~\ref{sec:cosmo} we will explore the cosmological constraints drawn from these results.

\subsection{Multiple Redshift Bins}\label{sec:results_zbins}

We perform a parallel analysis to that presented in the above \S~\ref{sec:anisotropic_results}, re-doing the full-AP fits in three overlapping redshift bins. We refer to them as {\it lowz}: $0.8\leq z \leq 1.5$ with effective redshift of $z_{\rm lowz}=1.19$; {\it midz} $1.2\leq z \leq 1.8$ with effective redshift of $z_{\rm midz}=1.50$; and {\it highz} $1.5\leq z \leq 2.2$ with effective redshift $z_{\rm highz}=1.83$. Table~\ref{table:resultsz} displays the best-fitting results for these three redshift bins. The measurements and best-fitting models were also gvien in Fig.~\ref{Pkz:plot} and briefly discussed in \S~\ref{sec:redshiftbins}. The approach of dividing the full redshift range into overlapping redshift bins is complementary to the single broad redshift bin analysis presented above. Although some large-scale signal is lost when dividing the sample, the three redshift bin analysis has the advantage of capturing the redshift evolution of parameters, such as the structure growth factor or the galaxy bias. In the following section we will compare the multipole $z$-bin approach with the single $z$-bin, showing the different power when constraining cosmological parameters. An alternative approach for analysing redshift dependent quantities without sub-dividing the full redshift range was proposed by \cite{Zhuetal:2015} and developed specifically for redshift space distortions in \cite{Ruggerietal:2017}. Also in \cite{Ruggerietal:2018a} this technique was tested on the DR14Q \textsc{ez} mocks and it is presented for the same DR14Q dataset in the companion papers, \cite{Ruggerietal:2018b,Zhaoetal:2017}. 

The first fit to the {\it midz} redshift bin produced a high value of $\chi^2$, $122/(84-7)$, which is a more than $3\sigma$ fluctuation, and none of the studied mocks have such high $\chi_{\rm min}^2$ (the highest value for the best-fitting $\chi^2$ from the mocks on the {\it midz} redshift bin is 107) when they are analysed in the same manner as the DR14Q dataset. Examining Fig.~\ref{Pkz:plot} (purple symbols for {\it midz} redshift bin), reveals that the origin of this high $\chi^2$ is two $>3\sigma$ outliers, one in the quadrupole ($3.2\sigma$ offset), the second one in the hexadecapole ($4\sigma$ offset) at $k\simeq0.11\,h{\rm Mpc}^{-1}$. Although one outlier at $\sim3\sigma$ is expected given the number of degrees of freedom ($84-7$), two $>3\sigma$ offsets are very unlikely ($<0.1\%$). We believe that such deviations are caused by an uncorrected observational systematic of unknown origin. We have checked the shape of the hexadecapole for the two extra redshift estimates, and both present these two features at similar significance, $3.2\sigma$ and $3.6\sigma$ for the $z_{\rm MgII}$ and $z_{\rm PCA}$ redshifts estimates, respectively.  Therefore, using a different redshift estimate does not modify the high $\chi^2$ issue ($\chi^2_{\rm PCA}=107$ and $\chi^2_{\rm MgII}=103$). We also have investigated these two features in the NGC and SGC patches separately. For example, the feature in the  hexadecapole is equally present in the NGC and SGC patches, with measurements of $P_{\rm NGC}^{(4)}(k=0.115\,h{\rm Mpc}^{-1})=-3444\pm1608\,[h^{-1}{\rm Mpc}]^3$ and $P_{\rm SGC}^{(4)}(k=0.115\,h{\rm Mpc}^{-1})=-5355\pm2005\,[h^{-1}{\rm Mpc}]^3$(and $P_{\rm N+S}^{(4)}(k=0.115\,h{\rm Mpc}^{-1})=-5161\,\pm1246[h^{-1}{\rm Mpc}]^3$ in the combined NGC+SGC sample), whereas the prediction from the mean of the 1000 realisations of the \textsc{ez} mocks is $-116\pm40\,[h^{-1}{\rm Mpc}]^3$. This result translates into a $2\sigma$ deviation for the NGC, and a $2.6\sigma$ deviation for the SGC, which are not particularly high if they are analysed individually. However, when both patches are combined, the fluctuation rises up to the reported $4\sigma$.  In the full redshift range, $0.8\leq z \leq 2.2$, such a systematic is probably diluted among the other two redshift bins, which reduces the tension between the model and the measurement, providing a consistent $\chi_{\rm min}^2=81$ as reported in Table~\ref{table:full_results_data}. In order to test the impact of these systematics in the parameters of the model, we remove these two frequencies in the corresponding multipole and redo the fitting process. After vetoing just these two $>3\sigma$ outliers the $\chi^2$ is reduced to 91, confirming, that the origin of the high $\chi^2$ is produced by these frequencies at $k\simeq 0.11\,h\,{\rm Mpc}^{-1}$. We check that the mean values of the model are not affected by more than $0.33\sigma$ statistical shifts, where $f\sigma_8$ is the most affected parameter. We leave for a future work the study and characterisation of this systematic effect. From this point we proceed our analysis using the vetoed {\it midz} sample.

Table~\ref{table:resultsz} displays the measurements of the cosmological parameters of interest for the different redshift bins. Some of these parameters indicate a redshift evolution across the three redshift bins. Of particular interest is $\sigma_P$, whose magnitude increases with redshift by $3\sigma$. The $\sigma_P$ parameter  partially captures the uncertainty on the redshift estimation, as described before at the end of  \S\ref{sec:qpmmocks}. Certainly, since the redshift error increases with redshift, $\sigma_P$ necessarily has to increase as well. The other parameters have a negligible dependence with redshift, such as $f\sigma_8$ and $b_1\sigma_8$. Although $f(z)$ and $b_1(z)$ may have a strong increasing dependence with redshift, $\sigma_8(z)$ decreases with redshift, which counterbalances their effect. The quantities $\alpha_\parallel$ and $\alpha_\perp$ represent deviations with respect to a fiducial model that do change with redshift. Therefore, although we do not detect an explicit redshift dependence on these parameters, the fiducial model does change with redshift, and the cosmological information we obtained by having three measurements of $D_A/r_s$ and $Hr_s$ instead of a single one is more interesting. 

Since the {\it midz} redshift bin fully overlaps with {\it lowz} and {\it highz}, a large correlation among the different parameters is expected. Fig.~\ref{fig:covzbins} displays the correlation among the cosmological parameters, $f\sigma_8(z)$, $D_A(z)/r_s(z_d)$ and $H(z)r_s(z_d)$ computed from the \textsc{ez} mocks. We use these correlation factors to compute the off-diagonal coefficient terms across $z$-bins of the $9\times9$ covariance matrix of the data for the cosmological parameters, $f\sigma_8(z)$, $D_A(z)/r_s(z_d)$ and $H(z)r_s(z_d)$ at $z_{\rm lowz}=1.19$, $z_{\rm midz}=1.50$ and $z_{\rm highz}=1.83$. For the off-diagonal coefficients terms belonging to the same $z$-bins we keep the cross-correlation value obtained by the Gaussian approximation to the data, which is consistent with the one obtained with the mocks.

\begin{figure}
\centering
\includegraphics[trim={120 70 0 0},clip=false,scale=0.4]{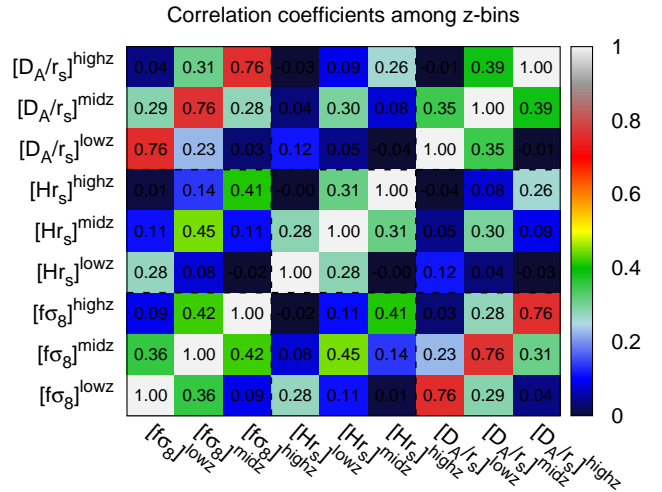}
\caption{Cross-covariance for the {\it lowz}, {\it midz} and {\it highz} redshifts bins for the $f\sigma_8$, $Hr_s$ and $D_A/r_s$ parameters derived from the mocks. The {\it midz} redshift bin fully overlaps with {\it lowz} and {\it highz} and therefore some correlation of parameters is expected. The typical cross correlation factors are $0.3-0.4$, in line with the fraction of overlapped volume. The {\it lowz} and {\it highz} datasets, however, are non-overlapping and therefore have small correlation ($\ll 0.1$) among their parameters.}
\label{fig:covzbins}
\end{figure}
As before, we present the results of this section in form of a data vector along with its covariance matrix. We run \textsc{mcmc}-chains to the individual overlapping redshift bins, as was done in the previous sections. In this case, we set up different priors to avoid secondary minima at $\alpha_\parallel$ and $\alpha_\perp$ outside  $0.8\leq \alpha_{\parallel,\,\perp}\leq 1.2$. These secondary minima arises as a consequence of having subsamples with smaller volumes compared to the single bin case. For e.g., we have identified a second minima in the {\it highz} sample, whose $\chi^2$ is  $\sim\chi^2_{\rm min}+2$, and which is located  at $f\sigma_8\simeq0.8$ and  $\alpha_\perp\simeq1.4$ (see Appendix~\ref{appendix:gaussian} for further details). Table~\ref{table:priors} summarises the different priors set on the different samples.
 \begin{table}
\caption{Flat priors ranges set on the three redshift-bin samples on the AP dilation scales. The priors on the other parameters are the same than those described in Table~\ref{table:generalpriors}.}
\begin{center}
\begin{tabular}{|c|c|c}
\hline
\hline
sample & $\alpha_\parallel$-flat prior & $\alpha_\perp$-flat prior \\
\hline
$0.8<z<2.2$ & $[0.00,\, 2.00]$ & $[0.00,\,2.00]$ \\
{\it lowz} & $[0.50,\, 1.50]$ & $[0.50,\,1.50]$ \\
{\it midz}& $[0.65,\, 1.50]$ & $[0.50,\,1.50]$ \\
{\it highz} & $[0.70,\, 1.70]$ & $[0.80,\,1.20]$ \\
\hline
\hline
\end{tabular}
\end{center}
\label{table:priors}
\end{table}%
 
 The resulting data vector taking the \textsc{mcmc} steps whose $\chi^2\leq\chi^2_{\rm min}+14.16$ along with the corresponding $9\times9$ covariance matrix is presented in Table~\ref{table:covariance}. The covariance is constructed using the diagonal terms, as well as the off-diagonal terms belonging to  same redshift bin, extracted from the \textsc{mcmc} chains computed at the three redshift bins using the same criteria described for the data-vector, combined with the cross correlation coefficients from different redshift-bins derived from the mocks and presented in Fig.~\ref{fig:covzbins}. The systematic error contribution is already included in the values of Table~\ref{table:systematics}.
\begin{figure}
\centering
\includegraphics[scale=0.5]{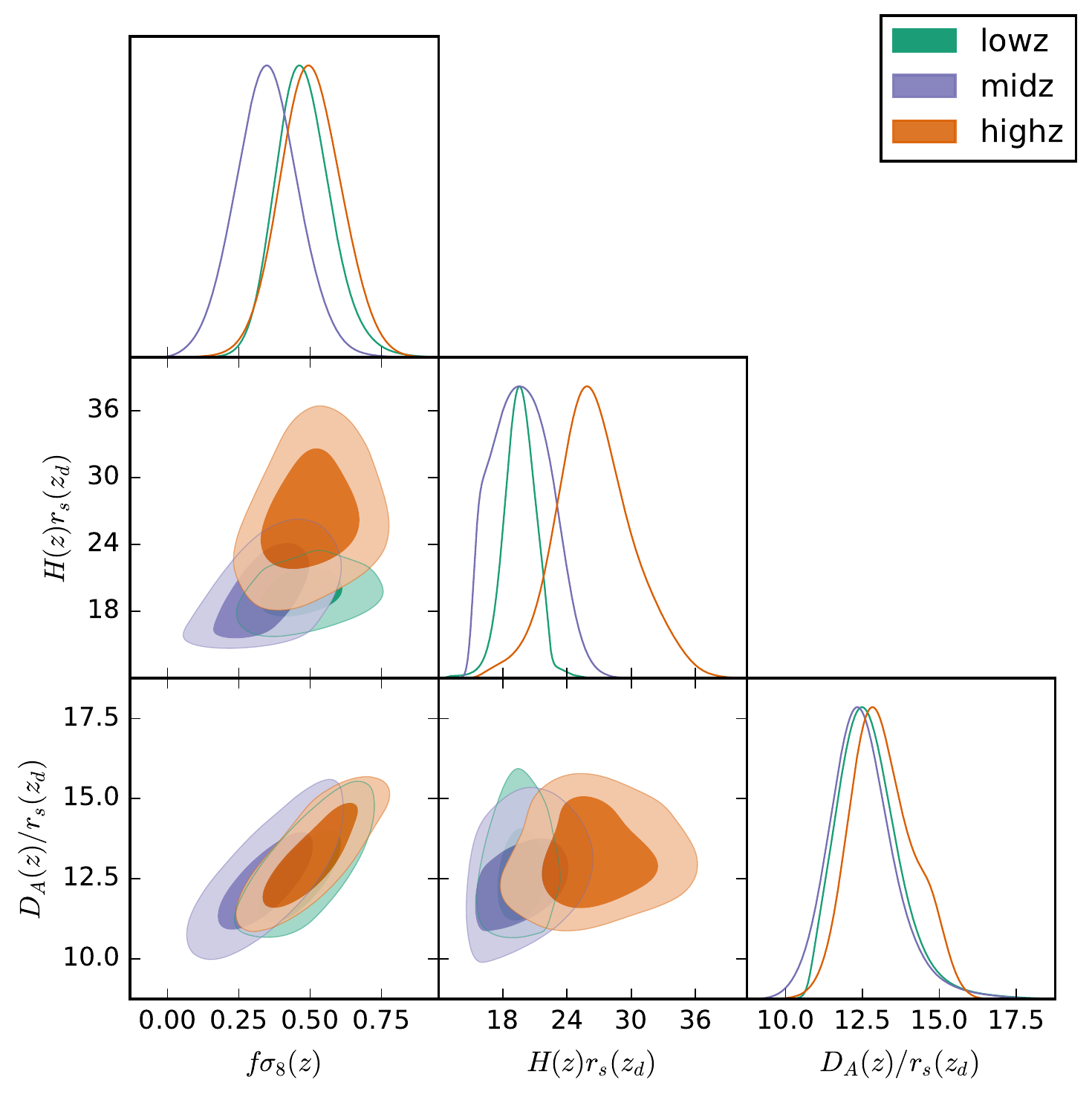}
\caption{Posterior likelihood contours from the DR14Q data corresponding to the different redshift bins: {\it lowz} $z_{\rm eff}=1.19$ (green contours); {\it midz} $z_{\rm eff}=1.50$ (purple contours); $z_{\rm eff}=1.83$ (orange contours); for $f\sigma_8$, $D_A/r_s$ and $Hr_s$ derived from the \textsc{mcmc} chains. In all cases the power spectrum monopole, quadrupole, and hexadecapole have been used. In the case of the {\it midz} two $k$-wave numbers have been vetoed as described in the main text. The priors set on the different parameters are displayed in Table~\ref{table:priors}.  The units of $H(z)r_s(z_d)$ are $10^3\,{\rm km\,s}^{-1}$.}
\label{fig:cosmozbins}
\end{figure}

Fig.~\ref{fig:cosmozbins} displays the full non-Gaussian posterior corresponding to the different redshift bins on the cosmological parameters of interest. The contours have been produced using the full set of \textsc{mcmc}-chains and not the Gaussian approximation provided by Table~\ref{fig:cosmozbins} (see Appendix~\ref{appendix:gaussian} for the differences between the actual likelihood posterior shape and its corresponding Gaussian approximation). In Eq. \ref{plot:cosmo2} we display these measurements as a function of redshift along with other probes.

\begin{table*}
\caption{Data vector and covariance matrix elements, for the results derived from the three overlapping redshift bins, {\it lowz} (L super-index), {\it midz} (M super-index) and {\it highz} (H super-index). The first row displays the data vector measurements, $x_i$ in the same units used on the data-vector of Eq. \ref{datafullAP}. The second row displays the diagonal errors (with their systematic contribution), $\sigma_i$, of the covariance matrix corresponding to the data-elements above. Below, the cross-correlation among the data-vector elements are presented.}
\begin{center}
\begin{tabular}{ccccccccccc}
\hline
\hline
&  $[f\sigma_8]^{\rm L}$ & $[f\sigma_8]^{\rm M}$ & $[f\sigma_8]^{\rm H}$ & $[Hr_s]^{\rm L}$ & $[Hr_s]^{\rm M}$ & $[Hr_s]^{\rm H}$ & $[D_A/r_s]^{\rm L}$ & $[D_A/r_s]^{\rm M}$ & $[D_A/r_s]^{\rm H}$ \\
\hline
$x_i$ & $0.4736$ & $0.3436$ & $0.4998$ & $19.6782$ & $19.8637$ & $26.7928$ & $12.6621$ & $12.4349$ & $13.1305$ \\
\hline
$\sigma_i$ & $0.0992$ & $0.1104$ & $0.1111$ & $1.5866$ & $2.7187$ & $3.5632$ & $0.9876$ & $1.0429$ & $1.0465$ \\
\hline
$[f\sigma_8]^{\rm L}$ & $1.0000$ & $0.3563$ & $0.0917$ & $0.3156$ & $0.1103$ & $0.0081$ & $0.7192$ & $0.2882$ & $0.0425$ \\
$[f\sigma_8]^{\rm M}$ & $-$ & $1.0000$ & $0.4244$ & $0.0820$ & $0.5231$ & $0.1388$ & $0.2280$ & $0.7446$ & $0.3089$  \\
$[f\sigma_8]^{\rm H}$ & $-$ & $-$ & $1.0000$ & $-0.0239$ & $0.1083$ &  $0.2490$ & $0.0323$ & $0.2795$ & $0.7954$ \\
 $[Hr_s]^{\rm L}$ & $-$ & $-$ & $-$ & $1.0000$ & $0.2836$ & $-0.0005$ & $0.1024$ & $0.0385$ & $-0.0304$ \\
$[Hr_s]^{\rm M}$ & $-$ & $-$ & $-$ & $-$ & $1.0000$ & $0.3144$ & $0.0462$ & $0.3462$ & $0.0904$ \\
$[Hr_s]^{\rm H}$ & $-$ & $-$ & $-$ & $-$ & $-$ & $1.0000$ & $-0.0399$ & $0.0819$ & $0.0637$ \\
 $[D_A/r_s]^{\rm L}$ & $-$ & $-$ & $-$ & $-$ & $-$ & $-$ & $1.0000$ & $0.3490$ & $-0.0065$  \\
$[D_A/r_s]^{\rm M}$ & $-$ & $-$ & $-$ & $-$ & $-$ & $-$ & $-$ & $1.0000$ & $0.3890$ \\
$[D_A/r_s]^{\rm H}$ & $-$ & $-$ & $-$ & $-$ & $-$ & $-$ & $-$ & $-$ & $1.0000$ \\
\hline\hline
\end{tabular}
\end{center}
\label{table:covariance}
\end{table*}%

\begin{table}
\caption{Parameters of the model for the full AP-fit when the DR14Q $0.8\leq z\leq2.2$ sample is divided in three overlapping redshift bins. We report the results for a fiducial redshift estimate, $z_{\rm fid}$ for the NGC+SGC sample. The results of the {\it midz} sample correspond to the vetoed sample (see text). For all cases we used the \textsc{ez}-derived covariance and the `standard' specifications corresponding to the definition in \S\ref{sec:sysdata}.  The correlation among the different parameters are presented in Fig.~\ref{fig:covzbins}. The units of $\sigma_P$ are $[{\rm Mpc}\,h^{-1}]$. The errors shown only represent the statistical error budget}
\begin{center}
\begin{tabular}{|c|c|c|c}
\hline
\hline
N+S ($z_{\rm fid}$)  & $0.8\leq z \leq 1.5$ & $1.2\leq z \leq 1.8$ & $1.5 \leq z \leq 2.2$ \\
  \hline
 $b_1\sigma_8$  & $0.900\pm0.056$  & $0.945 \pm 0.059$ & $0.947 \pm 0.077$ \\
 $f\sigma_8$   & $0.440\pm0.087$  & $0.364 \pm 0.093$ & $0.468 \pm0.091$ \\
 $\alpha_{\parallel}$  & $0.994\pm0.075$  & $1.05 \pm 0.12$ & $0.98 \pm 0.13$ \\
 $\alpha_{\perp}$  & $1.027\pm0.075$  &  $1.014 \pm 0.069$ & $1.039\pm0.067$ \\
 $b_2\sigma_8$  & $0.66\pm0.85$  & $0.71\pm 0.60$ & $0.87 \pm0.55$ \\
 $\sigma_P$  & $4.11\pm0.58$  & $5.25\pm 0.68$ & $6.38\pm0.77$ \\
 $A_{\rm noise}$  & $9.7\pm5.6$  & $7.8 \pm 4.0$ & $4.4\pm3.3$ \\ 
\hline
 $\chi^2/{\rm d.o.f}$ & $71/(84-7)$ &  $91/(82-7)$ & $99/(84-7)$ \\
 \hline
 \hline
\end{tabular}
\end{center}
\label{table:resultsz}
\end{table}%

\subsection{Bias evolution}

In this section we aim to compare the results on the measured linear bias of the quasars with previous measurements. \cite{Laurentetal:2017} measured the quasar correlation function monopole on the redshift range $0.9\leq z \leq 2.2$ for the eBOSS DR13 quasar sample \citep{Albaretietal:2016} and obtained $b_1(z=1.55)=2.45\pm0.05$ when the full redshift range was considered as a single bin, and when the sample was divided in several redshift bins (see black symbols and lines of Fig.~\ref{fig:bias}). In this paper we have measured $b_1\sigma_8(z)$ in a similar redshift range using the DR14 which contains $\sim 80,000$ more quasars and approximately twice the DR13 effective volume. Simply taking the ratio of our $b_1\sigma_8$ measurement and the Planck cosmology prediction for $\sigma_8$, $\sigma_8^{\rm Planck}$ produces $b_1\sigma_8(z)/\sigma_8^{\rm Planck}(z)\equiv b_1(z)$. When the isotropic fit is performed (i.e., setting $\epsilon=0$) using the power spectrum monopole and quadrupole measurements, $b_1(z=1.52)=2.30\pm0.11$, whereas the full-AP analysis using the three power spectrum multipoles yields $b_1(z=1.52)=2.32\pm0.10$. Both results are consistent, demonstrating that the bias measurements and errors do not depend on the type of fit used, or whether the hexadecapole is added. Also, our results are in good agreement with those presented in \cite{Laurentetal:2017}. We believe that the reason our errors are larger than those from \cite{Laurentetal:2017} is  because we marginalise over a larger set of nuisance parameters, such as $b_2$ and $\sigma_P$. Fig.~\ref{fig:bias} displays the measurements by \cite{Laurentetal:2017} when they subdivide the full redshift range in four non-overlapping redshift bins (black symbols) along with the best-fitting model as a function of redshift (solid black lines for the best-fitting model and dashed lines for $1\sigma$ confidence level). The coloured symbols display the measurements we report in this paper: purple symbols when the full redshift range is considered as a single bin and orange symbols when the redshift range is divided in the three previously mentioned redshift bins. The triangle-symbols represent the measurements when the full-AP fits are performed, whereas the circle-symbols indicate the isotropic fit. In all the cases there is an excellent agreement among the two analyses, demonstrating the consistency among the DR13Q and DR14Q and the two bias models used.

\begin{figure}
\includegraphics[trim={40 40 0 0},clip=false,scale=0.33]{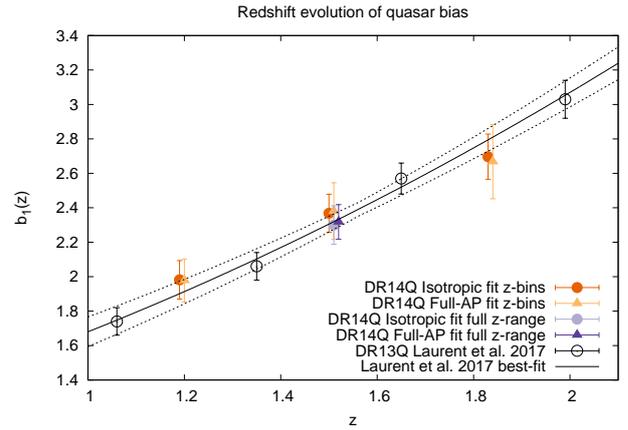}
\caption{Coloured symbols display the measured linear bias parameter for the DR14Q sample, as a function of redshift when the function $\sigma_8(z)$ from Planck cosmology is assumed, $b_1(z)\equiv[b_1\sigma_8(z)]/\sigma_8(z)^{\rm Planck}$, where $b_1\sigma_8(z)$ is the actual parameter measured in this paper. Orange symbols display the results when the DR14Q sample is divided in three overlapping redshift bins, {\it lowz}, {\it midz}, and {\it highz}. Purple symbols display the results when the full redshift range ($0.8\leq z \leq 2.2$) is considered as a single bin. Circles display the results when $b_1\sigma_8(z)$ is computed assuming $\epsilon=0$ (isotropic fit) and triangle symbols when this condition is relaxed (full-AP fit). Black empty symbols display the results found by \citealt{Laurentetal:2017} on the DR13Q sample using four different non-overlapping redshift bins, along with its best-fit (solid black lines and dashed black lines for $1\sigma$ uncertainties). For the three overlapping redshift bins, the correlation parameters are: {\it i)} for the isotropic case, $\rho_{\rm low-mid}=0.42$, $\rho_{\rm low-high}=0.04$, $\rho_{\rm mid-high}=0.42$; {\it ii)} for the full-AP case, $\rho_{\rm low-mid}=0.30$, $\rho_{\rm low-high}=-0.02$, $\rho_{\rm mid-high}=0.32$.}
\label{fig:bias}
\end{figure}

\section{Cosmological implications}\label{sec:cosmo}

In this section we compare and combine our cosmological results with other probes such as, the BOSS DR12 results \citep{Alametal2016,MdBetal:2017,Bautistaetal:2017} and CMB constraints from \cite{Planck15}.

\begin{figure*}
\includegraphics[scale=0.5]{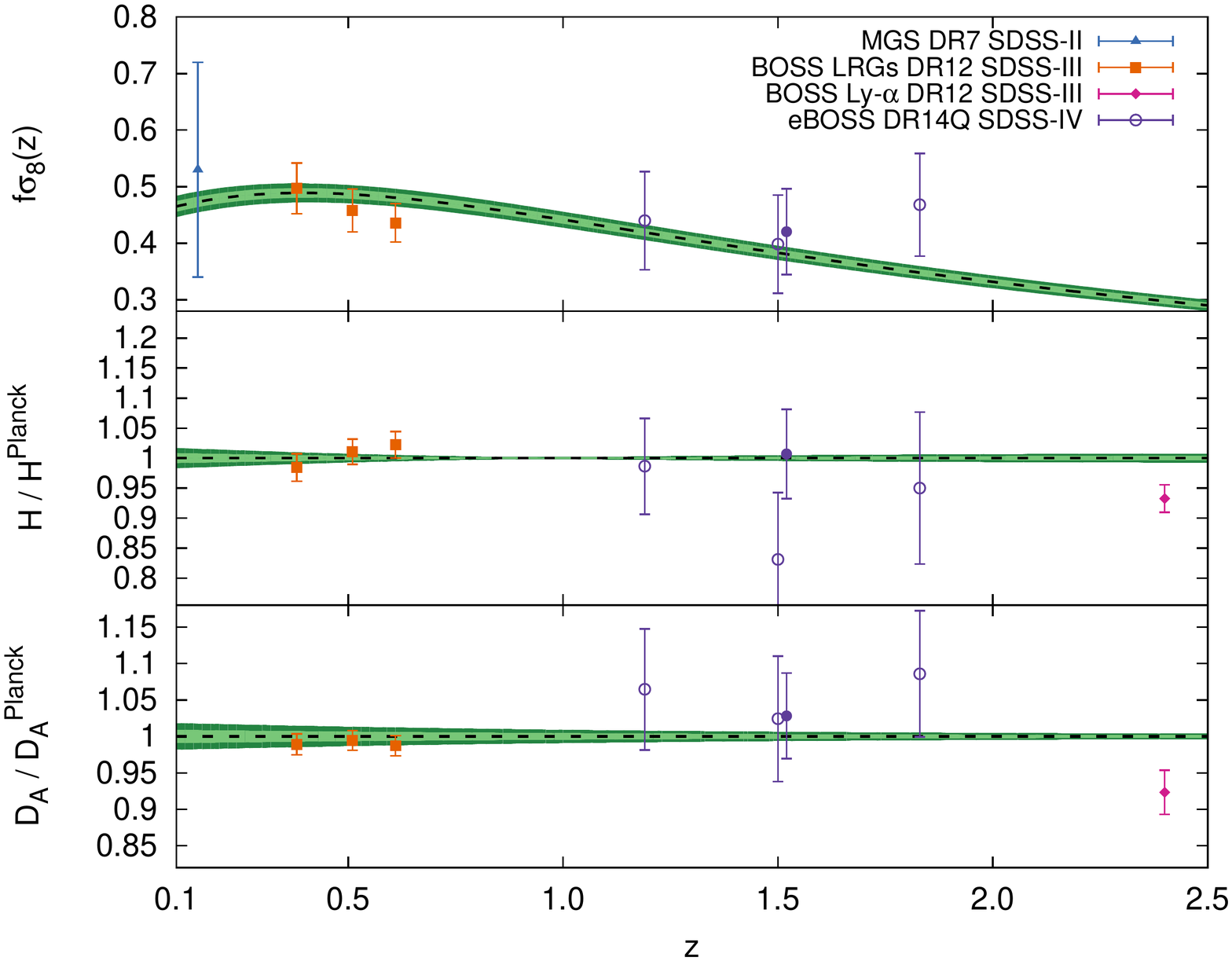}
\caption{The three panels show the redshift dependence of $f\sigma_8$, $H$ and $D_A$ inferred from a number of SDSS galaxy and quasar surveys. The black dashed lines along with the green bands display the predictions assuming a flat-$\Lambda$CDM Planck cosmology (\citealt{Planck15}). The blue triangle represents the RSD analysis of the SDSS-II MGS DR7 at $z_{\rm eff}=0.15$ (\citealt{Howlettetal:2015}); the orange squares display the BAO and RSD analyses from SDSS-III BOSS DR12 LRGs (\citealt{Alametal2016}) in the range $0.15\leq z \leq 0.75$; the magenta symbol represent the SDSS-III Ly-$\alpha$ measurement at $z_{\rm eff}=2.4$ from the auto- and cross-correlation analyses (\citealt{MdBetal:2017,Bautistaetal:2017}). The results derived from DR14Q SDSS-IV (this work) are represented by the purple symbols, where the filled symbols indicate the measurements from a single redshift bin analysis at $z_{\rm eff}=1.52$, and the empty symbols from the three overlapping redshift bins at $z_{\rm eff}=1.19$, $z_{\rm eff}=1.50$ and $z_{\rm eff}=1.83$. For clarity, the $D_A$ and $H$ quantities have been normalised to the fiducial prediction by the $\Lambda$CDM Planck cosmology. }
\label{plot:cosmo2}
\end{figure*}

The  panels of Fig.~\ref{plot:cosmo2} display the DR14Q measurement of $f\sigma_8$, $D_A$ and $H$ in purple circles, where the empty symbols represent the three overlapping redshift bin measurements of \S~\ref{sec:results_zbins}, and the filled symbols the measurement considering the full range as a single redshift bin as described in \S~\ref{sec:anisotropic_results}. Along with these measurements, we display the RSD results from the  Main Galaxy Sample (MGS) DR7 SDSS-II \citep{Howlettetal:2015}, the BOSS LRGs DR12 SDSS-III consensus results derived from RSD and BAO analyses \citep{Ataetal:2017}; and the measurement from the BOSS Lyman-$\alpha$  DR12 SDSS-III  auto- and cross-correlation result derived from BAO-only analyses \citep{MdBetal:2017,Bautistaetal:2017}. The black dashed line along with the green bands areas represent the $\Lambda$CDM-Planck prediction when a flat Universe is assumed \citep{Planck15}. The DR14Q measurements cover the, currently, unexplored region (in terms of $f\sigma_8$, $H$ and $D_A$ measurements) between redshifts 1 and 2, and are in fairly good agreement with the predictions from Planck.

\begin{table*}
\caption{Measurements on $\Omega_{m}$ and $\gamma$ produced by combining various datasets, when a flat-$\Lambda$CDM Universe has been assumed. These measurements are indicated by the corresponding colour contours of Fig.~\ref{plot:cosmo}. Using the eBOSS DR14Q at a single or at three overlapping redshift bins does not significantly affect the results, although the three redshift bin measurements have a slight larger constraining power on $\gamma$ if Planck results are not used. In all the cases the agreement of the measured cosmological parameters is in excelleng agreement with GR predictions.}
\begin{center}
\begin{tabular}{|c|c|c}
\hline
\hline
Dataset  / Model& $\Omega_{\rm m}$ & $\gamma$\\
\hline
eBOSS DR14Q 3$z$-bin + BOSS LRGs DR12 Cons. +  flat $\Lambda$CDM &  $0.313^{+0.040}_{-0.043}$ & $0.41\pm0.28$  \\
eBOSS DR14Q 1$z$-bin + BOSS LRGs DR12 Cons. +  flat $\Lambda$CDM &  $0.332^{+0.041}_{-0.045}$ & $0.34\pm0.31$  \\
\hline
eBOSS DR14Q 3$z$-bin + BOSS LRGs DR12 Cons. + Planck + flat $\Lambda$CDM & $0.3123^{+0.0072}_{-0.0074}$ & $0.55\pm0.19$ \\
eBOSS DR14Q 1$z$-bin + BOSS LRGs DR12 Cons. + Planck + flat $\Lambda$CDM & $0.3127^{+0.0075}_{-0.0071}$ & $0.54\pm0.19 $ \\
\hline
\hline
\end{tabular}
\end{center}
\label{table:cosmo}
\end{table*}%

\begin{figure}
\centering
\includegraphics[scale=0.33]{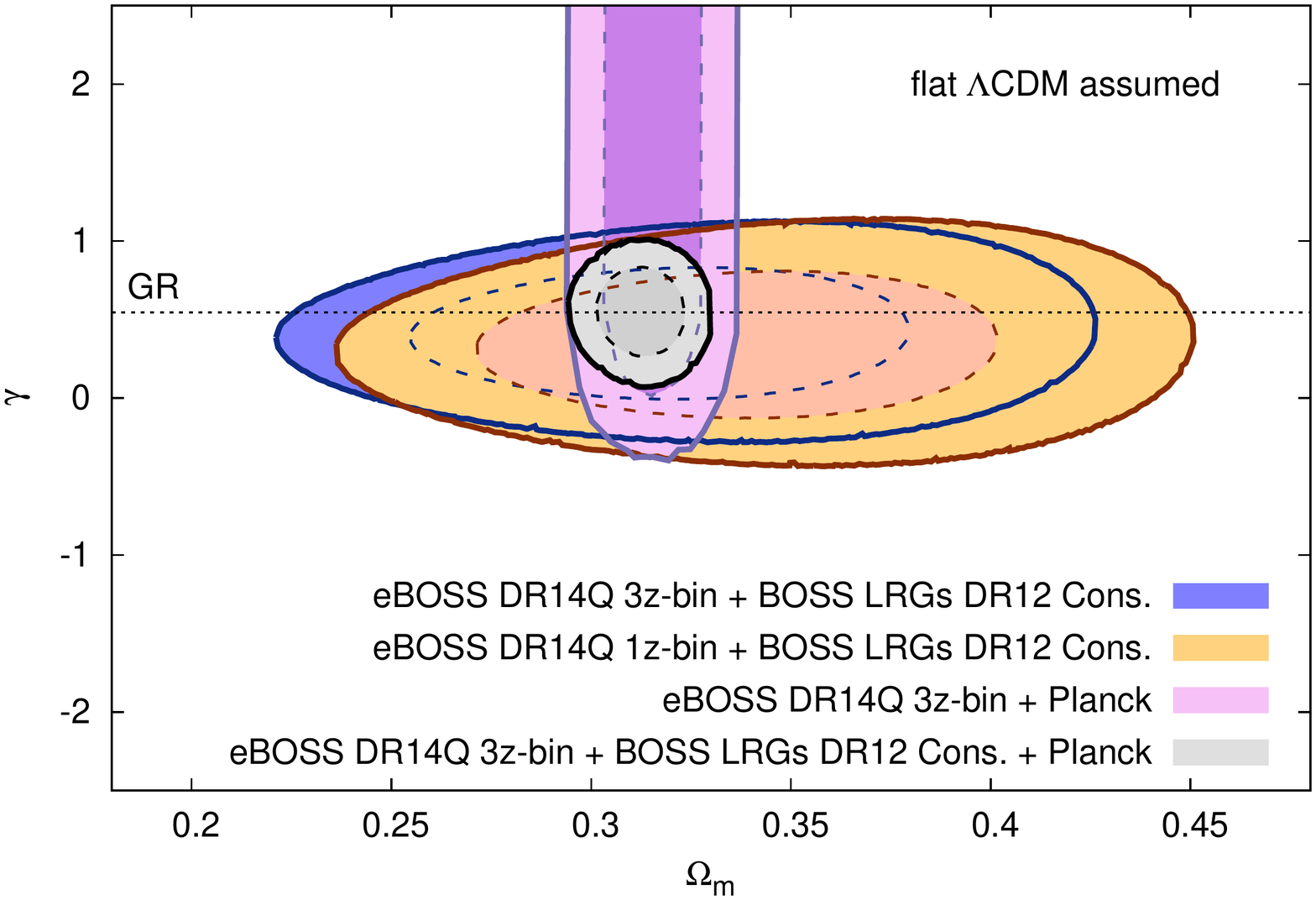}
\includegraphics[scale=0.33]{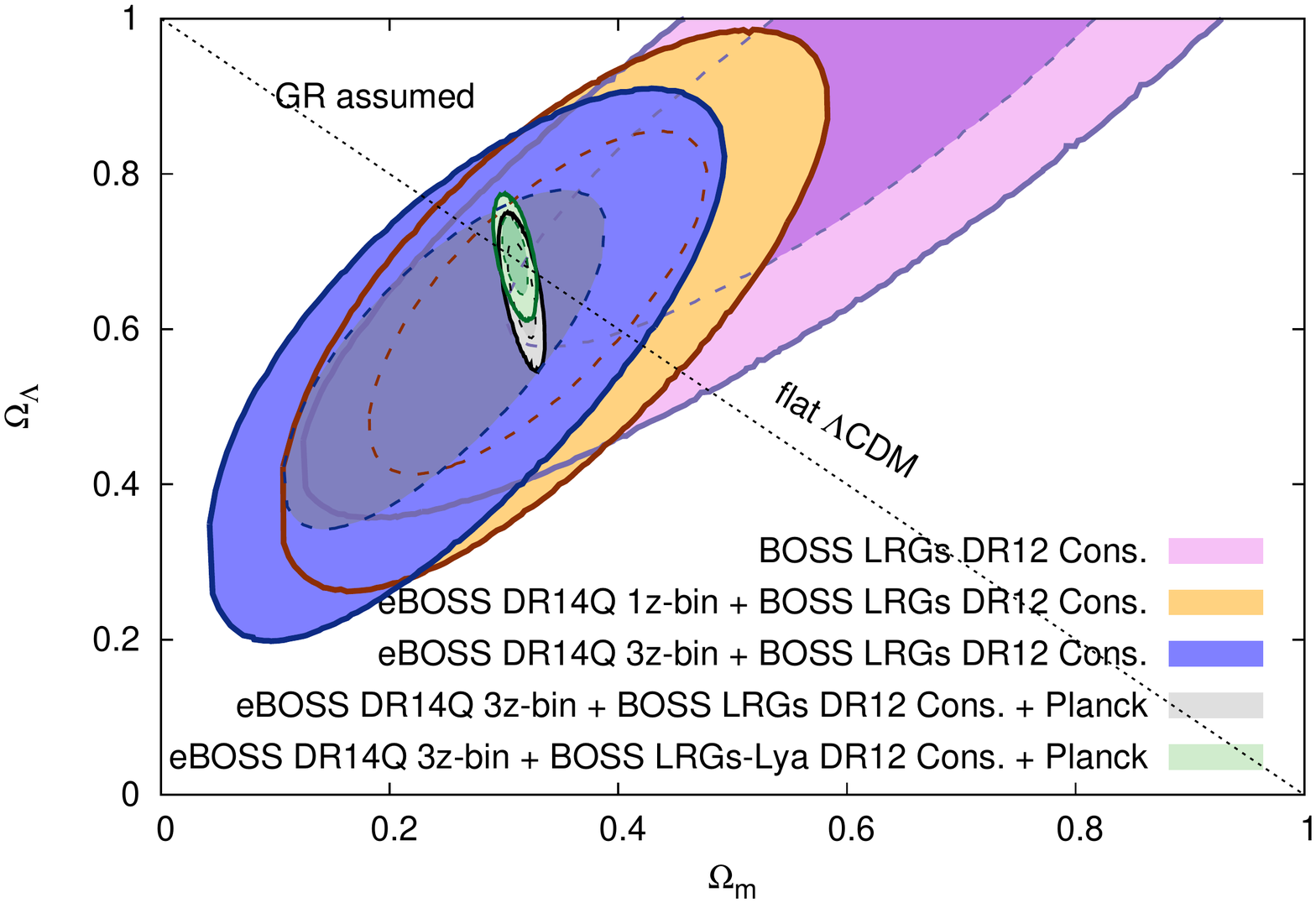}
\caption{{\it Top panel}: Constraints on the gravity model through the dependence between $\Omega_m$ and the $\gamma$ index, when a flat $\Lambda$CDM model is assumed. The black horizontal dashed line displays the prediction for GR. All results are consistent with $\Lambda$CDM-Planck cosmology + GR. {\it Bottom panel}: Constraints on the flatness of the Universe and through the relation between $\Omega_m$ and $\Omega_\Lambda$ when GR is assumed as the theory of gravity. The black dashed line indicates the prediction for a flat Universe, $\Omega_m+\Omega_\Lambda=1$. The colour contours display different probe combination among eBOSS DR14Q (this work), the BOSS DR12 LRGs (\citealt{Alametal2016}), BOSS DR12 Ly-$\alpha$ (\citealt{MdBetal:2017,Bautistaetal:2017}), and CMB data from \citealt{Planck15}. All results are consistent with a flat-$\Lambda$CDM Universe.}
\label{plot:cosmo}
\end{figure}

The methodology used to derive the cosmological parameters of this paper (as well as those of BOSS DR12) assumes General Relativity (GR) as the theory of gravity. In a $\Lambda$CDM scenario we relate the parameter for  the growth of structure, $f$, with the matter density of the Universe, $\Omega_m$, through the parametrisation, $f(z)=\Omega_m(z)^\gamma$ \citep{Kaiser:1987,Linder2005}, where $\gamma$ is the growth index; for GR $\gamma=0.55$. Therefore, determining $\Omega_m$ and $f$ through different physical processes allow us to perform a consistency test on the $\gamma$ parameter, which could potentially show departures from the GR prediction. We infer $f$ through the distortions of the peculiar velocities of the galaxies; and the value of $\Omega_m$ through the anisotropy generated by the AP effect and the Cosmic Microwave Background (CMB) data. In the top panel of Fig.~\ref{plot:cosmo} we perform a null-test of GR studying the dependencies between $\Omega_m$ and $\gamma$, where the contours display different combinations among eBOSS DR14Q, BOSS LRGs DR12 and Planck\footnote{In this paper we always use Planck cosmology to refer to those results on $H_0$, $\Omega_m$ and $\sigma_8$ derived from the TT+TE+EE+lowP,  fifth column from table 3 of \citealt{Planck15}.}, assuming a flat Universe model, $\Omega_\Lambda+\Omega_m=1$. For each dataset a Gaussian likelihood has been assumed, and the total likelihood has been constructed as the product of the individual likelihoods, as the different datasets are uncorrelated. For simplicity we do not exploit the Integrated Sachs-Wolfe effect (ISW) to place constraints on $\gamma$ using CMB measurements.

The constraints derived by combining LRG BOSS results with DR14Q provide a measurement of $\gamma$ with $60\%$ precision and $\Omega_m$ with  $\sim10\%$. These constraints become slightly better when we consider the DR14Q in three redshift bins (dark-blue contours), than in a single redshift bin (orange contours). The two top rows of Table~\ref{table:cosmo} display the results for these two cases. Combining the DR14Q results with Planck measurements (magenta contours) does not provide a competitive constraint on $\gamma$ due to the large errors of the $f\sigma_8$ measurements. Certainly, $\gamma$ regulates the amplitude of the $f\sigma_8$ parameter as a function of redshift, and it is particularly sensitive to $\gamma$ at low redshifts (see e.g. fig 12 from \citealt{RSDDR12}), where the BOSS LRG DR12 measurements dominate. Finally, we add all these three probes (grey contours) to obtain a $35\%$ measurement on $\gamma$, as shown in the two bottom rows of Table~\ref{table:cosmo}, $\gamma=0.55\pm0.19$. All the studied probe combinations are consistent with GR predictions. 

In the bottom panel of Fig.~\ref{plot:cosmo} we relax the flatness condition and fix $\gamma$ to be the predicted value by GR. We show the constraints on $\Omega_m$ and $\Omega_\Lambda$ when only the BOSS LRGs DR12 results are used (magenta contours). In this case, $\Omega_\Lambda$ and $\Omega_m$ present a large degeneracy which extends towards higher values of the $\Omega$. Adding eBOSS DR14Q data, in orange contours when only a single redshift bin is used, and in blue contours when the three redshift bins are considered, considerably breaks this degeneracy. The resulting constraints are $\{ \Omega_m,\,\Omega_\Lambda \}=\{0.322^{+0.095}_{-0.101},\, 0.64^{+0.15}_{-0.14} \,\}$ for BOSS LRGs DR12 + eBOSS DR14Q using a single redshift bin, and $\{ \Omega_m,\,\Omega_\Lambda\}=\{0.239^{+0.091}_{-0.098},\,0.57^{+0.15}_{-0.14} \,\}$ when three redshift bins are used. Both results are similar and are in good agreement with a flat-$\Lambda$CDM Universe with Planck best-fitting parameters. In both cases a Universe without Dark Energy ($\Omega_\Lambda=0$) is disfavoured by $4\sigma$ when only the BOSS LRGs DR12 and eBOSS DR14Q datasets are included. Adding the Planck and BOSS DR12 Ly-$\alpha$ results to these two datasets, provides tighter  constraints on the density of matter, $\Omega_m=0.3094^{+0.0076}_{-0.0080}$ ($2.5\%$ precision), and on the density of Dark Energy $\Omega_\Lambda =0.697^{+0.035}_{-0.032}$ ($0.5\%$ precision), again in full agreement with a Universe with no curvature, $\Omega_k=-0.007\pm0.030$.

\section{Consensus results}\label{sec:consensus}

The RSD analysis in this paper is based on the eBOSS DR14 quasar sample in the redshift range $0.8\leq z \leq 2.2$, using the power spectrum monopole, quadrupole and hexadecapole measurements on the $k$-range, $0.02\leq k\,[h{\rm Mpc}^{-1}]\leq 0.30$, shifting the centres of $k$-bins by fractions of $1/4$ of the bin size and averaging the four derived likelihoods. Applying the TNS model along with the 2-loop resumed perturbation theory, we are able to effectively constrain the cosmological parameters $f\sigma_8(z)$, $H(z)r_s(z_d)$ and $D_A(z)/r_s(z_d)$ at the effective redshift $z_{\rm eff}=1.52$, along with the remaining `nuisance' parameters, $b_1\sigma_8(z)$, $b_2\sigma_8(z)$, $A_{\rm noise}(z)$ and $\sigma_P(z)$, in all cases with wide flat priors. 

This work is released along with four other complementary RSD analyses based on the exact same sample, including identical weighting schemes (described in \S~\ref{weights:sec}), but using slightly different techniques and observables. The fiducial cosmology in which the sample has been analysed is also the same across papers. We briefly describe the other DR14Q works below.
\begin{itemize}
\item \cite{Houetal:2017} perform a RSD analysis using Legendre polynomial with order $\ell=0,2,4$ and clustering wedges. They use "gRPT" to model the non-linear matter clustering. For the RSD, they use a streaming model extended to one-loop contribution developed by \cite{Scoccimarro:2004} and \cite{TNS} and a non-linear corrected FoG term. Finally, the bias modelling adopted is the one described in \cite{Chanetal:2012}, which includes both local and non-local contribution.  A modelling for spectroscopic redshift error is also included.

\item \cite{Ruggerietal:2018b} present a RSD analysis using an optimised redshift-dependent weighting scheme presented in \cite{Ruggerietal:2017,Ruggerietal:2018a}. A Fourier space analysis is then applied, using evolving power spectrum multipoles to measure cosmological parameters alongside with its evolution across the redshift bin. 

\item \cite{Zarrouketal:2017} describe a RSD analysis using Legendre multipoles with $\ell=0,2,4$ and three wedges of the correlation function on the $s$-range from 16~$h^{-1}{\rm Mpc}$ to 138~$h^{-1}{\rm Mpc}$. They use the Convolution Lagrangian Perturbation Theory with a Gaussian Streaming model and demonstrate its applicability for dark matter halos of masses of the order of $10^{12.5}{\rm M}_\odot$ hosting eBOSS quasar tracers at mean redshift $z\simeq1.5$ using the \textsc{or} simulation. 

\item The combined BAO and RSD analysis presented in \cite{Zhaoetal:2017} takes only into account the power spectrum monopole and the quadrupole, in the $k$-range of $0.02\leq k\,[h{\rm Mpc}^{-1}]\leq 0.30$. The power spectrum template utilised is based on the regularised perturbation theory up to second order. With the optimal redshift weights, they constrain $D_{\rm A}, H$ and $f\sigma_8$ at four effective redshifts, $0.98$, $1.23$, $1.53$ and $1.94$.

\end{itemize}

 All these papers provide constraints on the same cosmological parameters, $f\sigma_8(z_{\rm eff})$ $D_{A}(z)/r_{\rm d}$, $H(z)r_d$ (at least) at the effective redshift $z_{\rm eff}=1.52$, and therefore their constrain can be easily compared.

\begin{figure}
\centering
\includegraphics[scale=0.43]{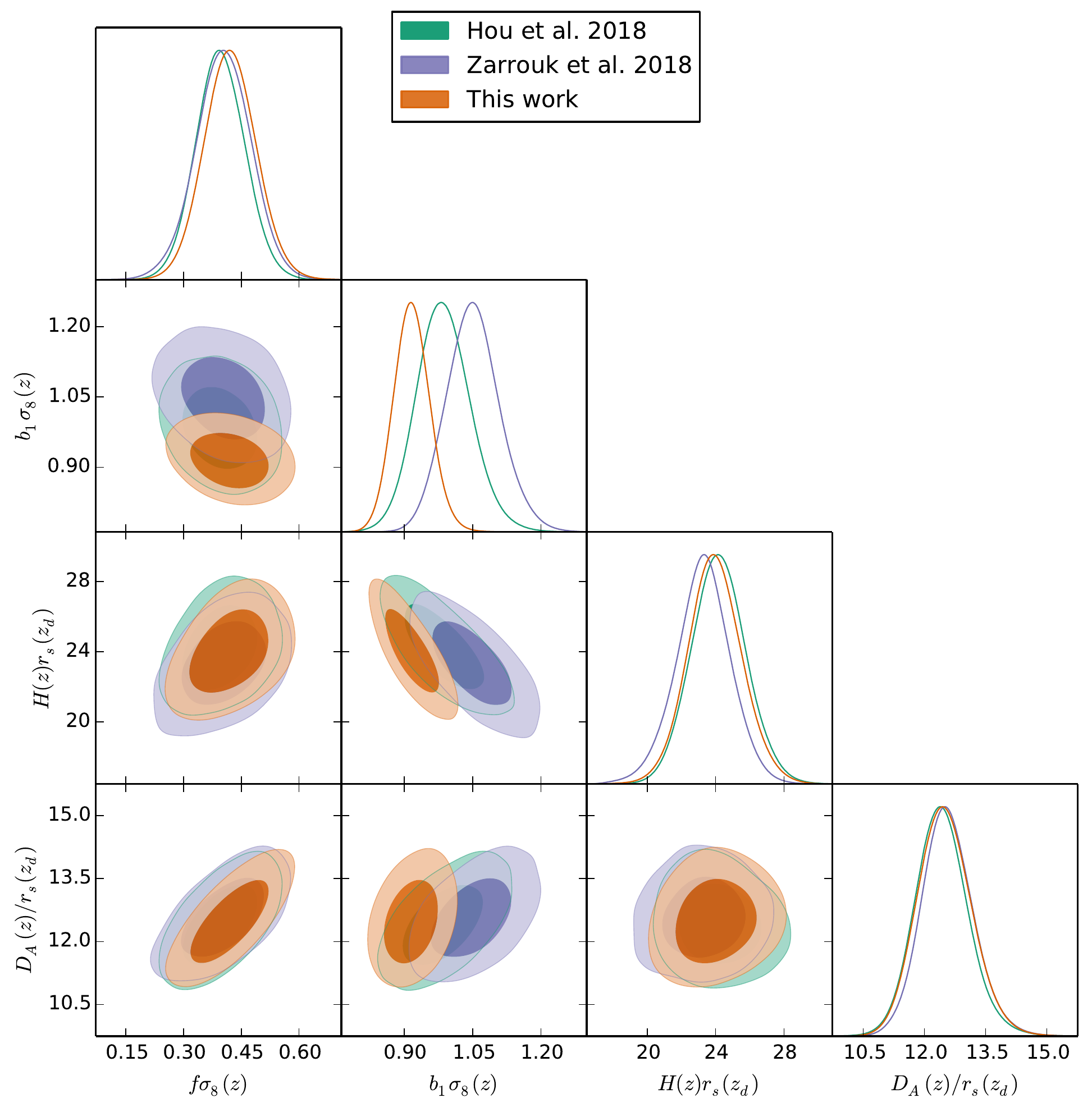}
\caption{Parameter contours for $b_1\sigma_8(z_{\rm eff})$, $f\sigma_8(z_{\rm eff})$, $D_A(z_{\rm eff})/r_s(z_d)$, and $H(z_{\rm eff})r_s(z_d)$ for the predictions by the companion papers using the same DR14Q dataset. Purple contours represent the prediction by \citealt{Zarrouketal:2017} analysis and green contours the predictions by \citealt{Houetal:2017}; both analyses use the configuration space multipoles (see text for details). Orange contours display the results presented in this work (using a single redshift bin).  All the three analyses produce consistent values of cosmological parameters, $f\sigma_8$, $D_A/r_s$, and $Hr_s$, both in the actual measurement as well as in the size and shape of the confidence level contours. The $\sim1\sigma$ difference on $b_1\sigma_8$ is caused by different prior conditions on the bias parameters and by the different bias models used (see text for a further description).}
\label{plot:consensus}
\end{figure}

Fig.~\ref{plot:consensus} displays the constraints represented as contours for those companion paper RSD analyses described above, which do not apply any redshift weighting scheme: \cite{Houetal:2017}, \cite{Zarrouketal:2017} (both using the three configuration space multipoles analyses) along with this work. We focus on the variables with higher interest, $f\sigma_8$, $D_A/r_s$ and $Hr_s$ along with the linear bias, $b_1\sigma_8$, all evaluated at $z_{\rm eff}=1.52$. The different analyses yield consistent results for the different measurements for the cosmological parameters using the methodologies described above. The obtained precision (which does not include the systematic error budget) is also comparable among the methods, with no significant difference among configuration space, and Fourier space methodologies. 

The $b_1\sigma_8$ panels show a $\sim1\sigma$ discrepancy between the Fourier space- and configuration space-based analyses. Further investigation has demonstrated that this behaviour is related to the different bias model assumptions used for the different papers. The configuration space model, as the one used by \cite{Zarrouketal:2017}, depends on two bias parameters, $F'$ and $F''$, which are eventually related to $b_1$ and $b_2$. However,  the 2-point correlation function displays a limited sensitivity on $F''$ so that this parameter is poorly constrained when fitting either Legendre multipoles with order $\ell=0,2,4$ or three wedges. \cite{Zarrouketal:2017} used mocks and N-body simulations to show that fixing the $F''$ parameter to the peak-background split  prediction improves the convergence of the fits without significantly shifting the cosmological parameters, $D_A$, $H$ and $f\sigma_8$. This prior on $F''$ does, however, have an effect on $F'$, and therefore on the derived $b_1\sigma_8$. Tests on the \textsc{or} mocks revealed  a reduction on $b_1\sigma_8$ best-fitting value by a factor $0.037$ when the described prior on $F''$ is applied.  Therefore, we conclude that the discrepancy among models in terms of $b_1\sigma_8$, at least for the configuration space model used by \cite{Zarrouketal:2017}, can be understood by difference in bias prescriptions and does not affect the cosmological parameters studied in this set of papers.

For complementary comparisons among the wedges approach, as well as the comparison among weighting versus non-weighting schemes, we refer the reader to \cite{Zarrouketal:2017}.

Two BAO additional analyses on the same DR14Q sample are released along with this paper: \cite{Wangetal:2017} and \cite{Zhuetal:2017}, which are complementary to the isotropic analysis recently presented by \cite{Ataetal:2017}. These two analyses utilise the redshift weights proposed in \cite{Zhuetal:2015} to compress the BAO information in the redshift direction onto a set of weighted correlation functions. These estimators provide optimised angular diameter distance and Hubble parameter measurements at all redshifts within the range of the quasar sample. Thus, this approach complements the traditional BAO analysis presented in \cite{Ataetal:2017} by providing a first BAO measurement of the Hubble parameter from this sample. 

\subsection{Consensus between RSD and isotropic BAO}

We compare the $\alpha_{\rm iso}$ values derived from the BAO analysis on the power spectrum monopole with those derived from the RSD full-AP analysis on the power spectrum monopole, quadrupole, and hexadecapole. 

Combining the values of $\alpha_\parallel$ and $\alpha_\perp$ from the RSD analysis presented in Table~\ref{table:full_results_data}, according to Eq. \ref{eq:aiso} produces, $\alpha^{\rm RSD}_{\rm iso}=1.003\pm0.035$, which corresponds to, $D^{\rm RSD}_V(1.52)/r_s(z_d)=26.27\pm0.93$. 

Similarly to the analysis in \cite{Ataetal:2017}, we perform a BAO analysis on the power spectrum monopole. Unlike the approach of \cite{Ataetal:2017}, we apply the weighting scheme described in \S~\ref{weights:sec}. The sole difference with the previous BAO analysis is our use of focal plane weights instead of the nearest neighbour weights used in \cite{Ataetal:2017}. However, neither of these weighting schemes has demonstrated any dependency on $\alpha_{\rm iso}$ (see Table~\ref{table:mocksparameters}), and therefore both approaches are expected to provide unbiased measurements. Performing the BAO fit in the range of scales $0.02\leq k\,[h{\rm Mpc}^{-1}]\leq 0.30$, yields $\alpha^{\rm BAO}_{\rm iso}=1.003\pm0.043$\footnote{This measurement slightly differs from the value found in  \citealt{Ataetal:2017} when the power spectrum alone was used in the scale range $0.02\leq k\,[h{\rm Mpc}^{-1}]\leq 0.23$, $\alpha_{\rm iso}=0.992 \pm0.040$ (see `$P(k)$ (combined)' in table 5 of the quoted paper). However, this small difference on $\alpha_{\rm iso}$ is not significant given the slight differences in terms of the analysis described in the main text.}, which corresponds to $D^{\rm BAO}_V(1.52)/r_s(z_d)=26.27\pm1.11$.  We do not provide more details on the BAO fit on this paper, as the results are similar to those presented in \citealt{Ataetal:2017}.

Both results are in excellent agreement, although, given the identical dataset we expect a high correlation between them. 
\begin{figure*}
\centering
\includegraphics[scale=0.3]{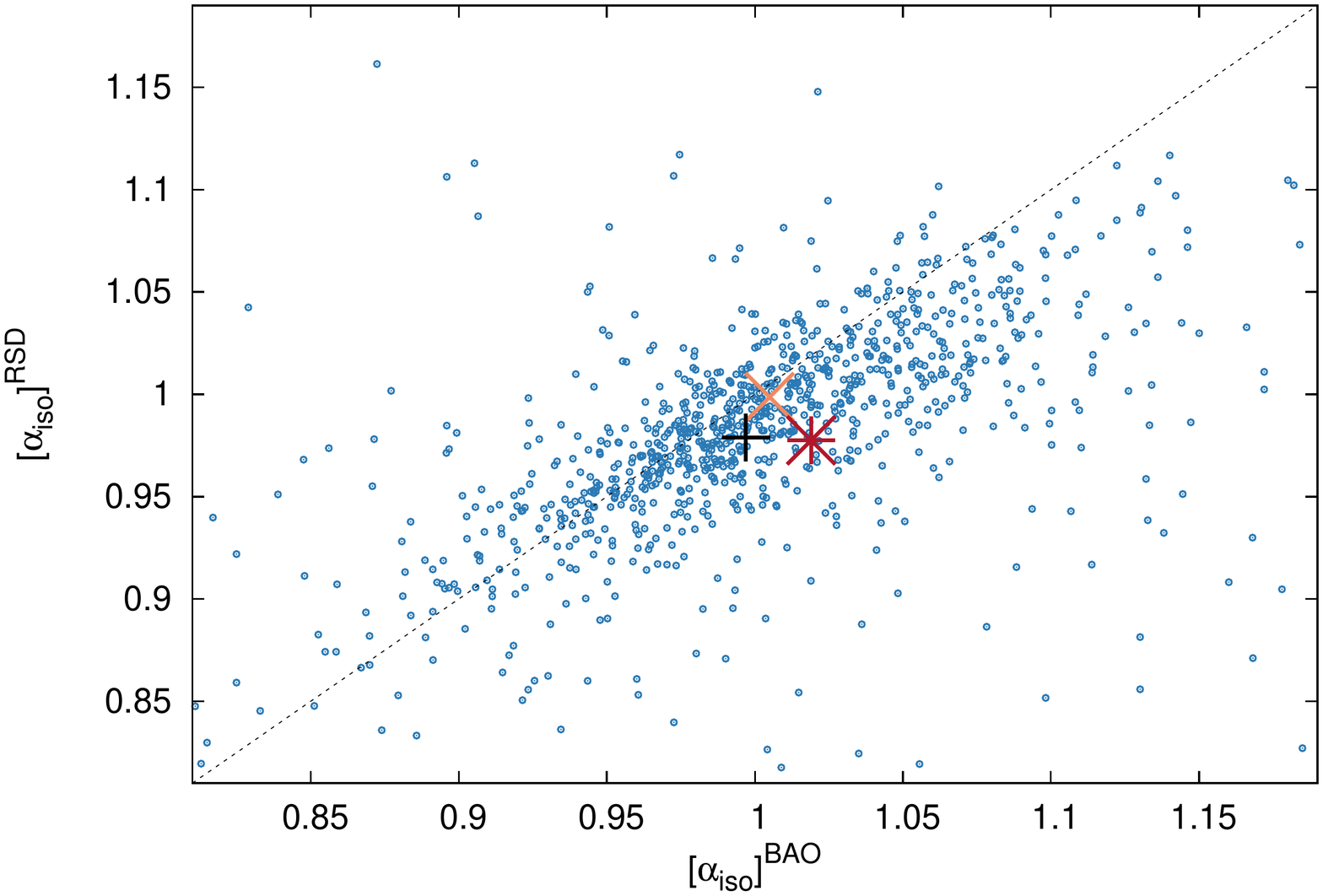}
\includegraphics[scale=0.3]{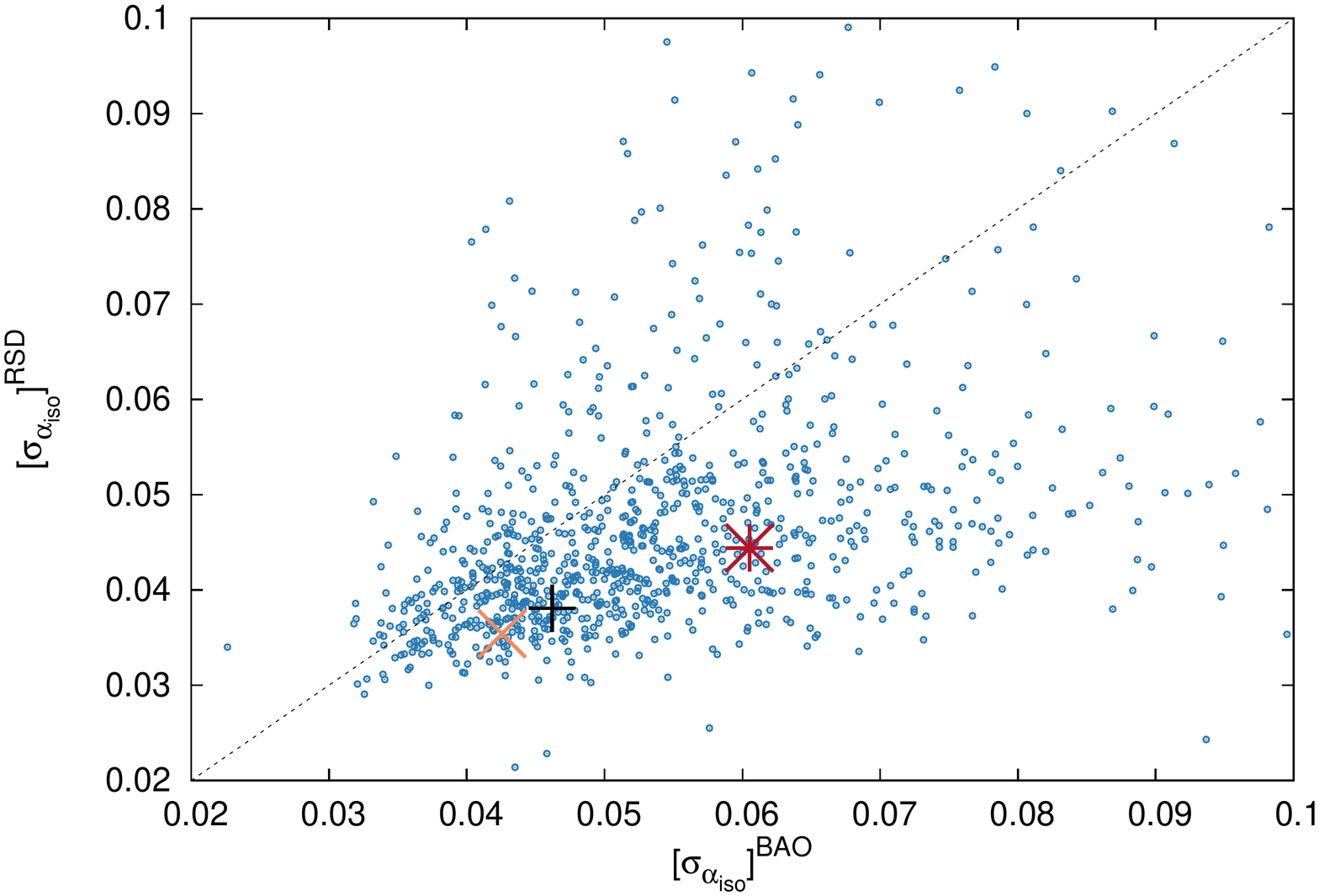}
\caption{Comparison between the isotropic BAO scale parameter, $\alpha_{\rm iso}$ (left panel) and its error, $\sigma_{\alpha_{\rm iso}}$ (right panel), inferred from a BAO fit on the power spectrum monopole, and from a RSD analysis using the power spectrum monopole, quadrupole, and hexadecapole. In the first case, $\alpha_{\rm iso}$ is inferred from the BAO peak position in the monopole, whereas in the former case $\alpha_{\rm iso}$ is inferred from $\alpha_\parallel$ and $\alpha_\perp$ through Eq. \ref{eq:aiso}. The blue dots represent the measurement for the 1000 \textsc{ez} mocks, whereas the coloured crosses represent the data when $z_{\rm fid}$ (orange cross), $z_{\rm MgII}$ (red cross), and $z_{\rm PCA}$ (black cross), are used as redshift estimates. The observed correlation between these mocks with $0.9\leq\alpha_{\rm iso}\leq1.1$ in both RSD and BAO measurements is $\rho=0.66$. The results obtained from the three datasets represent a typical realisation with respect to the mocks. }
\label{plot:BAORSD_comparison}
\end{figure*}
The left panel of Fig.~\ref{plot:BAORSD_comparison} displays the inferred $\alpha_{\rm iso}$ parameter: from a BAO analysis using $P^{(0)}$ in the $x$-axis; from the RSD analysis using the three described power spectrum multipoles in the $y$-axis. The symbols display the results for the 1000 \textsc{ez} mocks realizations and the data using the different redshift estimates, represented by the coloured symbols following the same colour-notation than in Fig.~\ref{scatter:ezmocks}. The right panel displays the comparison between the $1\sigma$ error using the same plot-notation. As expected, the correlation between the two techniques is visibly large,  and there is a correlation between the inferred errors in the right panel. The results from the mocks show that the RSD analyses using the mentioned three power spectrum multipoles tend to present a smaller error on $\alpha_{\rm iso}$ with respect to the BAO analysis on the monopole. For both cases, the quantities computed from the data are in good agreement with those observed from the mocks. 

We compute the correlation coefficient between $\alpha_{\rm iso}^{\rm BAO}$ and $\alpha_{\rm iso}^{\rm RSD}$. To be conservative, we only use those mocks with a clear detection of the peak, between $0.9\leq \alpha_{\rm iso}\leq 1.1$ in both BAO and RSD ($812$ out of $1000$ mocks fulfil these conditions) and find a correlation coefficient of $\rho=0.66$. We combine this coefficient with the measurement of $D_V$ from these two techniques. The consensus $D_V$ is defined as the weighted mean between BAO and RSD results, 
\begin{eqnarray}
D_V^{\rm cons}&=&D_V^{\rm RSD}w_{\rm RSD}+D_V^{\rm BAO}w_{\rm BAO},\\
\left(\sigma_{ D_V}^{\rm cons}\right)^2&=&w_{\rm RSD}^2\left(\sigma_{ D_V}^{\rm RSD}\right)^2+w_{\rm BAO}^2\left(\sigma_{ D_V}^{\rm BAO}\right)^2+\\
&+&2\rho\, w_{\rm RSD}\sigma_{ D_V}^{\rm RSD}w_{\rm BAO}\sigma_{ D_V}^{\rm BAO},
\end{eqnarray}
where the weights are normalised to be, $w_{\rm RSD}+w_{\rm BAO}=1$.
Applying the condition which minimises the variance of $D_V^{\rm cons}$, $\sigma_{ D_V}^{\rm cons}$,
\begin{equation}
 w_{\rm RSD}\equiv\frac{(\sigma_{ D_V}^{\rm BAO})^2-\rho\sigma_{ D_V}^{\rm BAO}\sigma_{ D_V}^{\rm RSD}}{(\sigma_{ D_V}^{\rm RSD})^2+(\sigma_{ D_V}^{\rm BAO})^2-2\rho\sigma_{ D_V}^{\rm BAO}\sigma_{ D_V}^{\rm RSD}}
\end{equation}
which is the usual inverse weighting scheme for correlated measurements. For the values of correlation and variance given above, $w_{\rm RSD}=0.76$ and $w_{\rm BAO}=0.24$. With these weights, the resulting consensus value for the isotropic BAO distance is  $D_V(1.52)/r_s(z_d)=26.27\pm0.90$, which shows a marginal improvement over the RSD result that dominates the consensus. 

\section{Conclusions}\label{sec:conclusions}

In this paper we perform a RSD analysis on the two-year data of SDSS-IV eBOSS quasar sample (DR14), which consists of $148,659$ quasars at $0.8\leq z \leq  2.2$, and measure the cosmological parameters: the logarithmic growth of structure times the amplitude of the dark matter fluctuations, $f\sigma_8$, the angular diameter distance of the sound horizon scale at drag redshift, $D_A/r_s$, and the Hubble parameter times the sound horizon scale at drag redshift, $Hr_s$, all at the effective redshift of $z_{\rm eff}=1.52$. We combine the measurements on the power spectrum monopole, quadrupole, and hexadecapole on the scale range $0.02\leq k\,[h{\rm Mpc}^{-1}] \leq 0.30$, with a theoretical model based on 2-loop resumed perturbation theory and TNS model for redshift space distortions and measure $f\sigma_8(z_{\rm eff})=0.420\pm0.076$,  $H(z_{\rm eff})=[162\pm 12]\, (r_s^{\rm fid}/r_s){\rm km s}^{-1}{\rm Mpc}^{-1}$ and  $D_A(z_{\rm eff})=[1.85\pm 0.11]\times10^3\,(r_s/r_s^{\rm fid}){\rm Mpc}$. These results include a systematic error budget which contains contributions from both observational and modelling systematics, extracted from realistic $N$-body mocks. Additionally, we perform a large number of systematic tests, and demonstrate that the cosmological results are robust and unbiased by the choice of parametrisation, e.g., the $k$-sampling of the data, the covariance matrix model, the redshift estimate used in the data, or the range of scales used in the fit. 

Additionally, we divide the full redshift range into three overlapping redshift ranges, $0.8\leq z \leq 1.5$, $1.2\leq z \leq 1.8$, $1.5\leq z \leq 2.2$, and measure the same quantities in each individual bin. Since the redshift bins overlap we also compute the covariance among the different parameters at different bins using a set of 1000 mocks. These results are presented along with their covariance matrix in Table~\ref{table:covariance}. We have found that for the intermediate redshift bin the best-fitting $\chi^2$ value is higher than any result found in the mocks. After removing the frequency $k\simeq0.11\,h{\rm Mpc}$ on the monopole and quadrupole the $\chi^2$ value is reduced to typical values. We have checked that this does not have a significant impact on the derived cosmological parameters. We leave for a future work the study and characterisation of this systematic effect. Finally, we combine the derived cosmological parameters with other complementary datasets, such as the cosmological measurements from the SDSS-III DR12 LRG BOSS sample and CMB measurements from Planck. When we perform a null-test of gravity, $\gamma=0.54\pm0.19$ for a flat-$\Lambda$CDM Universe, which is fully consistent with the GR predictions. Using the same datasets we relax the `flatness' condition and measure  $\Omega_\Lambda =0.697^{+0.035}_{-0.032}$ and $\Omega_m=0.3094^{+0.0076}_{-0.0080}$ assuming GR as the theory of gravity. Both measured $\Omega$ values are fully consistent with a flat-$\Lambda$CDM Universe, $\Omega_k=-0.007\pm 0.030$.

We have performed a comparison with the companion papers (\citealt{Houetal:2017} and \citealt{Zarrouketal:2017}) which offers complementary analysis on the same data sample and find an excellent agreement, both in the parameters measured as well on the errors and correlation among cosmological parameters. 

We now compare our results with those forecasted at the beginning of the survey in \cite{Zhaoetal:2016}. Table~\ref{table:forecast} displays, in the first column, the forecasted errors for a final area of $7500\,{\rm deg}^2$ in the redshift range $0.6\leq z \leq 2.2$ for the cosmological parameters of interest. The second column lists the scaling of those constraints to the volume of the current DR14Q sample. The scaling factor is 2.07, computed as the square-root of the ratio of volumes assuming a constant density of quasars across the redshift range. The third column presents the errors in this work for the $z_{\rm fid}$ redshift estimate when the power spectrum monopole, quadrupole, and hexadecapole are used. 
\begin{table}
\caption{Comparison between the statistical-only errors obtained in this paper, labeled as `This work', and those forecasted at the beginning of the survey (see table 4 of \citealt{Zhaoetal:2016}), labeled as `Forecast'. The column `Re-scaled Forecast' represents the `Forecast' column scaled by the square-root of the ratio of volumes between the one assumed by \citealt{Zhaoetal:2016} at the end of the eBOSS survey, and one corresponding to this survey. The scaling factor corresponds to 2.07 (see main text). The agreement found is very high in the AP-scale parameters, $D_A$, $H$ and $D_V$. On the other hand, the $f\sigma_8$ and $b_1\sigma_8$ error forecast were performed without marginalising with respect $D_A$ and $H$, and consequently, their errors are smaller. When this marginalisation is taken into account we find $\sigma_{f\sigma_8}/f\sigma_8=0.14$ and $\sigma_{b_1\sigma_8}/b_1\sigma_8=0.032$, which are in agreement with the findings of this work.}
\begin{center}
\begin{tabular}{|c|c|c|c}
\hline
& Forecast & Re-scaled Forecast & This work \\
\hline
\hline
$\sigma_{D_A}/D_A$ & 0.025 & 0.052 & 0.051 \\
$\sigma_{H}/H$ & 0.033 & 0.069 & 0.063 \\
$\sigma_{D_V}/D_V$ & 0.016 & 0.033 & 0.035 \\
$\sigma_{f\sigma_8}/f\sigma_8$ & 0.028 & 0.058 & 0.16 \\
$\sigma_{b_1\sigma_8}/b_1\sigma_8$ & 0.006 & 0.012 & 0.042 \\
\hline
\end{tabular}
\end{center}
\label{table:forecast}
\end{table}%
There is an excellent agreement between the errors of the scaled forecasted AP-parameters, $D_A$, $H$, and $D_V$, and those obtained in this work. The scaled forecasted errors for $f\sigma_8$ and $b_1\sigma_8$, however, present a 3.5 factor of disagreement with those measured. The reason for this is that the forecasts on $f\sigma_8$ and $b_1\sigma_8$ presented by \cite{Zhaoetal:2016} are performed without marginalising over $D_A$ or $H$.  When this marginalisation is taken into account, we obtain re-scaled errors for $f\sigma_8$ and $b_1\sigma_8$ of 0.140 and 0.032, respectively. These results are just $\sim20\%$ larger than those we report in this paper. Such small differences could originate from the idealised theoretical model used in the Fisher forecast or by large statistical fluctuations on the uncertainties (as found in the mocks). We conclude that the current analysis on the first two years of data from eBOSS quasar sample is in full agreement with the initial forecasts, and, consequently, the forecasted precision by the end of survey will be likely achieved in the final data release of quasars in 2019-2020.

This work, alongside the above quoted companion papers, for the first time measures the cosmological parameters, $f\sigma_8$, $D_A$, and $H$ using the full-shape analysis of power spectrum multipoles of eBOSS DR14 quasars as dark matter tracers, demonstrating the feasibility of this new dark matter tracer, not only for Lyman-$\alpha$-based analyses, but also in terms of galaxy-clustering to infer cosmological parameters. Previous works have begun to explore the $>0.8$ redshift range using Emission Line Galaxies (ELGs) (FastSound\footnote{The Subaru FMOS galaxy redshift survey, \url{http://www.kusastro.kyoto-u.ac.jp/Fastsound/}}, \citealt{Okumuraetal:2015}) and from a multisample of galaxies (VIPERS\footnote{The VIMOS Public Extragalactic Redshift Survey (VIPERS), \url{http://vipers.inaf.it/}.}, \citealt{Mohammadetal:2017}), and performed $f\sigma_8$ measurements using the full shape of the monopole and quadrupole. This paper, along with the companion papers, improve in terms of precision, but also extends the inferred cosmological parameters from a single $f\sigma_8$ measurement without marginalisation to a multipole $\{ f\sigma_8, D_A/r_s, Hr_s \}$ set of marginalised parameters. For instance, \cite{Okumuraetal:2015} and \cite{Mohammadetal:2017} measure $f\sigma_8$ with $\sim25\%$ precision at a fixed $D_A$ and $H$, whereas in this paper we find $f\sigma_8$ with $18\%$ precision, fully marginalising over $D_A$ and $H$, and $\sim 10\%$ when setting $H\times D_A$ to a fiducial value. We expect these errors to be reduced by a factor of $\sim 2$ by the completion of the eBOSS survey. 
The quasars sample as dark matter tracer represents only one aspect of the eBOSS program. Separate RSD and BAO analyses of the eBOSS LRGs and ELGs samples will fill in the $z\sim0.8$ region with more cosmological measurements in the next year (see \citealt{Bautistaetal:2018} for the first BAO measurement using the DR14 LRG sample), helping to complete the cosmological distance ladder measurements from $z\sim0$ to $z\sim3$ presented in Fig.~\ref{plot:cosmo2}.

Future galaxy spectroscopic surveys such as the ground-based Dark Energy Spectroscopic Instrument (DESI\footnote{DESI, \url{http://desi.lbl.gov/}}, \citealt{DESI1,DESI2}) and the space missions such as  {EUCLID}\footnote{EUCLID, \url{https://www.cosmos.esa.int/web/euclid/}} \citep{Euclid},  will after the year 2020 extensively probe  the intermediate redshift range $1\leq z \leq 2$, providing cosmological measurements with unprecedented precision. The eBOSS-related papers represent the first step in obtaining measurements at this previously unexplored region.

\section*{Acknowledgements} 
This work has been done within the Labex ILP (reference ANR-10-LABX-63) part of the Idex SUPER, and received financial state aid managed by the Agence Nationale de la Recherche, as part of the programme Investissements d'avenir under the reference ANR-11-IDEX-0004-02.

SH and KH work was supported under the U.S. Department of Energy contract DE-AC02-06CH11357.

GR acknowledges support from the National Research Foundation of Korea (NRF) through Grant No. 2017077508 funded by the Korean Ministry of Education, Science and Technology (MoEST), and from the faculty research fund of Sejong University in 2018.

Funding for SDSS-III and SDSS-IV has been provided by the Alfred P. Sloan Foundation and Participating Institutions. Additional funding for SDSS-III comes from the National Science Foundation and the U.S. Department of Energy Office of Science. Further information about both projects is available at www.sdss.org. SDSS is managed by the Astrophysical Research Consortium for the Participating Institutions in both collaborations. In SDSSIII these include the University of Arizona, the Brazilian Participation Group, Brookhaven National Laboratory, Carnegie Mellon University, University of Florida, the French Participation Group,
the German Participation Group, Harvard University, the Instituto de Astrofisica de Canarias, the Michigan State / Notre Dame / JINA Participation Group, Johns Hopkins University, Lawrence Berkeley National Laboratory, Max Planck Institute for Astrophysics, Max Planck Institute for Extraterrestrial Physics, New Mexico State University, New York University, Ohio State University, Pennsylvania State University, University of Portsmouth, Princeton University, the Spanish Participation Group, University of Tokyo, University of Utah, Vanderbilt University, University of Virginia, University of Washington, and Yale University. 

The Participating Institutions in SDSS-IV are Carnegie Mellon University, Colorado University, Boulder, Harvard-Smithsonian Center for Astrophysics Participation Group, Johns Hopkins University, Kavli Institute for the Physics and Mathematics of the Universe Max-Planck-Institut fuer Astrophysik (MPA, Garching), Max-Planck-Institut fuer Extraterrestrische Physik (MPE), Max-Planck-Institut fuer Astronomie (MPIA Heidelberg), National Astronomical Observatories of China, New Mexico State University, New York University, The Ohio State University, Penn State University, Shanghai Astronomical Observatory, United Kingdom Participation Group, University of Portsmouth, University of Utah, University of Wisconsin, and Yale University.

This work made use of the facilities and staff of the UK Sciama High Performance Computing cluster supported by the ICG, SEPNet and the University of Portsmouth. This research used resources of the National Energy Research Scientific Computing Centre (NERSC), a DOE Office of Science User Facility supported by the Office of Science of the U.S. Department of Energy under Contract No. DE-AC02-05CH11231. 

This research used resources of the Argonne Leadership Computing Facility, which is a DOE Office of Science User Facility supported under contract DE-AC02-06CH11357.





%
%
%


\def\jnl@style{\it}
\def\aaref@jnl#1{{\jnl@style#1}}

\def\aaref@jnl#1{{\jnl@style#1}}

\def\aj{\aaref@jnl{AJ}}                   
\def\araa{\aaref@jnl{ARA\&A}}             
\def\apj{\aaref@jnl{ApJ}}                 
\def\apjl{\aaref@jnl{ApJ}}                
\def\apjs{\aaref@jnl{ApJS}}               
\def\ao{\aaref@jnl{Appl.~Opt.}}           
\def\apss{\aaref@jnl{Ap\&SS}}             
\def\aap{\aaref@jnl{A\&A}}                
\def\aapr{\aaref@jnl{A\&A~Rev.}}          
\def\aaps{\aaref@jnl{A\&AS}}              
\def\azh{\aaref@jnl{AZh}}                 
\def\baas{\aaref@jnl{BAAS}}               
\def\jrasc{\aaref@jnl{JRASC}}             
\def\memras{\aaref@jnl{MmRAS}}            
\def\mnras{\aaref@jnl{MNRAS}}             
\def\pra{\aaref@jnl{Phys.~Rev.~A}}        
\def\prb{\aaref@jnl{Phys.~Rev.~B}}        
\def\prc{\aaref@jnl{Phys.~Rev.~C}}        
\def\prd{\aaref@jnl{Phys.~Rev.~D}}        
\def\pre{\aaref@jnl{Phys.~Rev.~E}}        
\def\prl{\aaref@jnl{Phys.~Rev.~Lett.}}    
\def\pasp{\aaref@jnl{PASP}}               
\def\pasj{\aaref@jnl{PASJ}}               
\def\qjras{\aaref@jnl{QJRAS}}             
\def\skytel{\aaref@jnl{S\&T}}             
\def\solphys{\aaref@jnl{Sol.~Phys.}}      
\def\sovast{\aaref@jnl{Soviet~Ast.}}      
\def\ssr{\aaref@jnl{Space~Sci.~Rev.}}     
\def\zap{\aaref@jnl{ZAp}}                 
\def\nat{\aaref@jnl{Nature}}              
\def\iaucirc{\aaref@jnl{IAU~Circ.}}       
\def\aplett{\aaref@jnl{Astrophys.~Lett.}} 
\def\apspr{\aaref@jnl{Astrophys.~Space~Phys.~Res.}}
\def\bain{\aaref@jnl{Bull.~Astron.~Inst.~Netherlands}} 
\def\fcp{\aaref@jnl{Fund.~Cosmic~Phys.}}  
\def\gca{\aaref@jnl{Geochim.~Cosmochim.~Acta}}   
\def\grl{\aaref@jnl{Geophys.~Res.~Lett.}} 
\def\jcp{\aaref@jnl{J.~Chem.~Phys.}}      
\def\jgr{\aaref@jnl{J.~Geophys.~Res.}}    
\def\jqsrt{\aaref@jnl{J.~Quant.~Spec.~Radiat.~Transf.}}
\def\memsai{\aaref@jnl{Mem.~Soc.~Astron.~Italiana}}
\def\nphysa{\aaref@jnl{Nucl.~Phys.~A}}   
\def\physrep{\aaref@jnl{Phys.~Rep.}}   
\def\physscr{\aaref@jnl{Phys.~Scr}}   
\def\planss{\aaref@jnl{Planet.~Space~Sci.}}   
\def\procspie{\aaref@jnl{Proc.~SPIE}}   
\def\jcap{\aaref@jnl{J. Cosmology Astropart. Phys.}}

\let\astap=\aap
\let\apjlett=\apjl
\let\apjsupp=\apjs
\let\applopt=\ao

\newcommand{\etal}{et al.\ }

\newcommand{\mpc}{\, {\rm Mpc}}
\newcommand{\kpc}{\, {\rm kpc}}
\newcommand{\hmpc}{\, h^{-1} \mpc}
\newcommand{\ihmpc}{\, h\, {\rm Mpc}^{-1}}
\newcommand{\ikms}{\, {\rm s\, km}^{-1}}
\newcommand{\kms}{\, {\rm km\, s}^{-1}}
\newcommand{\hkpc}{\, h^{-1} \kpc}
\newcommand{\lya}{Ly$\alpha$\ }
\newcommand{\lyb}{Lyman-$\beta$\ }
\newcommand{\lyaf}{Ly$\alpha$ forest}
\newcommand{\lr}{\lambda_{{\rm rest}}}
\newcommand{\bF}{\bar{F}}
\newcommand{\bS}{\bar{S}}
\newcommand{\bC}{\bar{C}}
\newcommand{\bB}{\bar{B}}
\newcommand{\vdF}{{\mathbf \delta_F}}
\newcommand{\vdS}{{\mathbf \delta_S}}
\newcommand{\vdf}{{\mathbf \delta_f}}
\newcommand{\vdn}{{\mathbf \delta_n}}
\newcommand{\vdC}{{\mathbf \delta_C}}
\newcommand{\vdX}{{\mathbf \delta_X}}
\newcommand{\xrei}{x_{rei}}
\newcommand{\lrmin}{\lambda_{{\rm rest, min}}}
\newcommand{\lrmax}{\lambda_{{\rm rest, max}}}
\newcommand{\lmin}{\lambda_{{\rm min}}}
\newcommand{\lmax}{\lambda_{{\rm max}}}
\newcommand{\hi}{\mbox{H\,{\scriptsize I}\ }}
\newcommand{\heii}{\mbox{He\,{\scriptsize II}\ }}
\newcommand{\vp}{\mathbf{p}}
\newcommand{\vq}{\mathbf{q}}
\newcommand{\vxperp}{\mathbf{x_\perp}}
\newcommand{\vkperp}{\mathbf{k_\perp}}
\newcommand{\vrperp}{\mathbf{r_\perp}}
\newcommand{\vx}{\mathbf{x}}
\newcommand{\vy}{\mathbf{y}}
\newcommand{\vk}{\mathbf{k}}
\newcommand{\vR}{\mathbf{r}}
\newcommand{\tdtwo}{\tilde{b}_{\delta^2}}
\newcommand{\tstwo}{\tilde{b}_{s^2}}
\newcommand{\tbthree}{\tilde{b}_3}
\newcommand{\tadtwo}{\tilde{a}_{\delta^2}}
\newcommand{\tastwo}{\tilde{a}_{s^2}}
\newcommand{\tabthree}{\tilde{a}_3}
\newcommand{\vnabla}{\mathbf{\nabla}}
\newcommand{\tpsi}{\tilde{\psi}}
\newcommand{\vv}{\mathbf{v}}
\newcommand{\fnl}{{f_{\rm NL}}}
\newcommand{\tfnl}{{\tilde{f}_{\rm NL}}}
\newcommand{\gnl}{g_{\rm NL}}
\newcommand{\orderfour}{\mathcal{O}\left(\delta_1^4\right)}
\newcommand{\SDSSPF}{\cite{2006ApJS..163...80M}}
\newcommand{\PF}{$P_F^{\rm 1D}(k_\parallel,z)$}
\newcommand\ionalt[2]{#1$\;${\scriptsize \uppercase\expandafter{\romannumeral #2}}}%
\newcommand{\vxone}{\mathbf{x_1}}
\newcommand{\vxtwo}{\mathbf{x_2}}
\newcommand{\vRot}{\mathbf{r_{12}}}
\newcommand{\cm}{\, {\rm cm}}

\bibliographystyle{mnras}
\bibliography{DR14_PK_RSD.bib}

\appendix

\section{Isotropic-$\alpha$ approximation}\label{sec:APtest}

In this appendix we examine the assumption of $\alpha_{\rm iso}=\alpha_\parallel^{1/3}\alpha_\perp^{2/3}$ used in some sections of the paper. \cite{Rossetal:2015} derive the analytic formulae for the parameter combination among  $\alpha_\parallel$ and $\alpha_\perp$  inferred from the different power spectrum and correlation function $\mu$-moments given by Legendre polynomials, showing explicitly what is being measured by each. Briefly, when the variable $\alpha_F^{m+n}=\alpha_\parallel^m\alpha_\perp^n$ is defined,  the $m$ and $n$ values which provide the degenere directions of these parameters given the observed multipole at linear order are reported. For the power spectrum monopole, $m=1/A\left(1/3+2\beta/5+\beta^2/7\right)$ and $n=1/A\left(2/3+4/15\beta+2/35\beta^2\right)$, where $\beta\equiv b_1/f$ and $A\equiv 1+2/3\beta + 1/5\beta^2$. For the $b_1$ and $f$ values obtained from the mean of the \textsc{ez} mocks, $b_1=2.25$ and $f=0.924$,  $\beta=0.4107$. Thus, the measurement of the power spectrum monopole in redshift space provides a degenerate direction corresponding to $\alpha_F=\alpha_\parallel^m\alpha_\perp^n$ with $m=0.399$ and $n=0.601$, which differs slightly from the expected values when the redshift space distortions are removed (in the reconstruction process for example), $m=0.333$ and $n=0.667$. The measurement of the $\mu^2$-moment of the power spectrum (in some sense equivalent to the quadrupole) provides $m=0.336$ and $n=0.1856$. The top panel of Fig.~\ref{fig:APtest} shows the degenerate direction for these two cases described above, $1=\alpha_\parallel^m \alpha_\perp^n$, along with the $\alpha_\parallel$ and $\alpha_\perp$ measurement from the individual \textsc{ez} mocks. Visually, the $\mu$-square moment case is disfavoured with respect to the monopole cases (with or without RSD), suggesting that when both monopole and quadrupole are added, the total signal remains dominated by the monopole, as the signal-to-noise ratio is higher.  The lower panel presents the histogram of the quantity $\alpha_\parallel^m\alpha_\perp^n-{\alpha|_{\epsilon=0}}^{m+n}$ computed from the same mocks, where we denote $\alpha|_{\epsilon=0}$ as the value of $\alpha$ computed when $\alpha_\parallel$ and $\alpha_\perp$ are set to the same value. The histogram displays the degree of distortion compared to the isotropic case (both $\alpha$  being equal) with the full-AP test. Again, the case for $m=0.399$ and $n=0.601$ ($P^{(0)}$ no-RSD) presents a distribution with lower dispersion than the one by $m=0.336$ and $n=0.1856$ ($P^{(2)}$ RSD) . We conclude that the quantity $D_V$ is well constrained when $\alpha_\parallel=\alpha_\perp$. 

\begin{figure}
\centering
\includegraphics[scale=0.3]{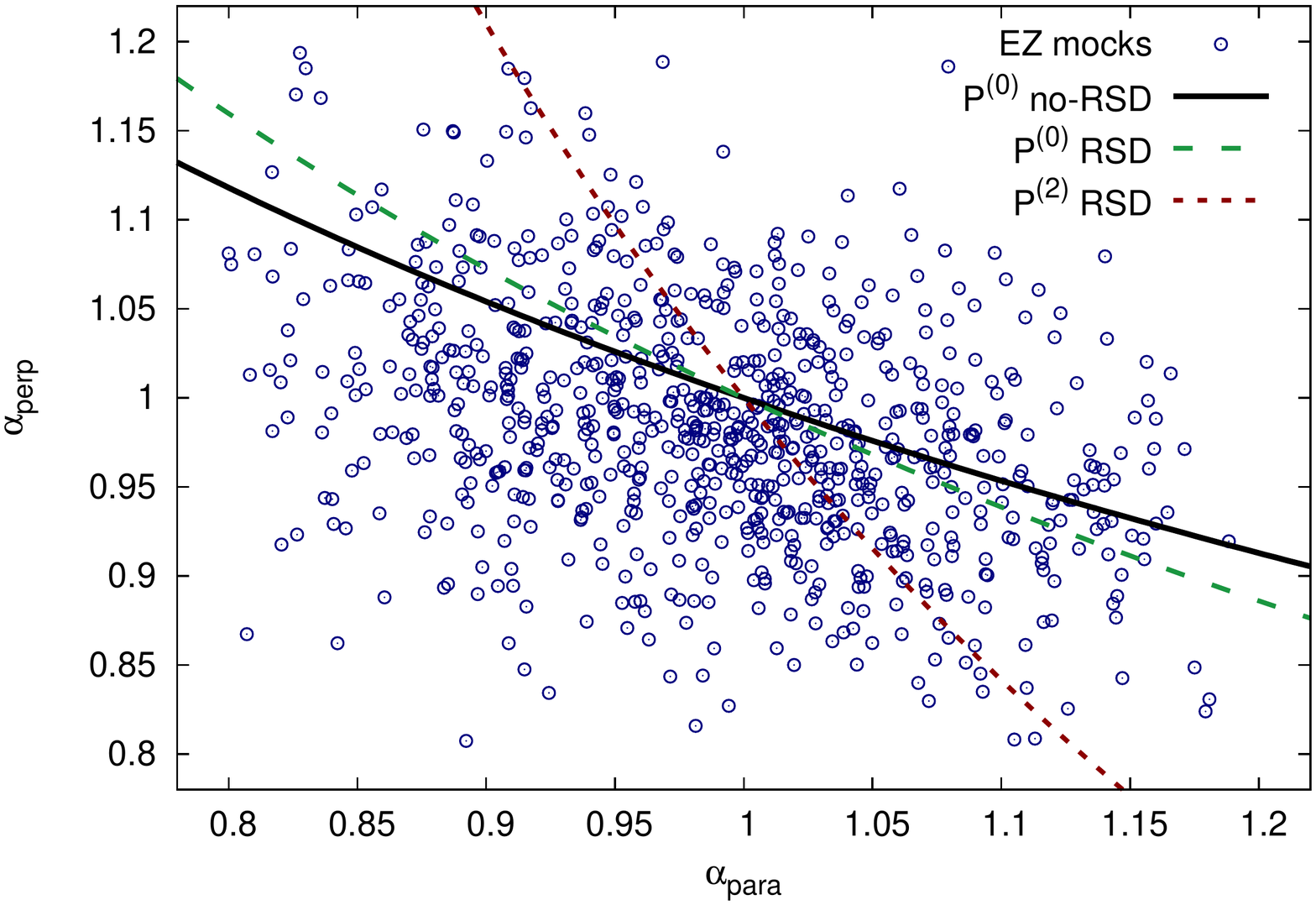}
\includegraphics[scale=0.3]{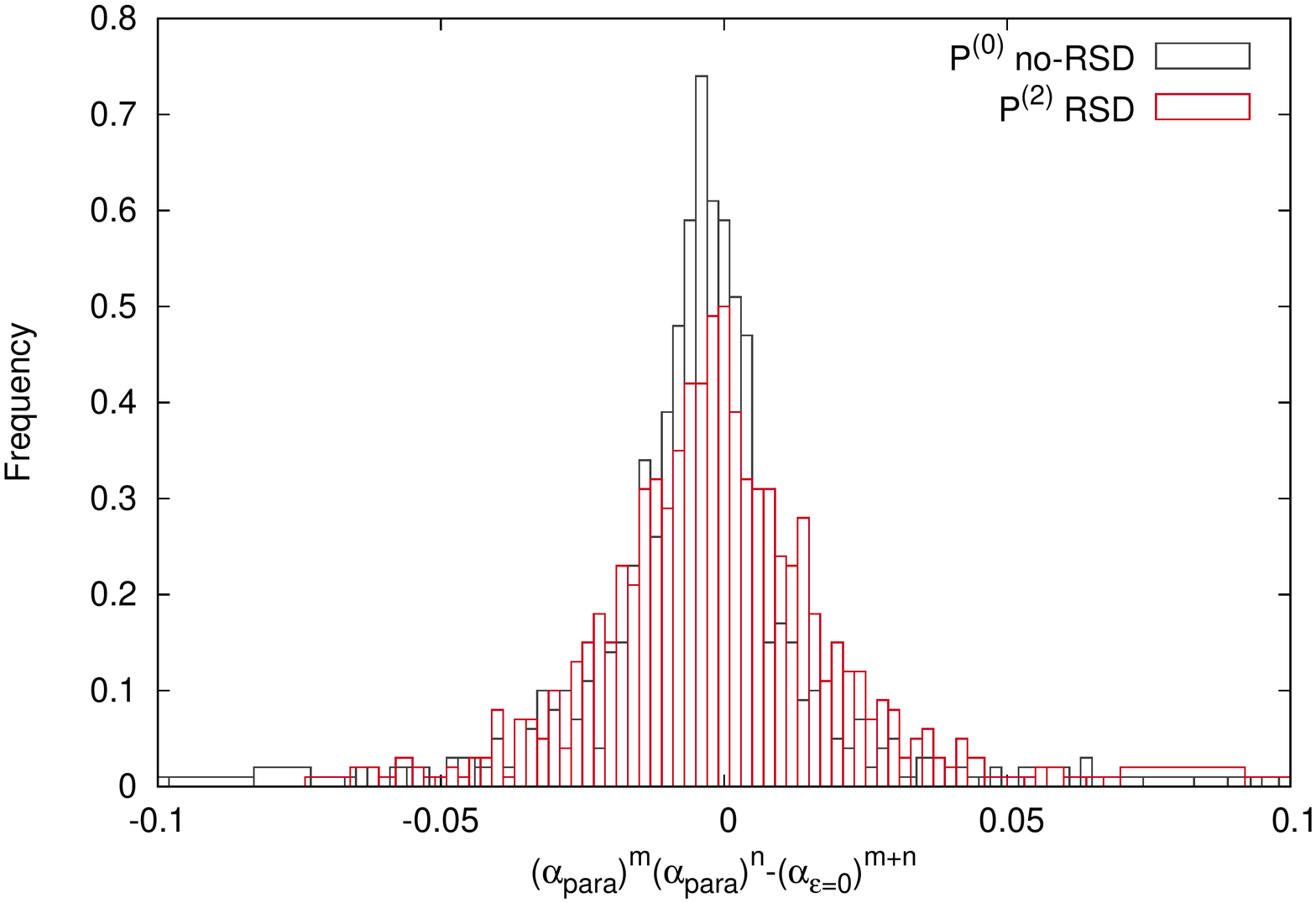}
\caption{{\it Top panel:} \textsc{ez} mock measurements for $\alpha_\parallel$ and $\alpha_\perp$ for the full-AP test (blue symbols) along with the degeneracy directions predicted by \citealt{Rossetal:2015} when only the monopole is used after (black solid line) and before (green dashed line) the RSD are removed, and when the $\mu^2$-moment is used (red dotted line). The mocks display a degenrate direction on $\alpha=\alpha_{\parallel}^m\alpha_\perp^{n}$ which is close to the one predicted by monopole with RSD removal, which corresponds to $\alpha_{\rm iso}$. {\it Bottom panel:} Histogram of the quantity $\alpha_\parallel^m\alpha_\perp^n-{\alpha|_{\epsilon=0}}^{m+n}$ computed from the mocks above, where ${\alpha|_{\epsilon=0}}$ is the $\alpha$ parameter obtained when we set $\alpha_\parallel$ to be equal to $\alpha_\perp$. The black bands display the results for $m$ and $n$ when $m=1/3$ and $n=2/3$, corresponding to the black solid line in the top panel figure, and the red bands for $m=0.336$ and $n=0.1856$, corresponding to the red dotted line in the top panel. }
\label{fig:APtest}
\end{figure}

\section{Survey Geometry}\label{sec:surveygeometry}

 We write the masked power spectrum multipoles as a Hankel Transform (HT) of the masked $\ell$-multipole of the correlation function, $\hat{\xi}^{(\ell)}$, 
\begin{equation}
\hat{P}^{(\ell)}(k)=4\pi(-i)^\ell\int dr\, r^2\hat{\xi}^{(\ell)}(r)j_\ell(kr),
\end{equation}
where $j_\ell$ is the spherical Bessel function of order $\ell$. $\hat{\xi}^{(\ell)}(r)$ can then be written in terms of the correlation function $\ell$-multipoles, corresponding to the inverse HT of the un-masked power spectrum theoretical model, 
\begin{eqnarray}
\hat{\xi}^{(0)}&=& \xi^{(0)}W_0^2+\frac{1}{5}\xi^{(2)}W_2^2+\frac{1}{9}\xi^{(4)}W_4^2\\
\nonumber\hat{\xi}^{(2)}&=& \xi^{(0)}W_2^2+\xi^{(2)}\left [ W_0^2+\frac{2}{7}W_2^2+\frac{2}{7}W_4^2 \right]\\
&+&\xi^{(4)}\left [\frac{2}{7}W_2^2+\frac{100}{693}W_4^2+\frac{25}{143}W_6^2\right]\\
\nonumber\hat{\xi}^{(4)}&=& \xi^{(0)}W_4^2+\xi^{(2)}\left [\frac{18}{35} W_2^2+\frac{20}{77}W_4^2+\frac{45}{143}W_6^2 \right]\\
\nonumber&+&\xi^{(4)}\left [W_0^2+\frac{20}{77}W_2^2\frac{162}{1001}W_4^2+\frac{20}{143}W_6^2+\right.\\
&+&\left.\frac{490}{2431}W_8^2\right]
\end{eqnarray}
We neglect the contribution of higher-than-hexadecapole terms into the monopole and quadrupole signal, as they are known to be negligible. The $W_i$ functions contain all the information on the radial and angular distribution selection functions and can be computed through a pair count of the random catalogue, 
\begin{equation}
W_\ell^2(r)\propto\sum_{ij} \frac{RR(r)}{r^2}\mathcal{L}_\ell(\mu_{r}),
\label{window:eq}
\end{equation}
where the normalisation factor is the same for all $W_\ell$ and is chosen such $W_0^2\rightarrow1$ in the limit $r\rightarrow0$. 
\section{Effect of spectroscopic weights in the power spectrum multipoles}\label{ap:weights}
\begin{figure*}
\includegraphics[scale=0.2]{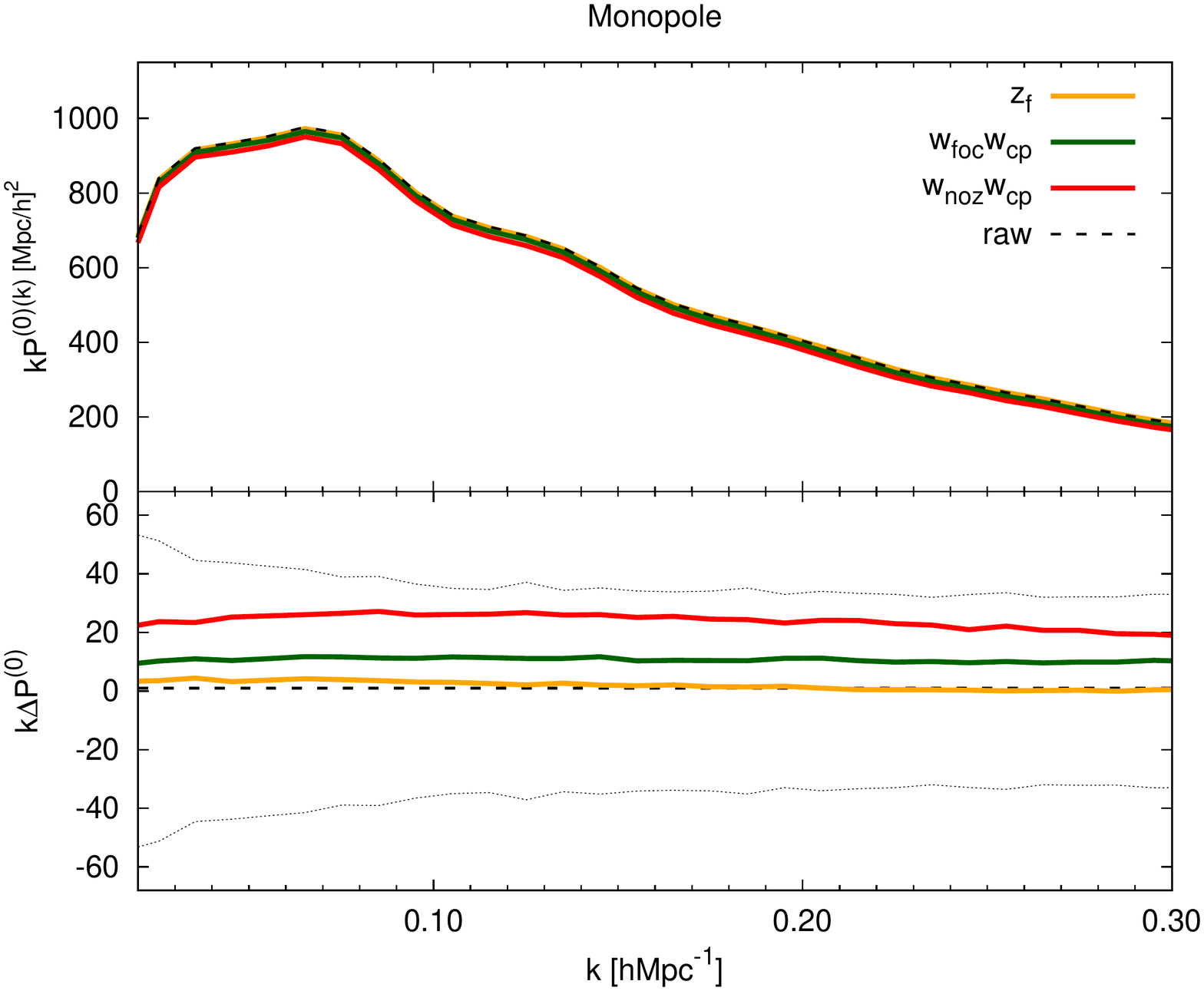}
\includegraphics[scale=0.2]{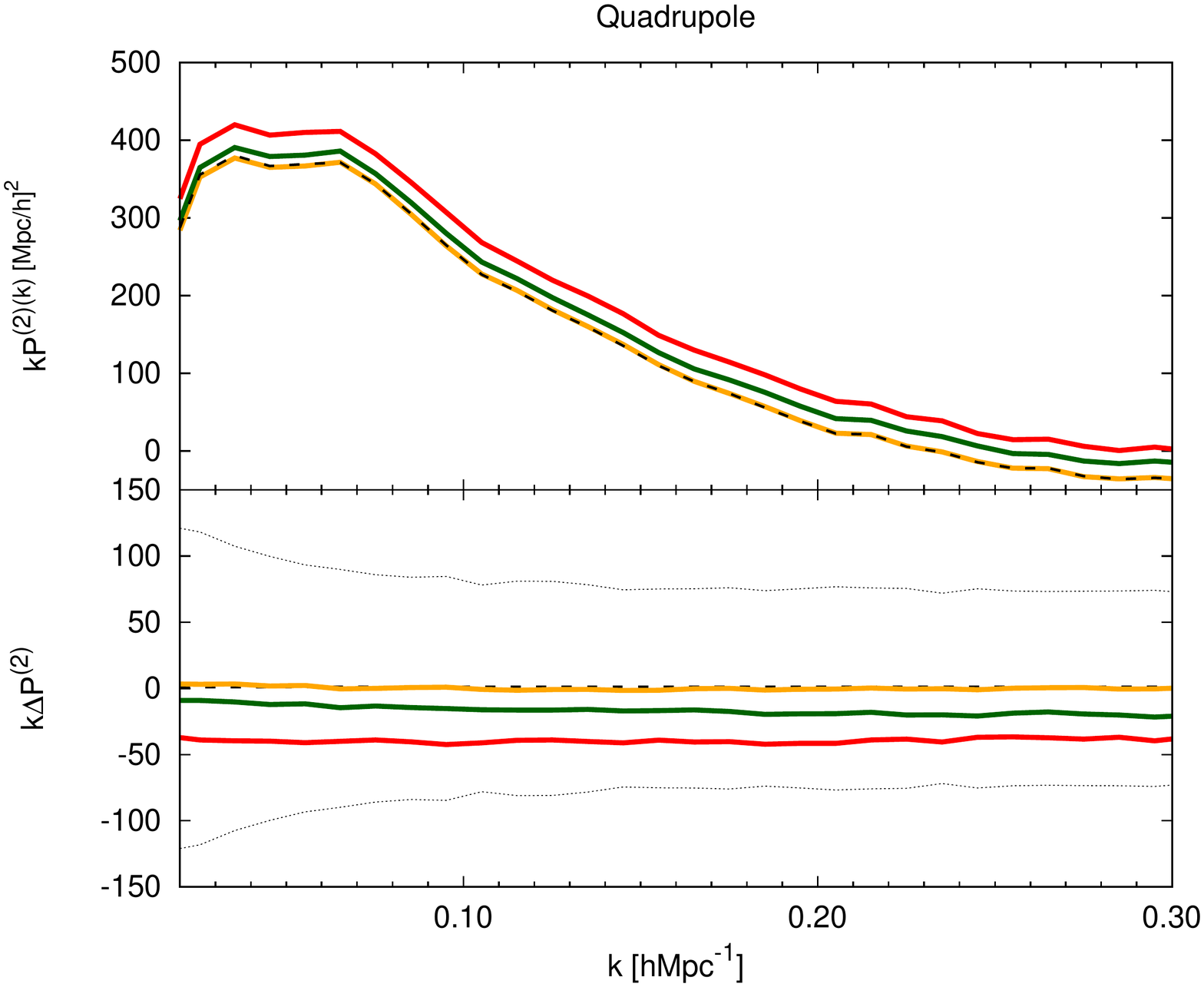}
\includegraphics[scale=0.2]{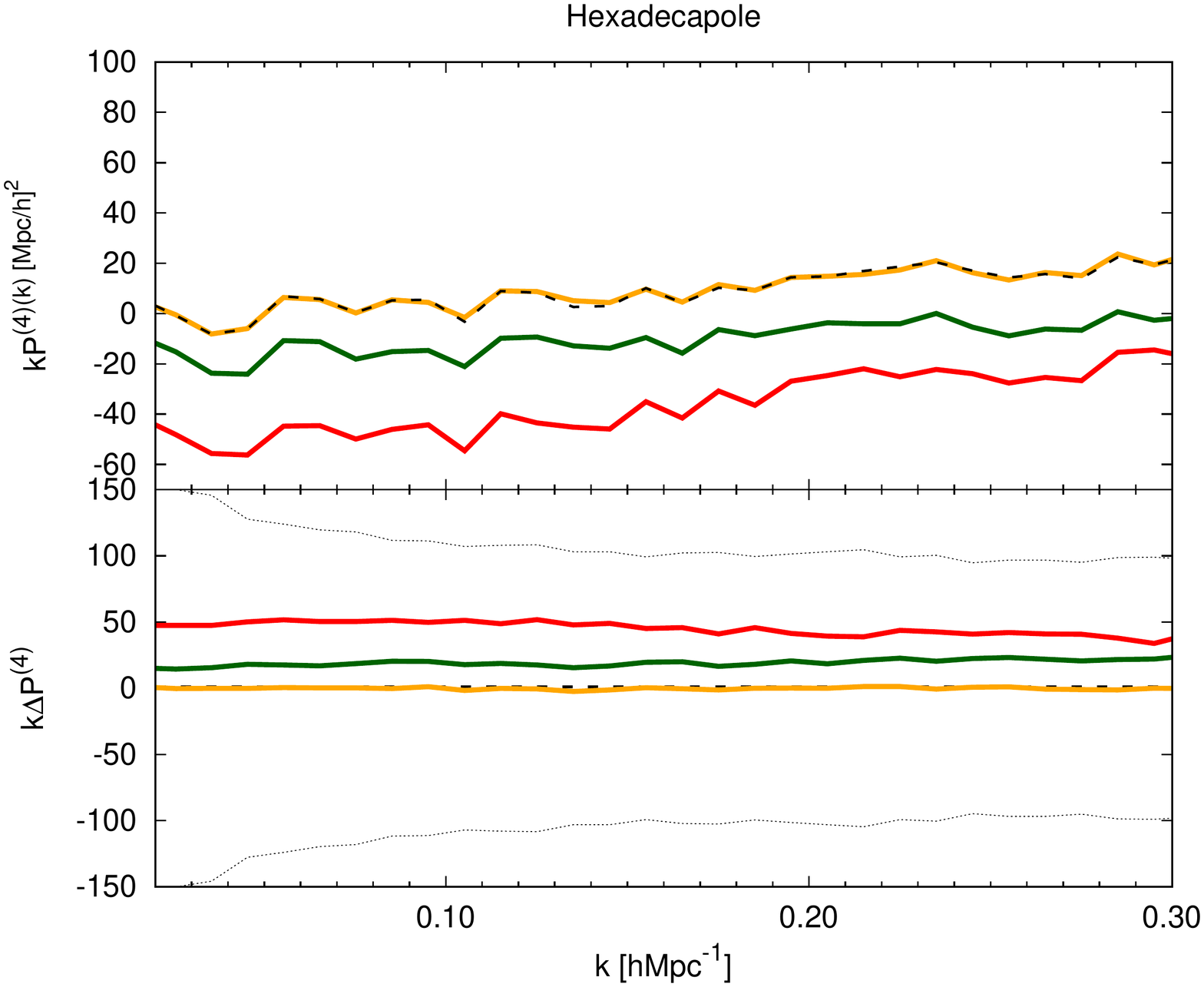}
\caption{Effect of the different spectroscopic weights in the power spectrum monopole (left panel), quadrupole (middle panel) and hexadecapole (right panel) on the mean of 1000 realisations of the \textsc{ez} mocks. The top sub-panels display the quantity $kP^{(\ell)}$ for the different weighting schemes (see text and Table~\ref{table:mocksparameters} for reference) in different cases. The bottom sub-panels display the difference with respect to the  `raw' case, where no observational effect has been applied. The black dotted lines correspond to the expected $1\sigma$ statistical error corresponding to the DR14Q sample and computed from the {\it rms} of the 1000 realisations.}
\label{plot:weigheffectsonP}
\end{figure*}

Table~\ref{table:mocksparameters} lists the results of our test of the impact of the weights $w_{\rm foc}$ and $w_{\rm cp}$ on the cosmological parameters of interest. In this appendix we describe the effect of such weights in the clustering amplitude and shape of the power spectrum monopole, quadrupole and hexadecapole. In order to measure the potential deviations caused by the spectroscopic weights, we compute the power spectrum multipoles on the \textsc{ez} mocks before such effects are applied (`raw' measurements of Table~\ref{table:mocksparameters}) and take the mean value over the 1000 realisations. We adopt this measurement as a clustering reference. We apply the different weighting schemes on individual mocks and compare their mean with the reference mean. The results are shown in Fig.~\ref{plot:weigheffectsonP}, where the reference measurement is presented as dashed black lines in the top sub-panels. $z_f$ represents the correction of the redshift failures through the $w_{\rm foc}$ weights according to Eq. \ref{eq:wfoc}.  $w_{\rm foc}w_{\rm cp}$ corresponds to the case where both fibre collisions and redshift failures are applied following the prescription described in \S~\ref{sec:spectroweights}. Finally, $w_{\rm noz}w_{\rm cp}$ corresponds to the case where both fibre collisions and redshift failures are included, but in this case the redshift failures have been corrected using the near-neighbour technique. The top sub-panels present the actual power spectrum multipole measurement, $kP^{(\ell)}$, and the bottom sub-panels the difference with respect to the reference case, $k\Delta P^{(\ell)}\equiv kP^{(\ell)}_i-kP^{(\ell)}_{\rm raw}$. In the top sub-panels the error-bar is not indicated as it would to small to be distinguished from the actual lines.  In the bottom sub-panels, the black dotted lines represent the expected $1\sigma$ statistical error for the data DR14Q sample, and has been computed as the {\it rms} of the 1000 realisations of the mocks (those with the $w_{\rm foc}w_{\rm cp}$ weighting scheme). 

 $z_f$ tests the isolated effect of the focal plane weights (without the fibre collisions) through $w_{\rm foc}$. Both Table~\ref{table:mocksparameters} and Fig.~\ref{plot:weigheffectsonP} demonstrate that $w_{\rm foc}$ perfectly accounts for the redshift failures, being able to recover the original power spectrum signal for the three studied multipoles.  Conversely, when the nearest neighbour technique is applied (both for correcting the fibre collision and the redshift failures), a spurious anisotropic signal is introduced in such a way that the monopole and hexadecapole are under-estimated, whereas the quadrupole is over-estimated. This anisotropic signal contaminates and biases the measurement of $f\sigma_8$, as discussed in \S~\ref{sec:anisofitsonmocks}. In this case, we observe that the spurious signal is higher when the nearest neighbour technique is applied to correct the redshift failure weights. The systematics associated with the inaccuracy when correcting the  fibre collision effect through the nearest neighbour technique are discussed in \S~\ref{sec:anisofitsonmocks}. 
 
\section{Full Covariance matrices}\label{ap:cov}

In this section we compare the covariance matrices for the power spectrum multipoles, $\ell=0,\,2,\,4$ when they are estimated from 400 realisations of the \textsc{qpm} mocks and the 1000 realisations of the \textsc{ez} mocks. Fig.~\ref{plot:covariance1} displays the off-diagonal elements (cross-correlation coefficients) of the covariance, when the $k$-binning is linear between $0.01\,h{\rm Mpc}^{-1}$ up to $0.40\,h{\rm Mpc}^{-1}$. The left(right) panel displays the terms computed from the \textsc{qpm}(\textsc{ez}) mocks. The scale for the correlation coefficients has been defined in the range $0$ to $0.16$ in order to stress the off-diagonal signal. The diagonal elements are by definition 1 and lie out of the scale. In general, both matrices are dominated by their diagonal component, as the off-diagonal cross-correlation terms are typically small, $<0.1$. The different degree of noise from the \textsc{qpm}- and \textsc{ez}-derived covariances is caused by the different number of realisations from which the two covariances are computed. 

\begin{figure*}
\centering
\includegraphics[scale=0.3]{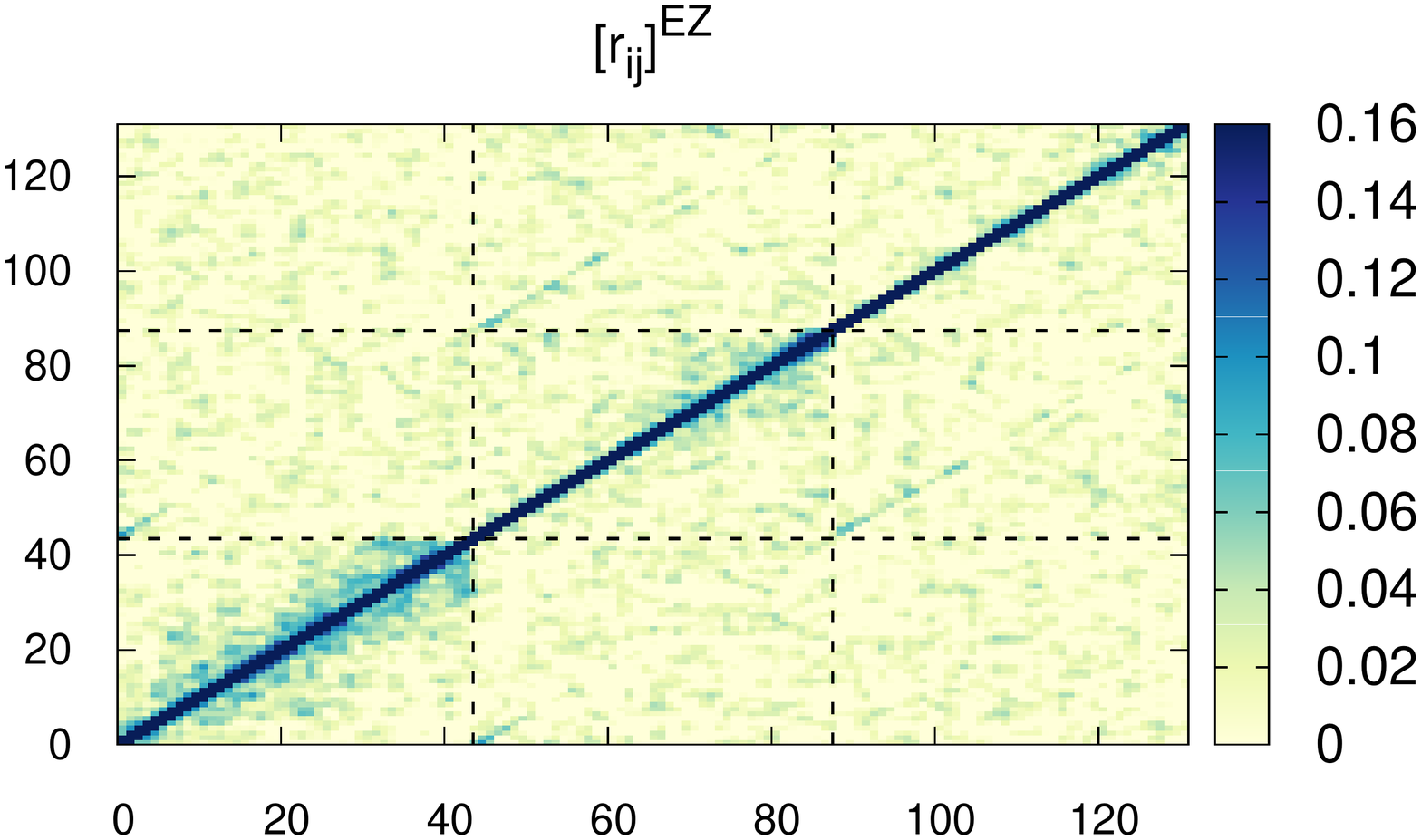}
\includegraphics[scale=0.3]{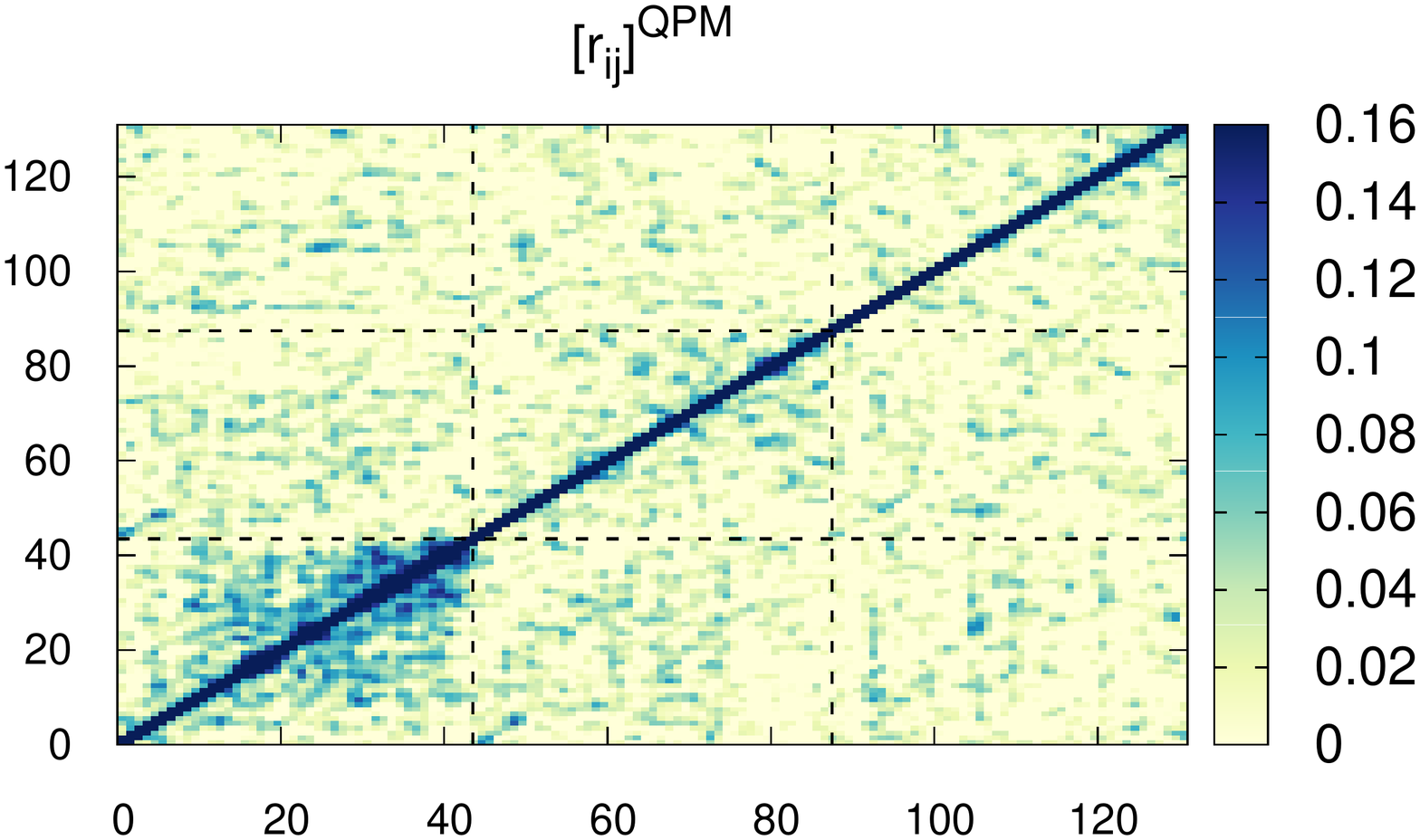}
\caption{Cross-correlation terms, $r_{ij}\equiv \sigma_{ij}/\sqrt{\sigma_{ii}^2\sigma^2_{jj}}$, corresponding to the covariance matrices inferred from 400 realisations of \textsc{qpm} mocks and 1000 realisations of \textsc{ez} mocks in the left and right panels, respectively. For each panel, from the left to the right, the monopole, quadrupole, and hexadecapole $k$-bins terms are represented and separated by the dashed black lines. For each of the multipoles, the covariance elements correspond to a linear $k$-binning between $0.01\leq k\,[h{\rm Mpc}^{-1}\leq 0.40$. The colour scale has been adjusted to highlight the off-diagonal terms. The terms in the diagonal have by definition a cross-correlation term of 1 and lie out of the scale. The \textsc{qpm}-derived covariance present higher values of the off-diagonal terms, specially on the monopole, whereas the \textsc{ez}-derived covariance is more diagonal, but this effect is minor compared to the differences observed in the diagonal terms, as displayed by Fig.~\ref{plot:covariance2}.  }
\label{plot:covariance1}
\end{figure*}

Fig.~\ref{plot:covariance2} displays the ratio of the diagonal elements of the two covariances for the three studied power spectrum multipoles in different colours. Both covariances are in agreement, although the \textsc{qpm}-derived elements tend to be $\sim5\%$ larger than those of the \textsc{ez}-derived covariance for the monopole and quadrupole. However, this trend is not maintained on the hexadecapole, neither for the monopole at small $k$. The impact of the covariance choice in the parameter estimation of the data is discussed in \S~\ref{sec:sysdata}.

\begin{figure}
\centering
\includegraphics[scale=0.3]{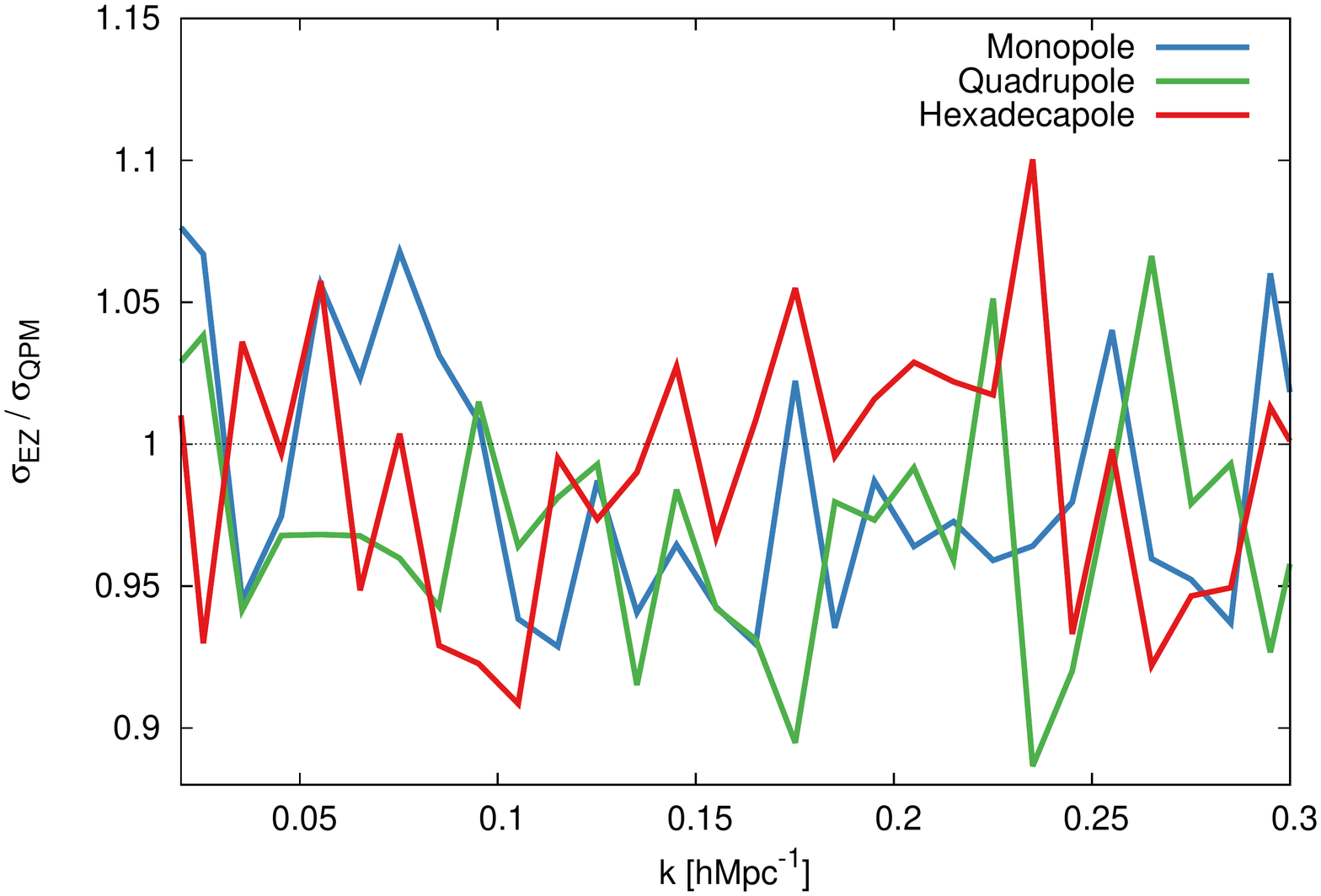}
\caption{Ratio of the square root of the diagonal terms of the covariance matrices, $\sigma_{ii}$, inferred from $400$ realisations of \textsc{qpm} mocks and 1000 realisations of \textsc{eq} mocks, $\sigma_{ii}^{\rm qpm} / \sigma_{ii}^{\rm ez}$, for the monopole (in blue), quadrupole (in green), and hexadecapole (in red). As a general trend for the monopole and quadrupole, the \textsc{qpm}-derived errors tend to be five times larger than in the \textsc{ez}-derived errors.}
\label{plot:covariance2}
\end{figure}

\section{Gaussian Approximation of the Likelihood}\label{appendix:gaussian}

In this Appendix we compare the full \textsc{mcmc} contours resulting from the actual dataset with those resulting from the Gaussian approximation used in \S\ref{sec:isotropicfits}, \ref{sec:anisotropic_results} and \ref{sec:results_zbins} to compute the reported data-vectors and covariance matrices.  For all the cases the data vector is taken as the mean of the considered \textsc{mcmc} steps. Fig.~\ref{fig:gaussian_aprox} display the posterior-likelihood for the anisotropic fit when the full redshift range, $0.8\leq z \leq 2.2$ is considered (corresponding to \S\ref{sec:anisotropic_results}). The contours drawn from the \textsc{mcmc}-full chain are represented in purple. On the other hand, the green and orange contours correspond to the Gaussian approximation when: {\it i}) all the \textsc{mcmc} step chains are used to compute the Gaussian covariance and central data vectors (orange contours); {\it ii}) only those steps contained within $\chi^2\leq \chi^2_{\rm min}+14.16$ are used to compute the Gaussian covariance (green contours). The former case is the one used to compute the data-vector and covariance presented in Eq. \ref{datafullAPvalues} and \ref{covfullAPvalues} (without the diagonal systematic contribution). In this sense, Fig.~\ref{fig:gaussian_aprox} demonstrate the excellent agreement between the actual \textsc{mcmc} posterior likelihood surface and the reported Gaussian approximation.  In this case, full \textsc{mcmc} distribution do not present strong non-Gaussian tails, and consequently, the Gaussian predictions from {\it i}) and {\it ii}) are very similar.

\begin{figure}
\centering
\includegraphics[scale=0.5]{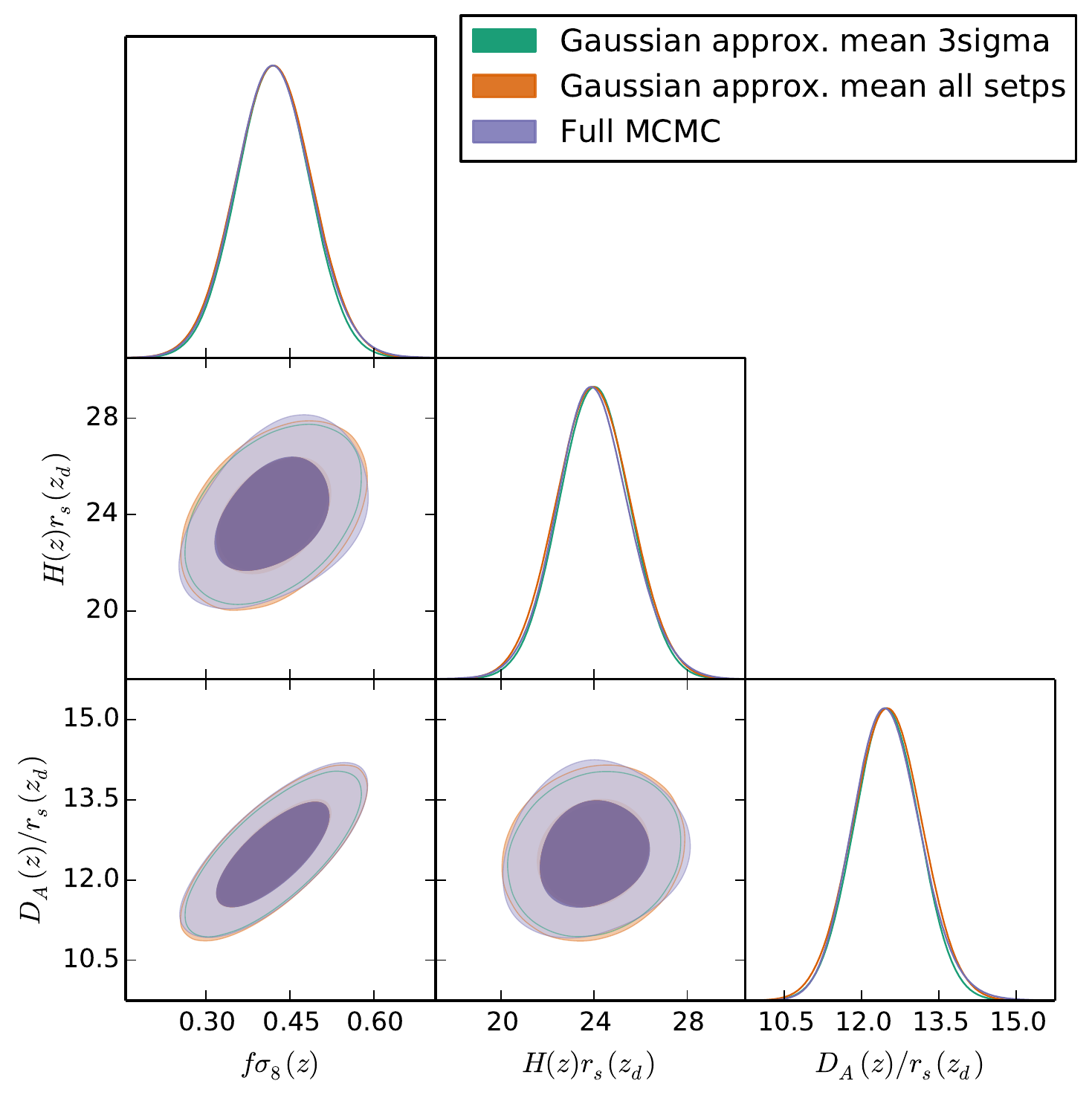}
\caption{Posterior likelihood for the cosmological parameters of interest corresponding the DR14Q dataset when the full redshift range, $0.8\leq z \leq 2.2$ is considered as a single redshift bin. The contours drawn from the full \textsc{mcmc} steps are represented in purple. In addition a Gaussian approximation to the full \textsc{mcmc} steps is also plotted when: {\it i}) all the \textsc{mcmc} step chains are used to compute the Gaussian covariance and central data vectors values; {\it ii}) only those steps within $\chi^2\leq \chi^2_{\rm min}+14.16$ are used to compute the Gaussian covariance parameters and data-vector values. The agreement between all three cases demonstrate the high degree of Gaussianity of the original sample drawn from the full  \textsc{mcmc} steps.  }
\label{fig:gaussian_aprox}
\end{figure}

\begin{figure*}
\centering
\includegraphics[scale=0.33]{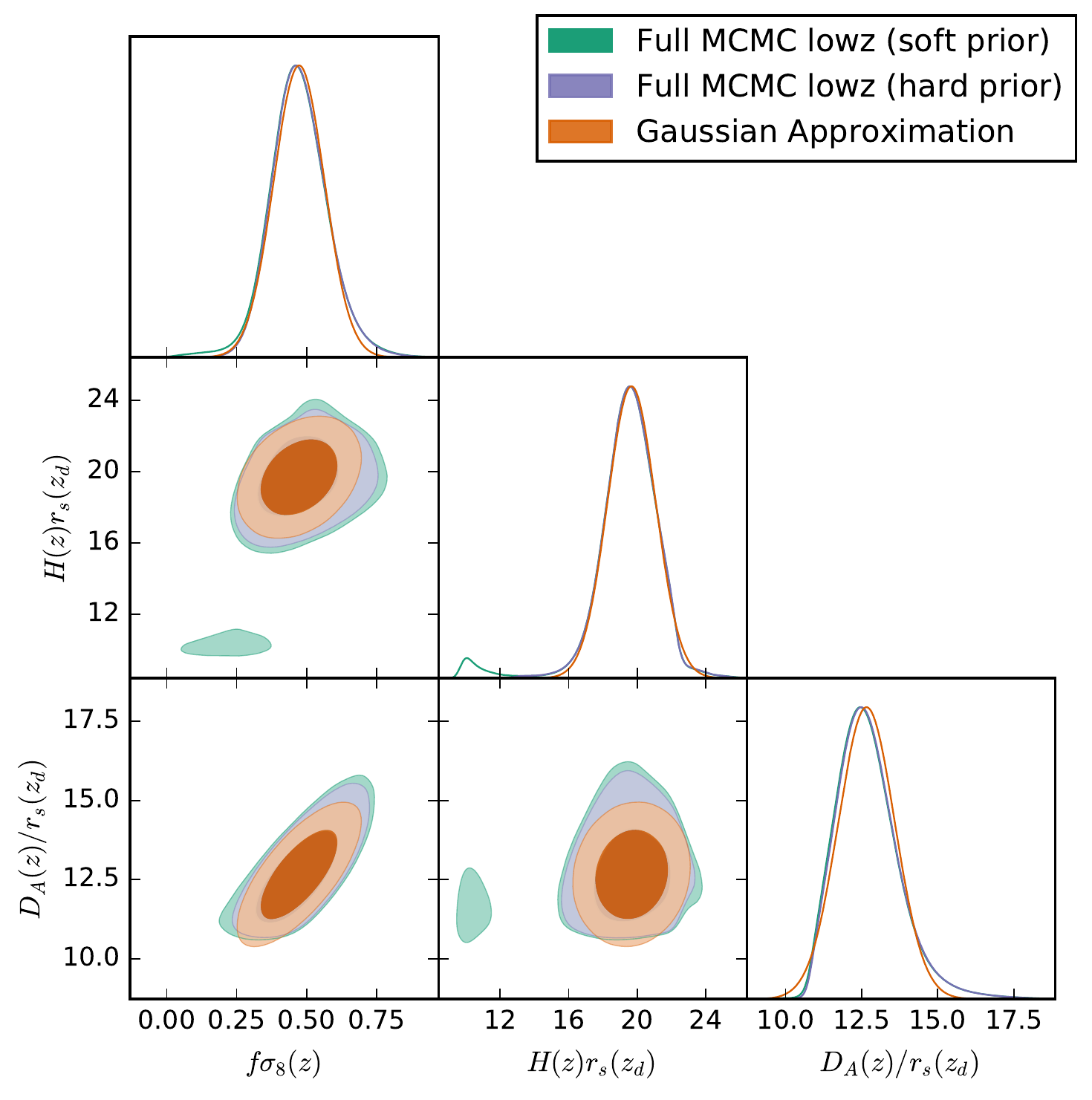}
\includegraphics[scale=0.33]{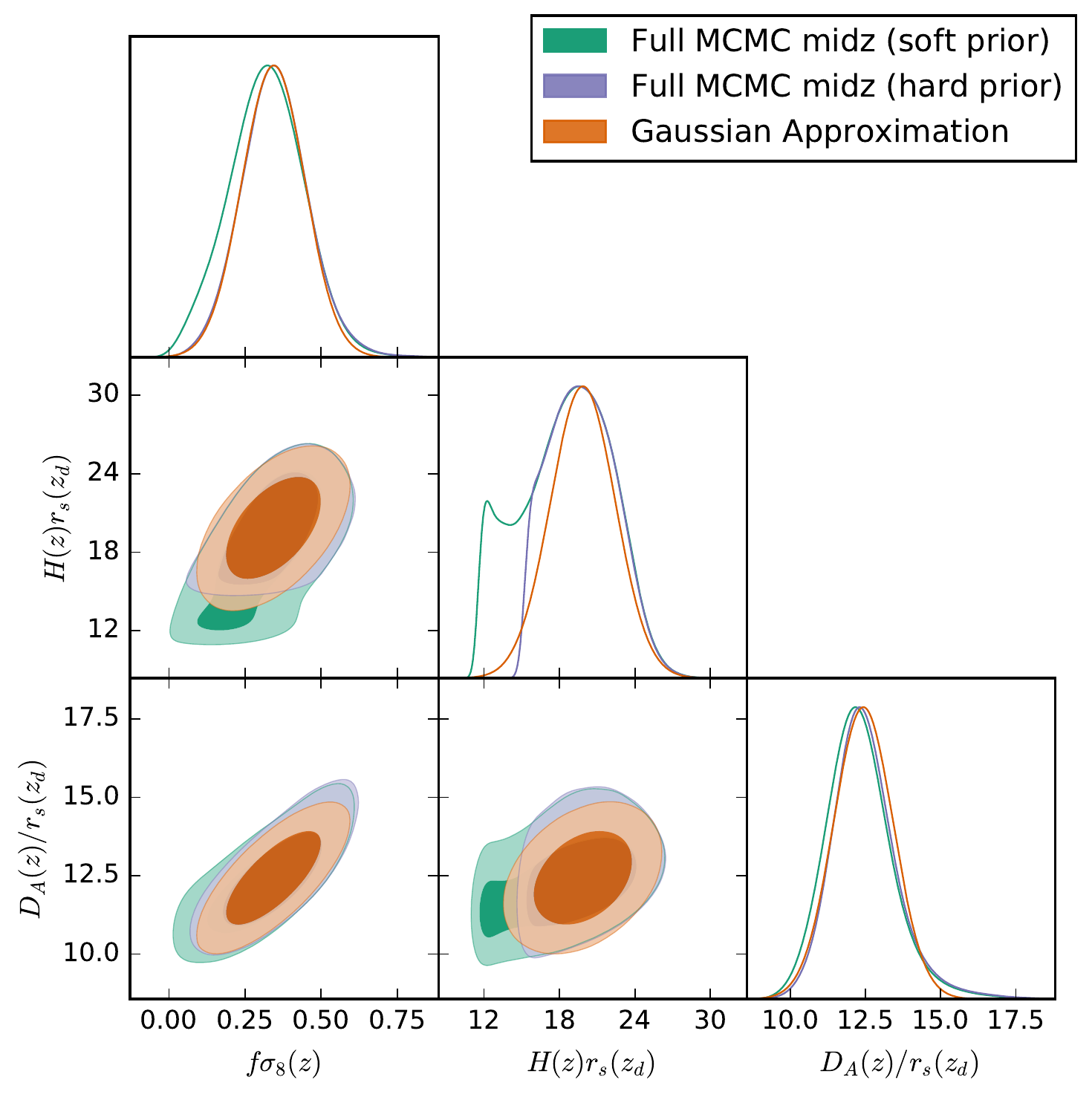}
\includegraphics[scale=0.33]{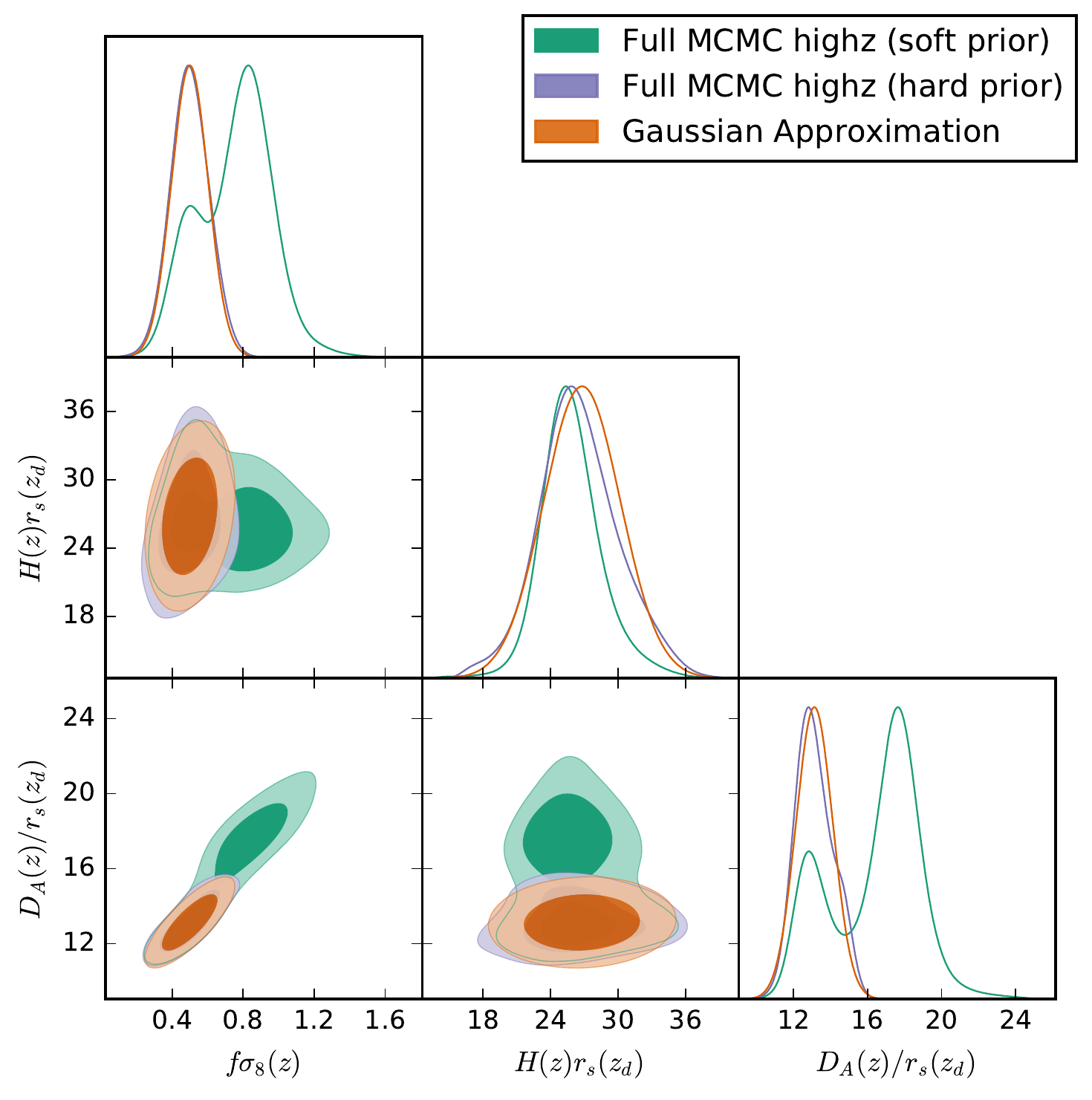}
\caption{Posterior likelihood for the cosmological parameters of interest corresponding the DR14Q dataset for the {\it lowz}, {\it midz} and {\it highz} redshift bins in the left, middle and right panels, respectively. Green contours display the results when a `soft' flat prior is applied  on the dilation scale factors: $0\leq \alpha_{\perp,\,\parallel}\leq 2$. In this case, secondary minima appear for some parameters, being the full distribution highly non-Gaussian. Purple contours display the results when a more restrictive `hard' flat prior (given by Table~\ref{table:priors}) is applied. In this case the secondary minima are cutoff and the \textsc{mcmc} steps describe a single-peaked distribution. Finally the orange contour display the Gaussian approximation following the approach described by {\it ii)} (see text) applied to the \textsc{mcmc} steps with the `hard' prior.   }
\label{fig:gaussian_aprox2}
\end{figure*}

On the other hand, Fig.~\ref{fig:gaussian_aprox2} display the posterior likelihood corresponding to \S\ref{sec:results_zbins}, when the DR14Q dataset is divided into three overlaping redshift bins: {\it highz} (right panel), {\it midz} (middle panel) and {\it lowz} (left panel). Green contours display the full \textsc{mcmc} steps when a very broad and flat prior is applied on $\alpha_\parallel$ and $\alpha_\perp$. We refer to this prior: $0\leq \alpha_{\perp,\,\parallel}\leq 2$ as 'soft prior'. On the other hand the purple contours result from applying the prior displayed by Table~\ref{table:priors} and we refer them as `hard prior'. These priors are defined to cutoff the secondary minima outside the range $0.8\leq \alpha_{\parallel,\,\perp}\leq 2.2$. The `hard prior' contours for the three redshift bins are over-plotted in Fig.~\ref{fig:cosmozbins}. Finally, the orange contours display the Gaussian approximation applying the {\it ii}) approach described above on the \textsc{mcmc} steps with the `hard prior' condition. These represent the covariance matrix given by Table~\ref{table:covariance}. Unlike the single bin case presented in Fig.~\ref{fig:gaussian_aprox}, the Gaussian approximation on the three overlapping redshift bins does not result in an excellent agreement. The reason is that when cutting off the dataset in three chunks, the errors increases and non-Gaussian tails and secondary minima appear as a result of shifting the BAO features into the noisy spectrum of the data. We envision that by the end of eBOSS, the data collected by the survey will be sufficiently large that these secondary minima will disappear without the necessity of applying these hard prior conditions. 

\bsp	
\label{lastpage}
\end{document}